\documentclass[twocolumn]{aastex631}

\usepackage{rotating}
\usepackage{ragged2e}
\usepackage{graphicx}
\usepackage{amsmath}
\usepackage{xurl}

\begin{document}

\title{Stellar Cycles in Fully Convective Stars and a New Interpretation of Dynamo Evolution}

\correspondingauthor{Zackery Irving}
\email{zi1g18@soton.ac.uk}

\author{Zackery A. Irving}
\affiliation{School of Physics and Astronomy, \\
University of Southampton, \\
University Road, Southampton SO17 1BJ, UK}
\affiliation{Center for Astrophysics | Harvard \& Smithsonian, \\
60 Garden Street, Cambridge, MA 02138, USA}

\author{Steven H. Saar}
\affiliation{Center for Astrophysics | Harvard \& Smithsonian, \\
60 Garden Street, Cambridge, MA 02138, USA}

\author{Bradford J. Wargelin}
\affiliation{Center for Astrophysics | Harvard \& Smithsonian, \\
60 Garden Street, Cambridge, MA 02138, USA}

\author{Jos\'e-Dias do Nascimento, Jr}
\affiliation{Center for Astrophysics | Harvard \& Smithsonian, \\
60 Garden Street, Cambridge, MA 02138, USA}
\affiliation{Univ. Federal do Rio G. do Norte, \\
UFRN, Dep. de F\'isica, \\
CP 1641, 59072-970, Natal, RN, Brazil}



\begin{abstract}

An $\alpha\Omega$ dynamo, combining shear and cyclonic convection in the tachocline, is believed to generate the solar cycle. However, this model cannot explain cycles in fast rotators (with minimal shear) or in fully convective stars (no tachocline); analysis of such stars could therefore provide key insights into how these cycles work. We reexamine ASAS data for 15 M dwarfs, 11 of which are presumed fully convective; the addition of newer ASAS-SN data confirms cycles in roughly a dozen of them, while presenting new or revised rotation periods for five. The amplitudes and periods of these cycles follow $A_{\rm cyc} \propto P_{\rm cyc}^{0.94 \pm 0.11}$, with $P_{\rm cyc}/P_{\rm rot} \propto {\rm Ro}^{-1.02 \pm 0.06}$ (where Ro is the Rossby number), very similar to $P_{\rm cyc}/P_{\rm rot} \propto {\rm Ro}^{-0.81 \pm 0.17}$ that we find for 40 previously studied FGK stars, although $P_{\rm cyc}/P_{\rm rot}$ and $\alpha$ are a factor of $\sim$20 smaller in the M stars. The very different $P_{\rm cyc}/P_{\rm rot}$-Ro relationship seen here compared to previous work suggests that two types of dynamo, with opposite Ro dependences, operate in cool stars. Initially, a (likely $\alpha^2$ or $\alpha^2\Omega$) dynamo operates throughout the convective zone in mid-late M and fast rotating FGK stars, but once magnetic breaking decouples the core and convective envelope, a tachocline $\alpha\Omega$ dynamo begins and eventually dominates in older FGK stars. A change in $\alpha$ in the tachocline dynamo generates the fundamentally different $P_{\rm cyc}/P_{\rm rot}$-Ro relationship.

\end{abstract}



\section{Introduction} \label{sec: intro}

Stellar cycles are (quasi-)periodic fluctuations in the intrinsic brightness of stars due to changing internal magnetic fields. Such cycles are most clearly manifest visually by the appearance of starspots, however are more easily observed by studying chromospheric ({Ca\,{\sc ii}} H and K) or coronal (X-ray) variability - both of which are more monotonically correlated with magnetic activity. Indeed, in the optical waveband the solar cycle has an amplitude of the order of just $0.1\%$, while the coronal (X-ray) variability varies by a factor of $\sim$6 \citep{Judge2003}. Sunspot observations first led to the discovery of the 11-yr solar cycle almost 200 years ago \citep{Schwabe1844}. Over half a century later, it was realised that sunspots are a manifestation of solar magnetic activity \citep{Hale1908, Hale1913}, and it is now understood that this 11-yr cycle is just one-half of the 22-yr polarity reversal cycle \citep{Hale1919, Babcock1959, Babcock1961}. However, while the solar cycle has an average period of 11 years, due to the quasi-periodic nature of stellar cycles it is not uncommon for a given cycle to have a period between 9-13 years \citep[see, for example,][]{Donahue1992}.

Presently, stellar cycles are believed to be a result of dynamo processes \citep[see, for example,][]{Parker1955, Baliunas1995, Dikpati1999, Kitchatinov1999}. A dynamo process is a physical mechanism in which kinetic energy is converted into magnetic energy via inductive effects of motions in an electrically conducting fluid, in a self-regenerating process. The solar cycle can be well described by an $\alpha \Omega$ dynamo, where shearing (due to differential rotation) transforms a poloidal field into a toroidal one ($\Omega$ effect), and cyclonic convection (due to Coriolis and magnetic forces) then restores the poloidal field with reversed polarity ($\alpha$ effect) \citep{Parker1955, Babcock1961, Leighton1969}. Comparing our Sun's cycle to the cycles of 4454 cool stars, \cite{Saikai2018} found that the solar dynamo cycle is not uncommon. Many of the finer details of these dynamo processes, however, remain poorly understood - even in the case of our Sun \citep[see, for example,][]{Ossendrijver2003, Charbonneau2010}.

Differential rotation (DR) is a key component of the solar $\alpha \Omega$ dynamo model, with current models suggesting that the $\Omega$ effect takes place in the tachocline layer at the base of the convective envelope \citep[e.g.,][]{Gilman1997, Dikpati1999}.  In the case of our Sun, DR causes the equator to rotate more quickly than the poles; this is referred to as solar-like DR. Some theoretical work \citep[e.g.,][]{Kitchatinov1999} and observations \citep[e.g.,][]{Barnes2005} suggest that DR has little or no dependence on rotation, while other observations indicate that the two are closely tied \cite[e.g.,][]{Donahue1996}. \cite{Saar2011} updated the work of \cite{Donahue1996} and confirmed that DR appears to scale almost linearly with rotation period, at least for relatively slow rotators. \cite{Saar2011} further argued that the results of \cite{Barnes2005} were misleading as they included close binaries, where large tidal forces can freeze out DR and cause solid-body
rotation.

For stars with very slow rotation, numerical simulations \citep[e.g.,][]{Karak2020, Kapyla2021} suggest that DR should be antisolar, where the poles rotate more quickly than the equator, although the point of transition from solar to antisolar DR is still debated.  Antisolar DR in the classical $\alpha \Omega$ dynamo model does not produce polarity reversal, and so such stars would not be expected to have solar-like cycles.  Nonetheless, there is strong evidence that some slowly rotating stars {\em do} have cycles \citep[see, for example,][]{S-M2016}.

Whether or not DR depends on rotation rate, dynamo processes in more rapidly rotating stars are expected to differ fundamentally from those in the Sun, possessing $\alpha^2 \Omega$ or even $\alpha^2$ dynamos \citep[e.g.,][]{Kitchatinov1999}, where the dynamo actions originate from surface shear or in the convection zone, respectively. \cite{Kitchatinov1999} suggested that these $\alpha^2$ dynamos should not result in cyclic activity, and could explain why \cite{Baliunas1995} found an absence of activity cycles in rapidly rotating young stars. However, more recent studies \citep[e.g.,][]{S-M2016, Lehtinen2016} have shown that activity cycles are observed in fast and slow rotating stars alike. In addition, it is difficult to explain the observed increase in magnetic activity with rotation under the assumption that DR is constant \citep[e.g.,][]{Durney1993, Saar2011, Wright2011}. It should be noted, however, that DR studies can only measure surface DR, not DR in the tachocline, and there is no guarantee that the two are the same.

A convective envelope is another key ingredient of the $\alpha \Omega$ dynamo. Convective envelopes surrounding a radiative core are a staple of cool stars earlier than type $\sim$M3.5, while late M-type stars are believed to be fully convective \citep{Chabrier1997}; as such, these stars cannot support solar-like $\alpha \Omega$ dynamos (since they have no tachocline). Instead, late M-type stars are expected to support $\alpha^2$ dynamos, which, as already discussed, are not expected to exhibit activity cycles. However, there is some evidence suggesting that $\alpha^2$ dynamos may exhibit activity cycles under certain conditions \citep[e.g.,][]{Rudiger2003, Gastine2012, Kapyla2013}. In addition to this, activity cycles have also been observed in several stars which are believed to be fully convective \citep[see, for example,][]{S-M2016, Wargelin2017}.

Interestingly, there is evidence suggesting that stellar activity cycle periods are (weakly) correlated with rotation period \citep[see, for example,][]{Noyes1984, Brandenburg1998, Saar1999, B-V2007, S-M2016, Saikai2018, Wright2018} - M-type stars included. In their two paper series, \cite{Brandenburg1998, Saar1999} found that FGK stars typically occupy one of three branches: termed the ``superactive" (S), ``active" (A), and ``inactive" (I) branches, with our Sun residing approximately between the A and I branches. \cite{Saar1999} also found hints of a tentative fourth, ``transitional" (T), branch connecting the A and S branches. Recent studies \citep[e.g.,][]{Lehtinen2016, Distefano2017, Saikai2018} have found further evidence for the existence of this T branch, while raising doubts about the A branch, which has become increasingly less clear as more cycles have been found.

To develop our understanding of the dynamo processes responsible for stellar activity cycles, active stars with a wide range of physical parameters should be observed. Active late M-type stars are of particular interest since their cycles cannot be adequately explained by current dynamo models. Due to how intrinsically faint M stars are, however, they were underrepresented in earlier surveys \citep[e.g.,][]{Wilson1978, Baliunas1995}. Furthermore, even in recent studies \citep[e.g.,][]{S-M2016, Wright2018}, rapidly rotating M stars are still poorly represented. Rapidly rotating late M stars could be especially revealing, as their deep convective zones combined with short rotation periods may allow for unique dynamos.

Herein, we reexamine the 15 latest type stars listed in \cite{S-M2016} (excluding GJ 526, discussed below), who analyzed $\sim$9 years of ASAS-3 \citep{Pojmanski1997, Pojmanski2002} photometric monitoring data and suggested that 13 of them, several of which are fully convective, showed evidence for cyclic activity. We use those same data and include new data from ASAS-SN \citep{Shappee2014, Kochanek2017}, approximately doubling the period of monitoring; in some cases, we also include TESS data to constrain shorter rotation periods. Note, however, that there is some uncertainty regarding the spectral types, and some stars may not be fully convective. GJ 526, for example, is listed in \cite{S-M2016} as M4, but M2 in SIMBAD and \cite{S-M2015}. Moreover, its V--K$_{\rm S}$ color (see Section \ref{sec: determining periods}) also indicates it is earlier than M4. In any case, GJ 526 is often too bright for ASAS-SN monitoring (g=7.6). A few stars may also be unresolved binary systems (see Section \ref{sec: contamination}), further complicating spectral type determinations, but roughly a dozen of our stars are almost certainly fully convective (see Section \ref{sec: convectiveness}).

While photometric measurements are not as tightly correlated with magnetic activity as other metrics (e.g. chromospheric line intensities and coronal X-ray emission), and can suffer from counterbalancing effects between starspots and the surrounding faculae, they are technologically easier to make. Furthermore, there is evidence suggesting that in active, rapidly rotating stars, chromospheric emission can become saturated \citep[e.g.,][]{Vilhu1984}, while starspot (and hence photometric) variability can continue to rise with increasing rotation \citep[e.g.,][]{Odell1995, Krishnamurthi1998}.

In Section \ref{sec: method}, we discuss the data used and the analysis performed to identify rotation periods and activity cycles in 15 M-type stars from their photometric time series. Results are presented in Section \ref{sec: results}, and our interpretation of these results is provided in Section \ref{sec: discussion}. Our final thoughts are then given in Section \ref{sec: conclusions}.

\section{Method}\label{sec: method}

\subsection{Data Collection}\label{sec: data collection}

ASAS-3 optical photometry data were downloaded from the ASAS All Star Catalogue.\footnote{\url{http://www.astrouw.edu.pl/asas/?page=aasc}} All ASAS-3 data use V-band filters, and individual measurements are ascribed one of four grades: A - best data; B - mean data; C - A and B with not measured indication; D - worst data. We opted to use only measurements with an A or B grade. For most stars in our sample, the MAG\_2 (four pixel) aperture data had the least scatter, and therefore the lowest empirical uncertainty; to be consistent, we therefore used this aperture for all our stars. We note, however, that the choice of aperture can sometimes have a significant effect on the inferred cycle period, and could explain (at least partly) why our results sometimes differ from those of \cite{S-M2016}.\footnote{We further note that the aperture columns are sometimes arranged inconsistently between different stars, and so care needs to be taken to ensure the same aperture is being used for all stars.} ASAS-3 has a FWHM PSF of typically 23", corresponding to approximately 1.5 pixels.

ASAS-SN optical photometry data were collected using the light curve server.\footnote{\url{https://asas-sn.osu.edu}} Early ASAS-SN data were gathered using V-band filters, while the most recent data use g-band. The ASAS-SN light curve server does not perform proper motion corrections, unlike the ASAS All Star Catalogue, so proper motion corrected coordinates had to be specified for each query. ASAS-SN has a 16" FWHM PSF, so we allowed for a proper motion of up to 1" either side of the input coordinates; extractions use a two pixel (16") aperture. We then compiled these proper motion corrected light curves into single light curves for each star once we had downloaded all available data.

TESS optical photometry data were downloaded using the \textsc{Lightkurve} Python package \citep{Lightkurve2018, astropy2018, astrocut2019, astroquery2019}, which accesses publicly available data directly from the Mikulski Archive for Space Telescopes (MAST). Only data products produced via their Science Processing Operations Center Science Analysis Pipeline \citep{Jenkins2016} were used. For each star, we also used the pipeline aperture.

\subsection{Data Preparation}\label{sec: data prep}

\subsubsection{ASAS-SN}\label{sec: ASAS-SN prep}

Initially, the early V-band and recent g-band data were disjointed in the compiled ASAS-SN light curves. We therefore needed to color correct these data. Using the empirical UBVR-uvgr relation found by \cite{Kent1985}, and improved by \cite{Windhorst1991}, we can convert from V- to g-band magnitudes using:
\begin{equation}
    {\rm V} = {\rm g} - 0.03 - 0.42({\rm g-r}),
    \label{eq: colour correction}
\end{equation}
where V is the V-band magnitude, g is the g-band magnitude, and ${\rm g-r}$ is the g--r color index, given by:
\begin{equation}
    {\rm g-r} = 1.02({\rm B-V}) - 0.22,
\end{equation}
where ${\rm B-V}$ is the B--V color index. However, M stars are relatively cool objects, and emit mostly towards the red end of the optical spectrum; as such, the B--V color index is not the best diagnostic for these stars. In addition, the relation given in Equation \ref{eq: colour correction} was derived using predominantly O--K stars. After color correcting V-band data using these formulae, residual offsets were therefore required. These offsets were computed by comparing the means of overlapping g- and V-band data.

Further internal cross-calibration is then required because ASAS-SN data are collected using multiple telescopes positioned around the world, with small but sometimes significant differences between nominally identical instruments. The ASAS-SN pipeline automatically adjusts for these differences, but color-dependent residuals may remain, as we found for our reddish M stars.

To correct for differences between different ASAS-SN telescopes, we first selected the telescope with the largest number of measurements, thus defining our reference data set. We then calibrated the remaining telescopes sequentially in descending order of relative overlap. To calibrate a telescope, we applied an offset such that the averages of overlapping data points between the telescope and the reference data set were equal. Once a telescope had been calibrated, it then became part of the reference data set for subsequent calibrations. Fortunately, we did not find any instances of telescopes with no overlapping data. However, if a particular telescope was found to show unusual behavior within a given light curve (relative to the other telescopes) then this telescope's data were removed. For example, telescope ``bA" showed an unusually large scatter for GJ 447, obscuring its cyclic modulations, but appeared perfectly normal for GJ 581, GJ 729, and GJ 849; this telescope's measurements were therefore removed in the former case, but kept in the latter three.

\subsubsection{TESS}\label{TESS}

Uncorrected TESS light curves are dominated by scattered light background caused by spacecraft motions. To reduce this background, we used \textsc{Lightkurve} to perform pixel level decorrelation (PLD), which has been shown to be effective at removing systematic noise from both Spitzer \citep{Deming2015} and K2 \citep{Luger2016, Luger2018} data. PLD works by creating masks of the object of interest and the surrounding background, and then using linear regression to model trends in the background. After subtracting this noise model from the uncorrected light curve, systematic noise is greatly reduced.

\subsubsection{Outlier Removal}\label{sec: outliers}

After performing the analysis described above on ASAS-SN and TESS data, we then removed outliers from all our data sets. For ASAS-3 data, we clipped each observing season to remove measurements more than two standard deviations from the mean.

For ASAS-SN data, we initially separated the data by telescope (and consequently, by filter). We then binned each filter's telescope-separated data according to the star's rotation period (since observing seasons were not always well-defined when data were separated by telescope). ASAS-SN typically makes between 1--3 observations of each region of the visible sky each night (3 being the nominal number); binning ASAS-SN data into bins of less than 2d therefore becomes volatile and unreliable due to the small number of data points per bin. Consequently, we chose to bin stars with rotation periods of less than 24d differently, instead using bin sizes of 10$P_{\textrm{rot}}$.\footnote{For a star with $P_{\rm rot}=24$d, this is approximately equal to an observing season.} In all other cases, we used the rotation period. We then clipped these data using a tolerance of two standard deviations from the mean. Next, we repeated this process using bin sizes of $P_{\rm rot}$ if $P_{\rm rot} \leq 24$d, or $P_{\rm rot}/12$ otherwise, with a tolerance of two and a half standard deviations from the mean (note we used a looser tolerance for these smaller bins).

After removing outliers from the telescope-separated ASAS-SN data, we recalibrated these telescopes and recorrected any residual offset between the V- and g-band data resulting from outlier removal. We then binned the entire calibrated light curve using bin sizes of $P_{\rm rot}$ if $P_{\rm rot} \leq 24$d, or $P_{\rm rot}/12$ otherwise, and removed measurements more than two and a half standard deviations from the mean. In total, this usually resulted in $\sim$10\% of the data being removed. Finally, we recalibrated these telescopes again and recorrected any residual offset between the V- and g-band data. 

Since TESS data have such high cadence, we simply binned these data into 0.1d bins and removed outliers using a tolerance of two and a half standard deviations from the mean.

\subsection{Binary Systems and Crowded Fields}\label{sec: contamination}

When assessing our results it is important to keep in mind that some target-star intensity extractions may suffer from contamination by binary companions or other stars that appear nearby in the sky. \cite{Winters2019} list data from an all-sky volume-limited (25pc) survey for stellar companions to 1120 M-dwarf primaries, and serves as our principal reference regarding binarity since it includes all our stars of interest except for Proxima (GJ 551). Two systems are listed as binaries: GJ 234 and GJ 896.

The GJ 234 pair has a semi-major axis of 1.1" \citep[e.g.,][]{Gatewood2003, Kervella2019} but contamination by the secondary is small, with estimated intensity ratios of 100:5.3 in the B band \citep{Gatewood2003}, 100:5.9 in V \citep{Henry1999}, and 100:25 in K \citep{Coppenbarger1994}.  GJ 896 A and B had a separation of 5.35" in 2004 \citep{Winters2019} with an estimated V magnitude of 10.29 for component A and 2.12 magnitudes dimmer for B (100:14 ratio). \cite{Davison2015} measured a separation of 7" and list the spectral types as M3.5+M4.0. These stars are mostly blended in ASAS and ASAS-SN data, and it is difficult to determine if the lack of an obvious cycle in GJ 896A is intrinsic, from contamination by the secondary star, or variable extraction efficiencies as the pair's center of emission moves around its center of mass and proper motion.

The multiplicity of GJ 273 is somewhat unclear, with \cite{Vrijmoet2022} listing it as a double based on Gaia data, while \cite{Winters2019} conclude that the putative companion is probably an unassociated background object. \cite{Astudillo-Defru2017} analyzed HARPS RV data and report that GJ 273 has four planets, without any mention of spectral evidence for a companion; in our analysis, we treat this star as a single.

Although GJ 273, along with GJ 317 and GJ 406, were not included in their study of binarity in nearby stars using data from HIPPARCOS and Gaia DR2, \cite{Kervella2019} calculated minimum (inclination dependent) masses for companions to the rest of the stars in our sample.  Apart from the two known binary systems, minimum masses were no larger than $0.25 M_{J}$ assuming 0 inclination and orbital radii of $r=1$ AU, with mass scaling as $\sqrt r$.  All our stars have clear rotational modulation, indicating that inclination cannot be too small (assuming rotational and orbital co-planarity), so it seems unlikely that any of the stars under study here (apart from GJ 234 and GJ 896) have stellar mass companions.  Many of these stars are also the subjects of RV studies, which would readily reveal binarity unless the orbital inclination was very small.  We particularly mention GJ 358 because of its remarkable changing cycle periods: \cite{Bonfils2013} explicitly treat this star as a single after excluding known and discovered spectroscopic binaries and visual pairs from their analysis of HARPS data.

Intensity extractions can also be affected by stars that appear nearby on the sky even if they are not physically associated.  GJ 406 has a very large proper motion (4.72"/yr) and has recently passed quite close to two objects that may provide contaminating flux. Likewise, Proxima lies in a very crowded part of the sky and is currently moving at 3.86"/yr within a group of stars that, although dimmer, may be bright enough to cause significant contamination within the ASAS-3 and ASAS-SN extraction regions. Those effects should be slowly varying but complicate interpretation of observations and may be at least partly responsible for the general brightening seen in Proxima's ASAS and ASAS-SN data.

\subsection{Separating Partially from Fully Convective Stars}\label{sec: convectiveness}

A further consideration to make when assessing our results is which stars are fully convective. As mentioned in Section \ref{sec: intro}, spectral type is often used to distinguish partially from fully convective M stars. However, spectral type determinations are not always a robust metric, and the boundary between sub-types is often blurred. Indeed, some of these difficulties were mentioned in Section \ref{sec: intro}, and discussed in detail in Section \ref{sec: contamination}. Fortunately, recent studies have alleviated some of the ambiguity in identifying fully convective M stars.

Using data from Gaia DR2, \cite{Jao2018} identified a gap in the color-magnitude diagram for the lower main sequence, which they proposed could be used to distinguish partially from fully convective M stars. \cite{vanSaders2012} first predicted this gap as the result of nonequilibrium $^3$He fusion prior to stars becoming fully convective; using numerical simulations, \cite{Feiden2021} were able to reproduce this gap. Using this distinction, specifically in the V--K$_{\rm S}$~vs.~M$_V$ plane, we identify 11 of our 15 M stars as being fully convective (see Table \ref{tab: optical cycle periods}). We note, however, that for a few high proper motion stars not included in Gaia DR2, we had to use DR3 parallaxes.

\subsection{Time Series Analysis Techniques}\label{sec: time series analysis techniques}

Herein, we predominately use traditional time series analysis techniques, specifically Lomb--Scargle (L--S) periodograms \citep{Lomb1976, Scargle1982} for two key reasons. Firstly, more advanced techniques, such as Gaussian processes (GPs), come with a significant computational expense. Depending on the number of covariance function hyperparameters, the number of mean function parameters, the priors, and the size of the data set itself, fitting a GP model can easily take several orders of magnitude times the time it takes to obtain an L--S periodogram. Secondly, as we discuss later, the literature shows that stellar cycles often appear to be well approximated by sines. We note, however, two exceptions: GJ 358 and GJ 551 (Proxima), whose cyclic modulations cannot adequately be described by one or two single-period sinusoids; these stars are therefore better suited to a GP analysis.

To compute the false alarm probabilities (FAPs) on our detections, we use the approximation proposed by \cite{Baluev2008}, often referred to as the Baluev estimate. We compared this method with a more robust bootstrap simulation method for half a dozen ASAS-3 and ASAS-SN light curves, and found that results were usually within 10\%, and many times within 5\%; the largest difference we found was $ {\rm FAP}_{\rm Baluev}:{\rm FAP}_{\rm bootstrap} \approx 0.79$.\footnote{Note, however, this result was obtained for GJ 285's ASAS-SN light curve, where the peak in the periodogram was an order of magnitude above either method's 0.1\% FAP power threshold.} Due to the computational expense of bootstrap simulations, without much improvement to performance, we prefer the Baluev estimate. In our work, we consider a FAP of $\leq 0.1\%$ to constitute a significant detection, the same threshold used in \cite{S-M2016}.

In theory, typical time series analysis techniques, such as fast Fourier transforms (FFTs) and L--S periodograms, may not necessarily be the best tools for the analysis of stellar cycles. Both are limited in that they model signals with sinusoids; while this may be a good model for idealized systems, it often fails in reality. Indeed, stellar cycles need not be sinusoidal in nature, nor strictly periodic.

To identify stellar cycles in unevenly sampled light curves, \cite{Olspert2018a} introduced a Bayesian generalized Lomb--Scargle periodogram with trend (BGLST); using synthetic quasi-periodic data, they showed that their BGLST performed better than the standard L--S periodogram. Moreover, in their follow-up paper, \cite{Olspert2018b} further showed that GPs can be used effectively in the analysis of stellar cycles - in their case using {Ca\,{\sc ii}} H and K data.

The greatest benefit of GPs over traditional time series analysis techniques, such as the L--S periodogram, is their flexibility. Much like L--S periodograms, GPs can be used with unevenly sampled data. However, GPs need not assume the data contain a periodic, sinusoidal signal. The assumptions of a GP are encoded in its mean and covariance functions, which, in theory, allow for any number of arbitrary assumptions to be made.

\cite{Olspert2018b}, in their study of stellar cycles, compared the performance of a Bayesian harmonic (i.e. sinusoidal) regression model, and GP models with periodic and quasi-periodic covariance functions. They found that where traditional methods suggest double cycles cycles, their quasi-periodic GP model often found only a single cycle; as the assumption of a quasi-periodic signal has more physical justifications, \cite{Olspert2018b} concluded that double cycles are rarer than initial results \citep[e.g.,][]{Wilson1978, Baliunas1995, Brandenburg1998, Saar1999} suggest.

However, while a quasi-periodic model is more physically justified, its results are often not much of an improvement over a simple harmonic model. \cite{Olspert2018b} identified harmonic signals in the light curves in 36 of stars; of these 36 stars, quasi-periodic signals were also identified in 26. Comparing these 26 quasi-periodic periods with their harmonic equivalents, 22 were comfortably within $3 \sigma$. This suggests that while stellar cycles need not be strictly periodic, nor sinusoidal, they can often be well approximated by sines.

\subsection{Determining Periods}\label{sec: determining periods}

\subsubsection{Rotation Periods from Photometric Time Series}\label{sec: optical rotation method}

\begin{figure}
    \centering
    \includegraphics[width=\columnwidth]{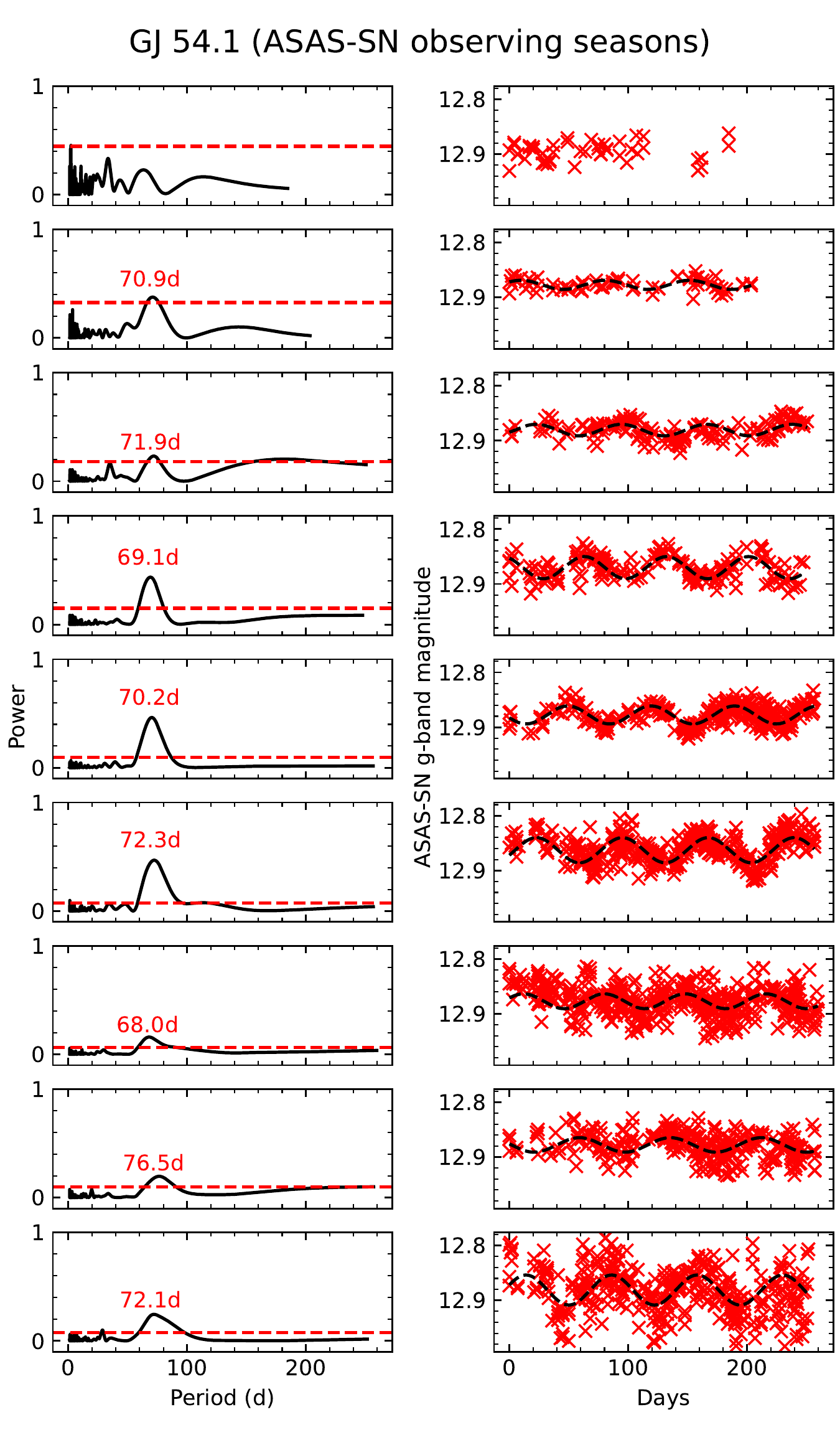}
    \caption{L--S periodograms (left) computed from ASAS-SN observing seasons (right) on GJ 54.1. Dashed red lines (left) show the power thresholds for FAPs of 0.1\%. Dashed black lines (right) show optimized sine fits using the period of the tallest peak in the corresponding periodogram.}
    \label{fig: GJ 54.1 observing seasons}
\end{figure}

To determine rotation periods we used L--S periodograms, as well as data from both ASAS-SN and TESS. ASAS-3 data are noisier and lower cadence than ASAS-SN data, and therefore provide no additional benefit for rotational studies; for these reasons, ASAS-3 data are not included. The first step in our analysis was to separate the ASAS-SN data by observing season. We then computed L--S periodograms for each of these seasons, and, provided the period of the peak was greater than 15d, used the tallest peak in each season's L--S periodogram to fit a sine model using the Levenberg--Marquardt (L--M) algorithm \citep[a sophisticated non-linear least-squares fitting algorithm, see][]{More1977} - as shown in Figure \ref{fig: GJ 54.1 observing seasons} for GJ 54.1. Note that there were often no significant peaks at $P_{\rm rot}/2$.

We discarded periodograms computed from seasons with low data density, or incomplete seasons,\footnote{In some cases, incomplete seasons were included if multiple rotations were clearly shown.} and identified significant peaks in each periodogram. After disregarding anomalous peaks, like the broad peak at approximately 180d in the third row of Figure \ref{fig: GJ 54.1 observing seasons}, we inferred the rotation period by computing the mean period of the remaining peaks, and estimated the uncertainty as the standard deviation on this mean.

However, if the rotation period of a star was sufficiently small (below $15$d), we used the higher cadence but shorter duration TESS data (TESS observation windows typically span $\sim$27d) to better constrain this period. If a star had multiple TESS observations, we found that in every case the inferred periods were identical to within the (tightly constrained) uncertainties.

\begin{figure}
    \centering
    \includegraphics[width=\columnwidth]{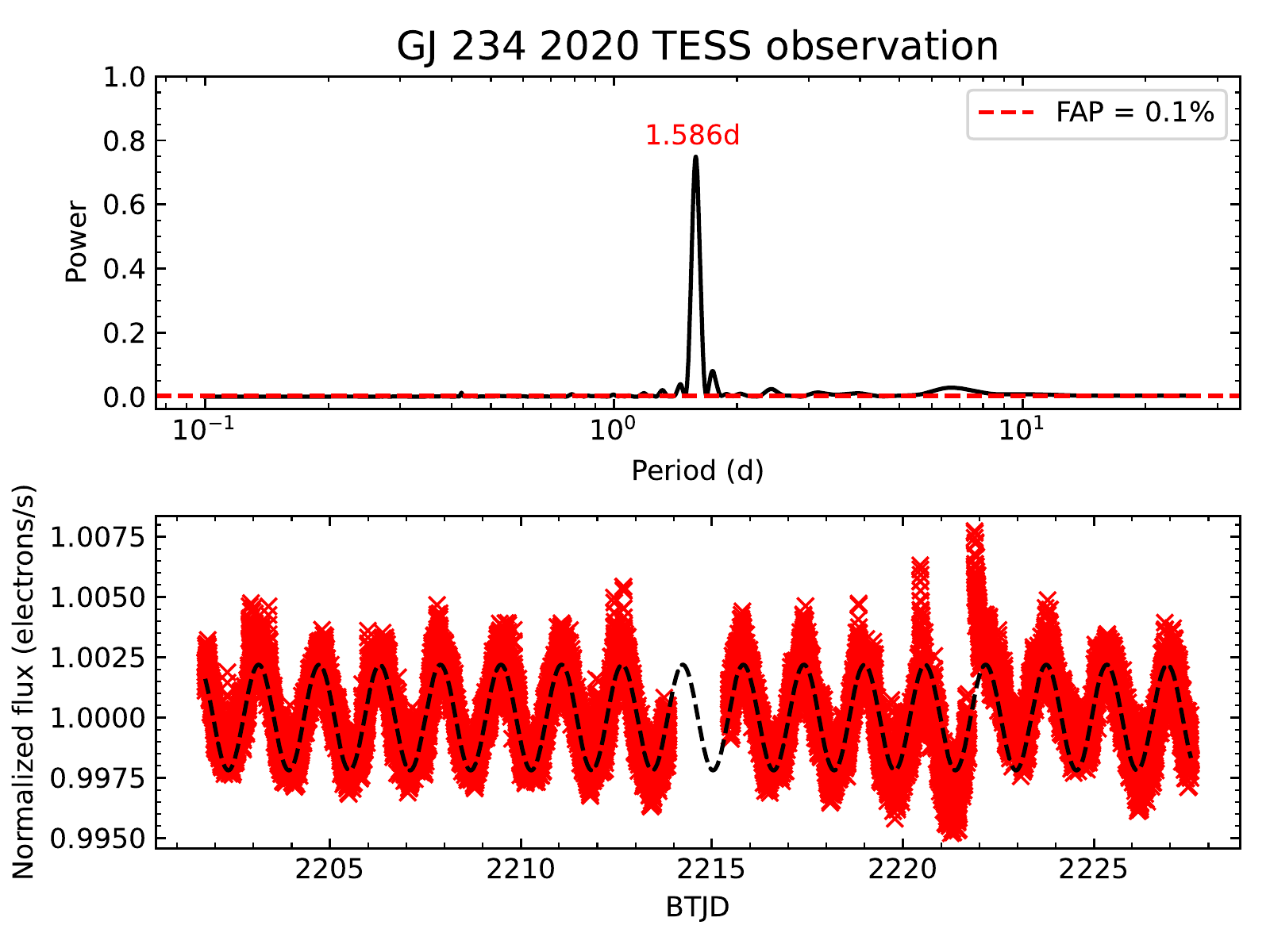}
    \caption{L--S periodogram (top) computed from the 2020 TESS observation of GJ 234 (bottom). The dashed red line (top) shows the power threshold for a FAP of 0.1\%; dashed black line (bottom) shows the optimized sine fit using the period of the tallest peak in the periodogram.}
    \label{fig: GJ 234 TESS}
\end{figure}

Our analysis using TESS data proceeded in much the same way as our analysis using ASAS-SN data; an example is shown in Figure \ref{fig: GJ 234 TESS} for GJ 234. Due to our currently limited understanding of TESS systematics, however, the identification of rotation periods is restricted to below approximately half a TESS observing window (hence our 15d cutoff) - see, for example, \cite{Anthony2022, Claytor2022}.

To infer the rotation period of a star from its TESS light curve, we fit a Gaussian to the tallest peak in its periodogram. Our estimate for the rotation period was thus the mean of this Gaussian, with the uncertainty given by its standard deviation. Since we were often limited to one or two TESS observations, following the same procedure as described above for ASAS-SN resulted in unrealistically small errors in some cases, and no errors in others; we therefore adopted this alternative method. We note that when we tested this approach on ASAS-SN data, it resulted in larger uncertainties due to the broad peaks in these periodograms. However, TESS observations only span $\sim$27d, and so these broadening effects (e.g., resulting from differential rotation and multiple star spot generations) are minimal. TESS is also a space-based observatory, with a much higher signal-to-noise ratio than ASAS-SN. Given these differences, we deem this alternative method to be appropriate. Results are presented in Table \ref{tab: optical cycle periods} and discussed in Section \ref{sec: results}.

\subsubsection{Cycle Periods from Photometric Time Series}\label{sec: optical cycle method}

To determine cycle periods we typically use L--S periodograms incorporating data from both ASAS-3 and ASAS-SN; exceptions to this are Proxima (for which we also have ASAS-4 data and use GPs), and GJ 358 (for which we also use GPs) - see Section \ref{sec: time series analysis techniques} for more details.

\begin{figure}
    \centering
    \includegraphics[width=\columnwidth]{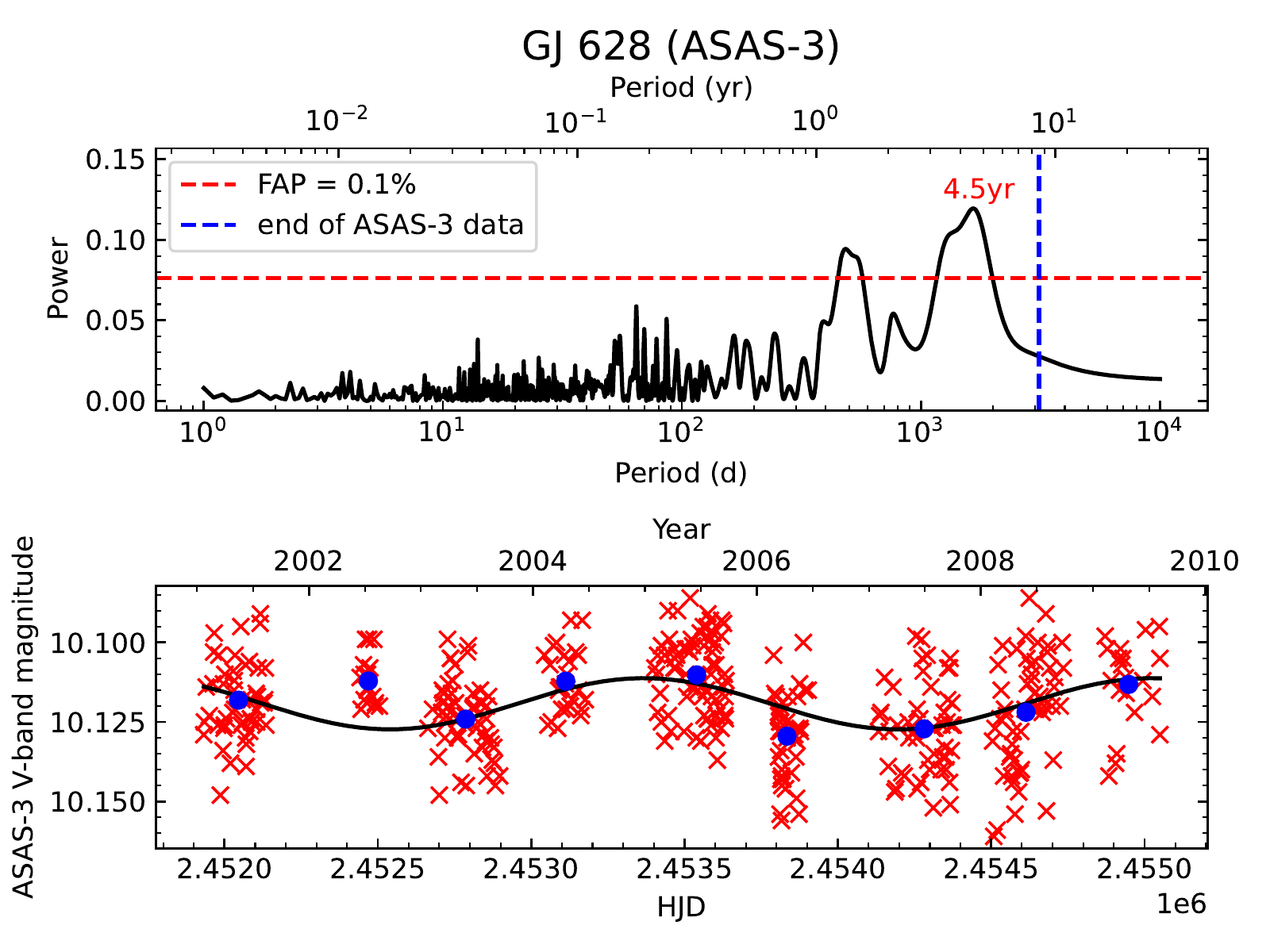}
    \caption{L--S periodogram (top) computed from ASAS-3 data on GJ 628 (bottom). The FAP $=0.1\%$ power threshold is shown by the dashed red line, while the dashed blue line shows the time-span of the data. The blue markers show the mean values (dates and magnitudes) for each observing season. The period of the labeled peak in the periodogram was used to create the sine fit in the bottom plot. The complete figure set (16 images) is available in the online journal (Figure Set 1).}
    \label{fig: GJ 447 ASAS3 sine fit}
\end{figure}

\begin{figure}
    \centering
    \includegraphics[width=\columnwidth]{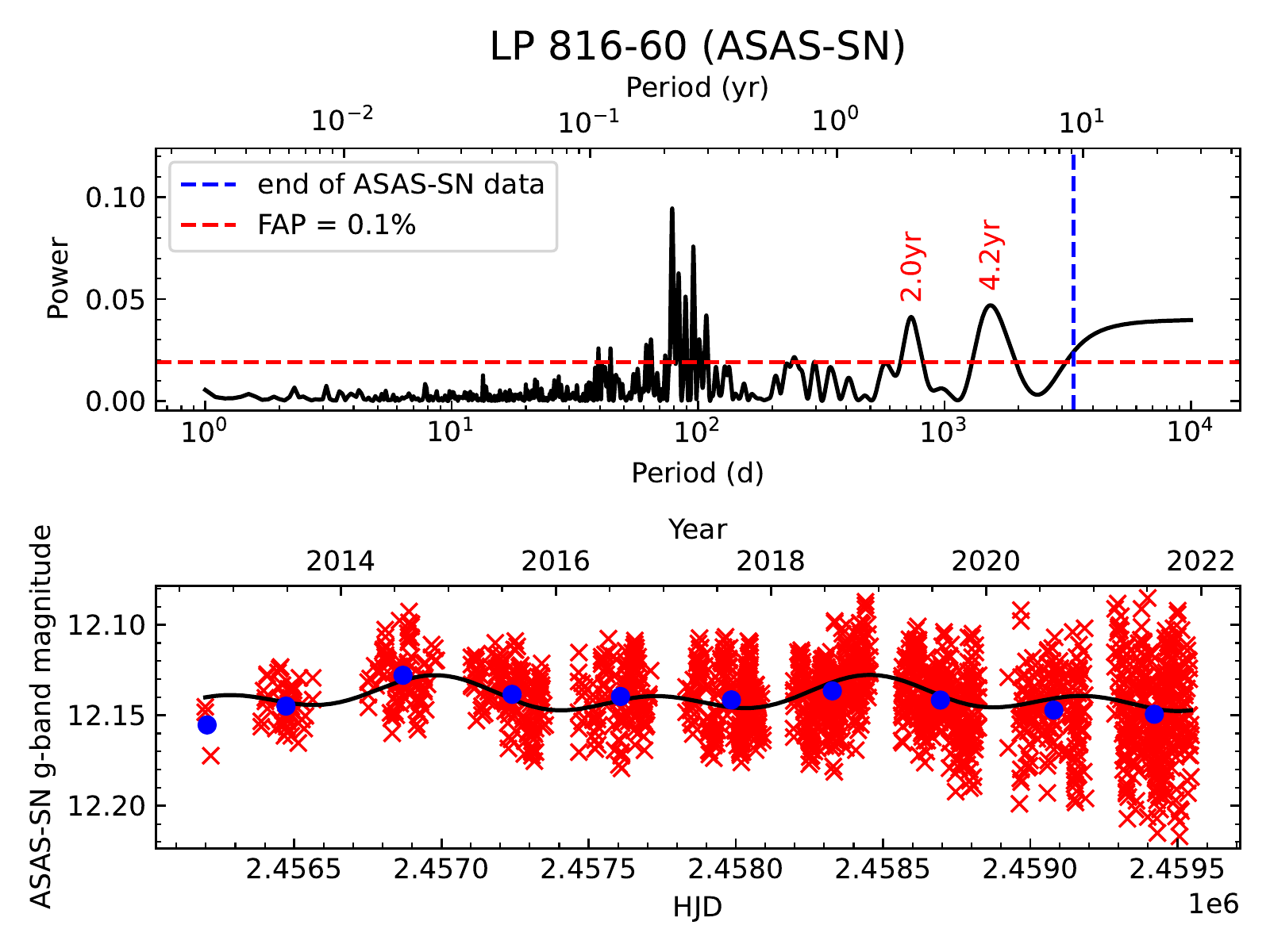}
    \caption{Periodogram and ASAS-SN data for LP 816-60. The complete figure set (15 images) is available in the online journal (Figure Set 2).}
    \label{fig: LP 816-60 ASAS-SN sine fit}
\end{figure}

For each star, we computed a L--S periodogram from its light curve. We then identified significant peaks in those periodograms with periods of more than 1.5 years, and used the periods of those peaks to fit sine functions to the light curve; see Figures \ref{fig: GJ 447 ASAS3 sine fit} and \ref{fig: LP 816-60 ASAS-SN sine fit} for examples. All ASAS-3/-4 light curves and associated periodograms (16 images) are provided in Figure Set 1; likewise for ASAS-SN (15 images) in Figure Set 2.

\figsetstart
\figsetnum{1}
\figsettitle{ASAS-3/-4}

\figsetgrpstart
\figsetgrpnum{1.1}
\figsetgrptitle{Periodogram and ASAS-3 data for GJ 234.}
\figsetplot{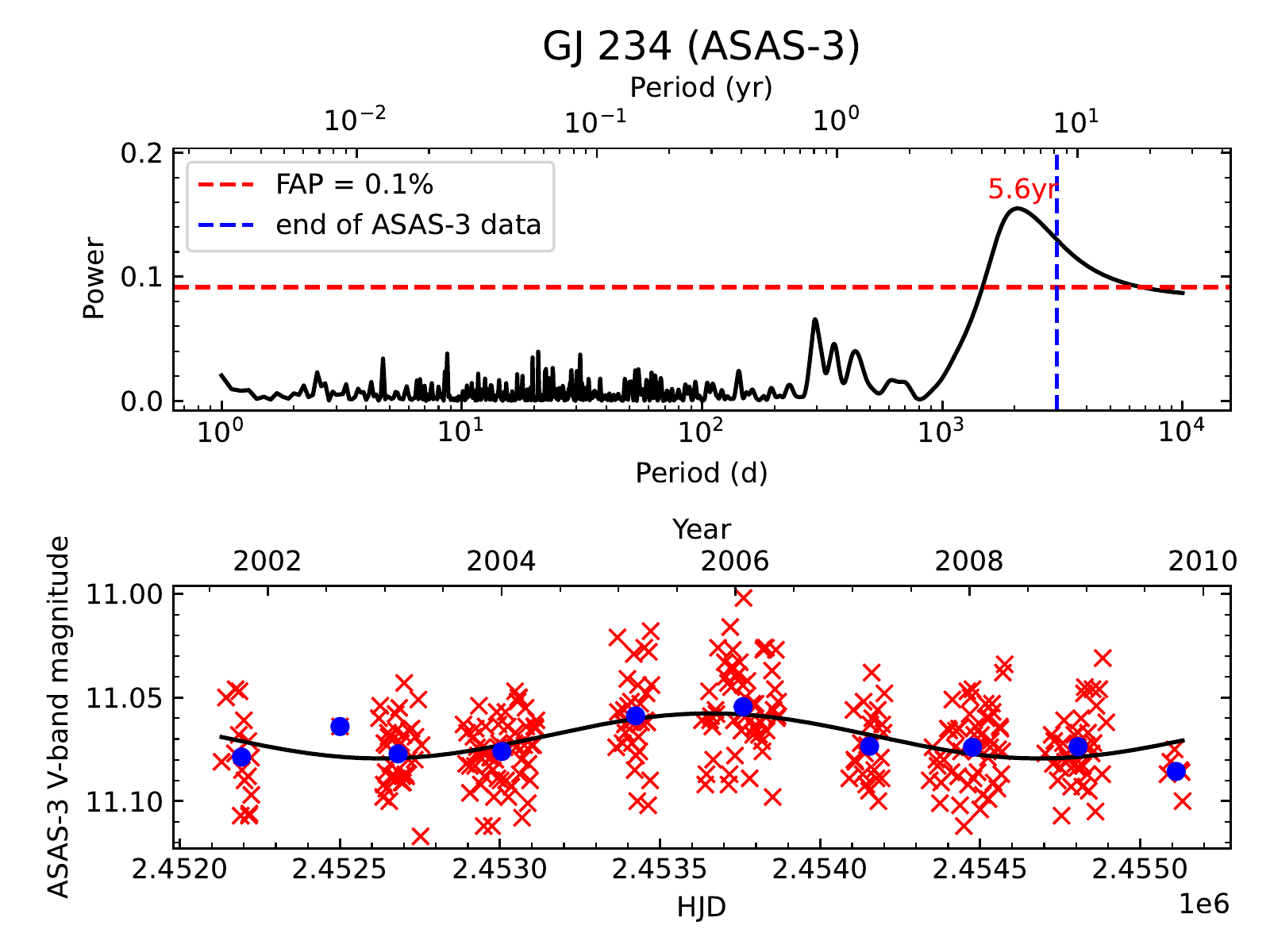}
\figsetgrpnote{Periodograms and ASAS-3/-4 light curves for our sample of M dwarfs.}
\figsetgrpend

\figsetgrpstart
\figsetgrpnum{1.2}
\figsetgrptitle{Periodogram and ASAS-3 data for GJ 273.}
\figsetplot{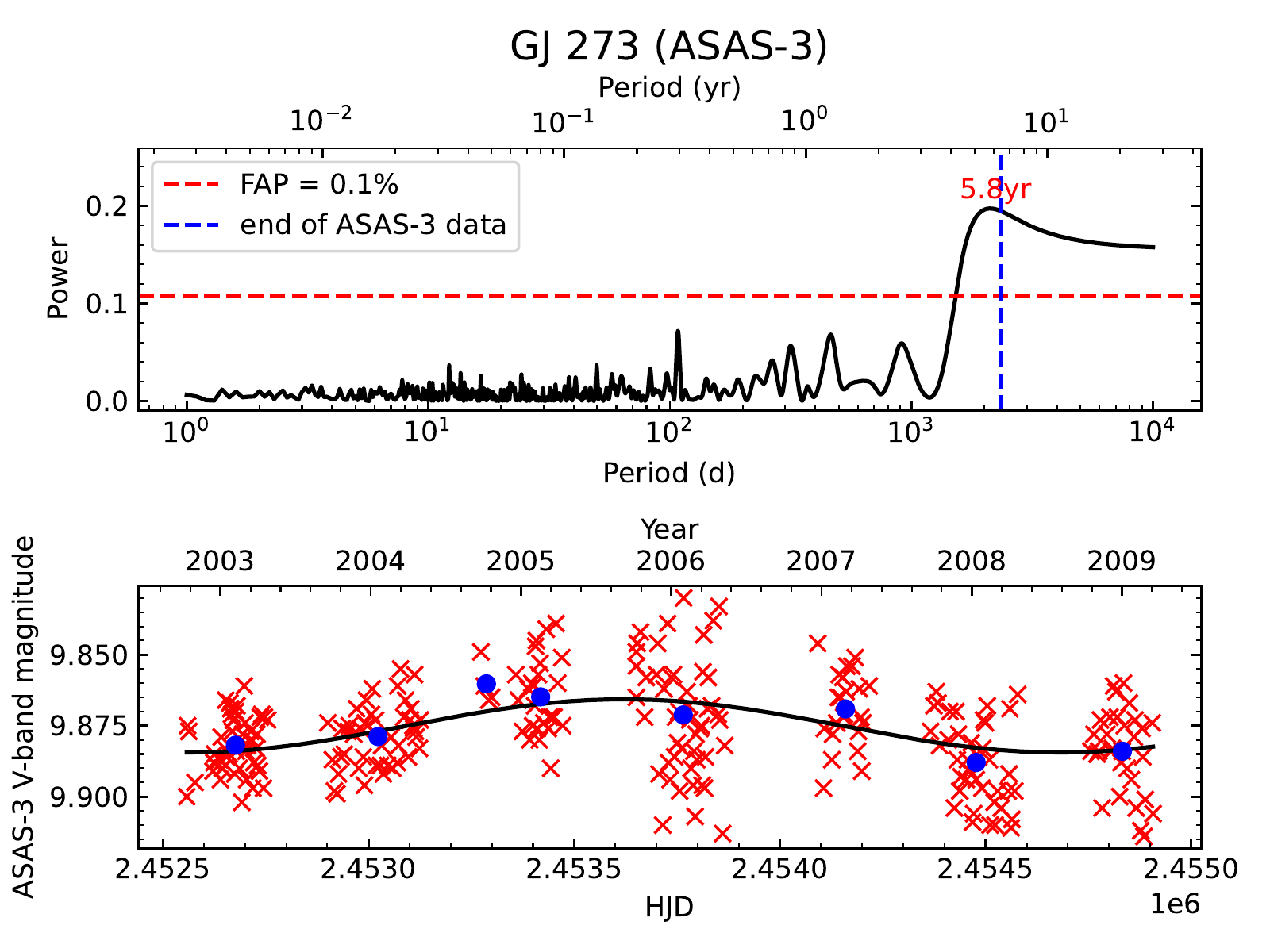}
\figsetgrpnote{Periodograms and ASAS-3/-4 light curves for our sample of M dwarfs.}
\figsetgrpend

\figsetgrpstart
\figsetgrpnum{1.3}
\figsetgrptitle{Periodogram and ASAS-3 data for GJ 285.}
\figsetplot{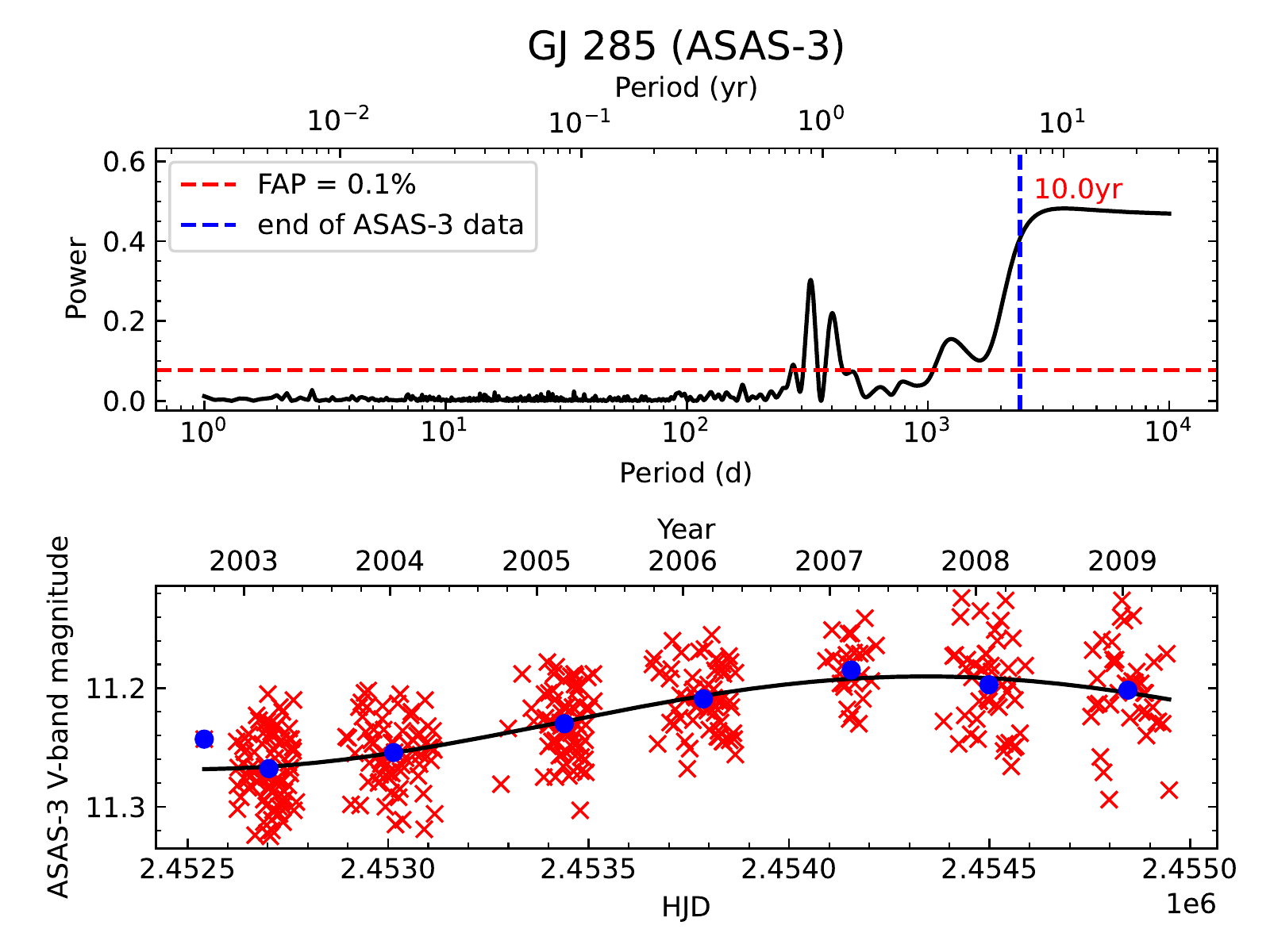}
\figsetgrpnote{Periodograms and ASAS-3/-4 light curves for our sample of M dwarfs.}
\figsetgrpend

\figsetgrpstart
\figsetgrpnum{1.4}
\figsetgrptitle{Periodogram and ASAS-3 data for GJ 317.}
\figsetplot{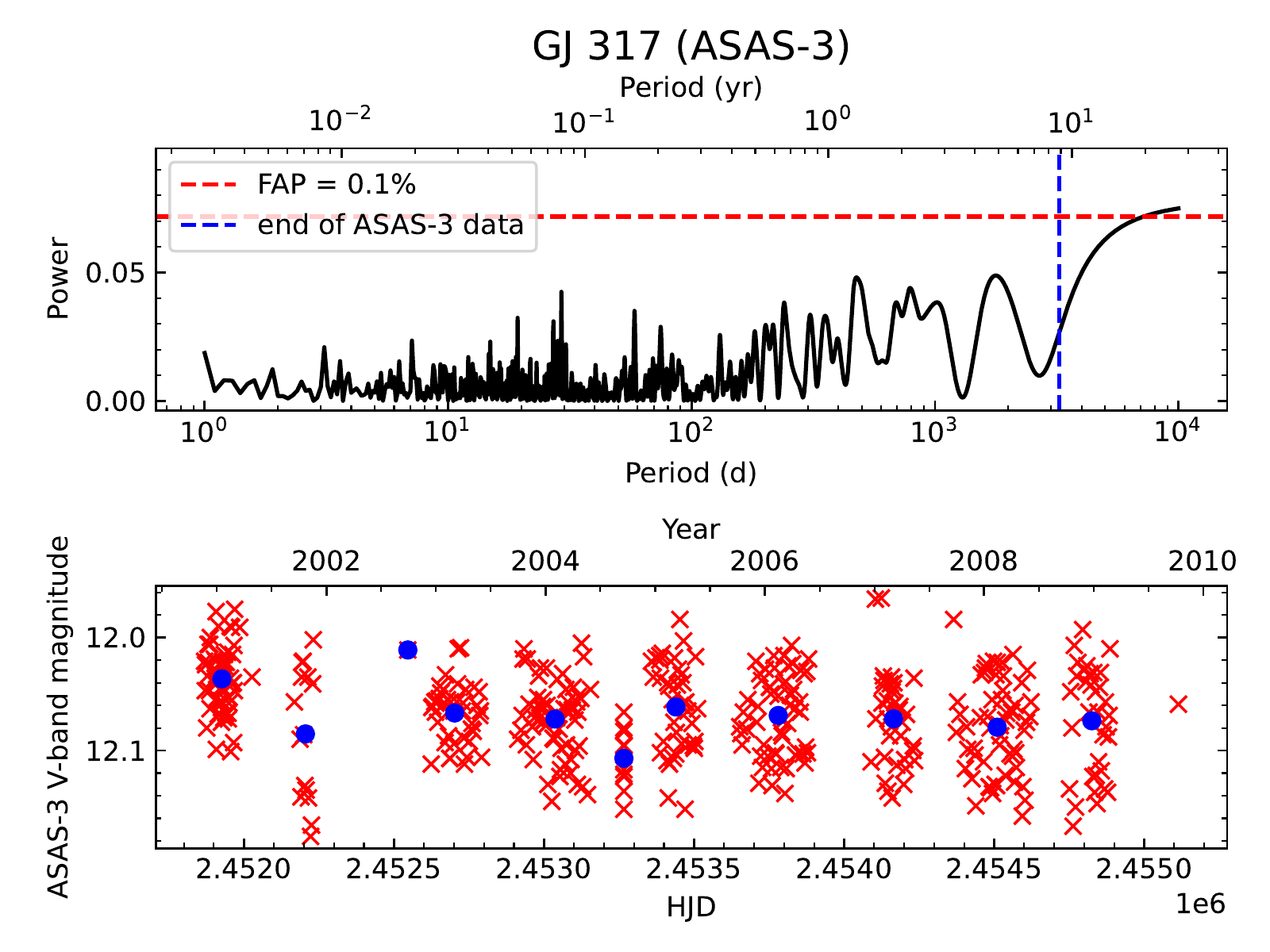}
\figsetgrpnote{Periodograms and ASAS-3/-4 light curves for our sample of M dwarfs.}
\figsetgrpend

\figsetgrpstart
\figsetgrpnum{1.5}
\figsetgrptitle{Periodogram and ASAS-3 data for GJ 358.}
\figsetplot{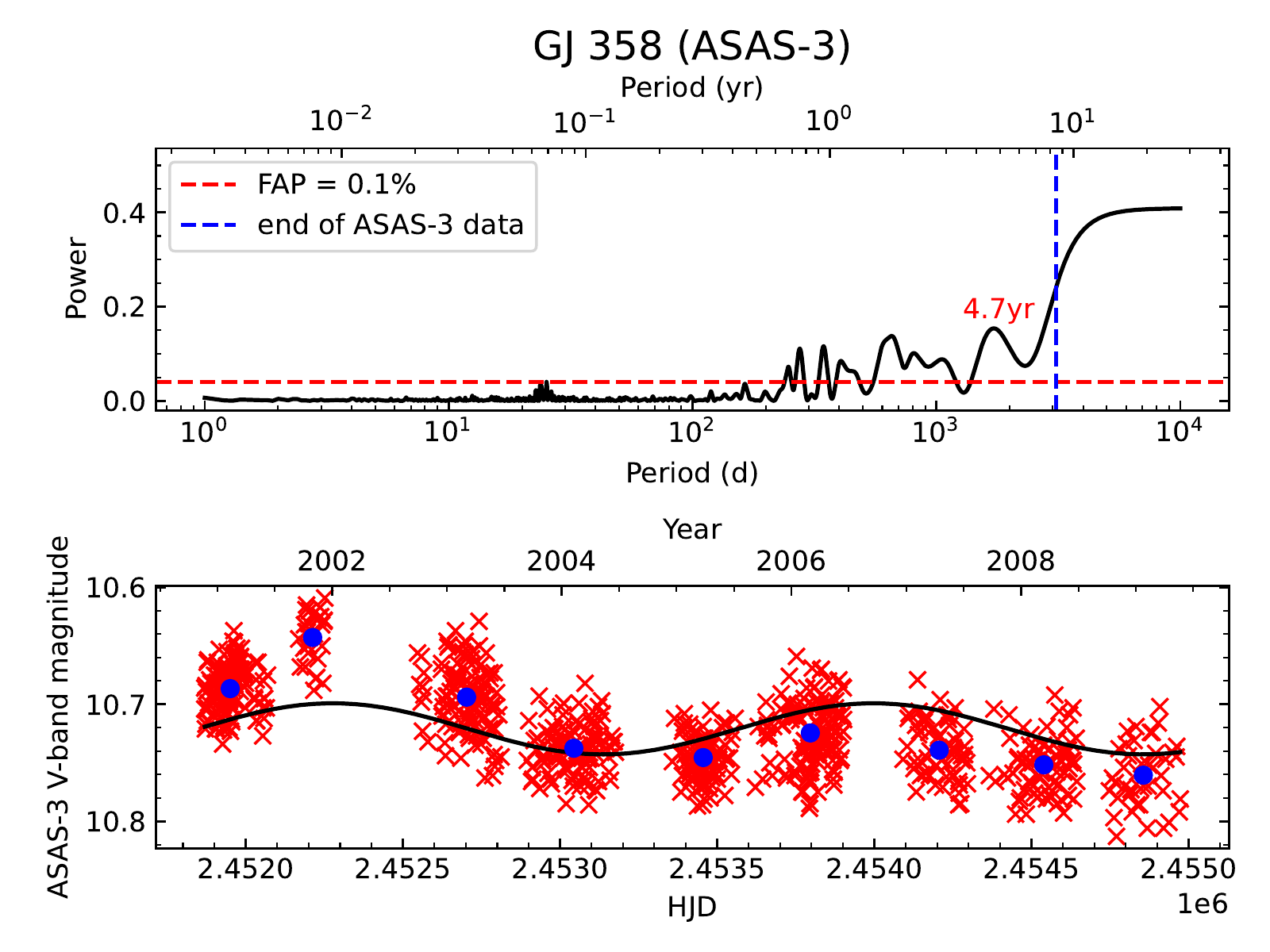}
\figsetgrpnote{Periodograms and ASAS-3/-4 light curves for our sample of M dwarfs.}
\figsetgrpend

\figsetgrpstart
\figsetgrpnum{1.6}
\figsetgrptitle{Periodogram and ASAS-3 data for GJ 406.}
\figsetplot{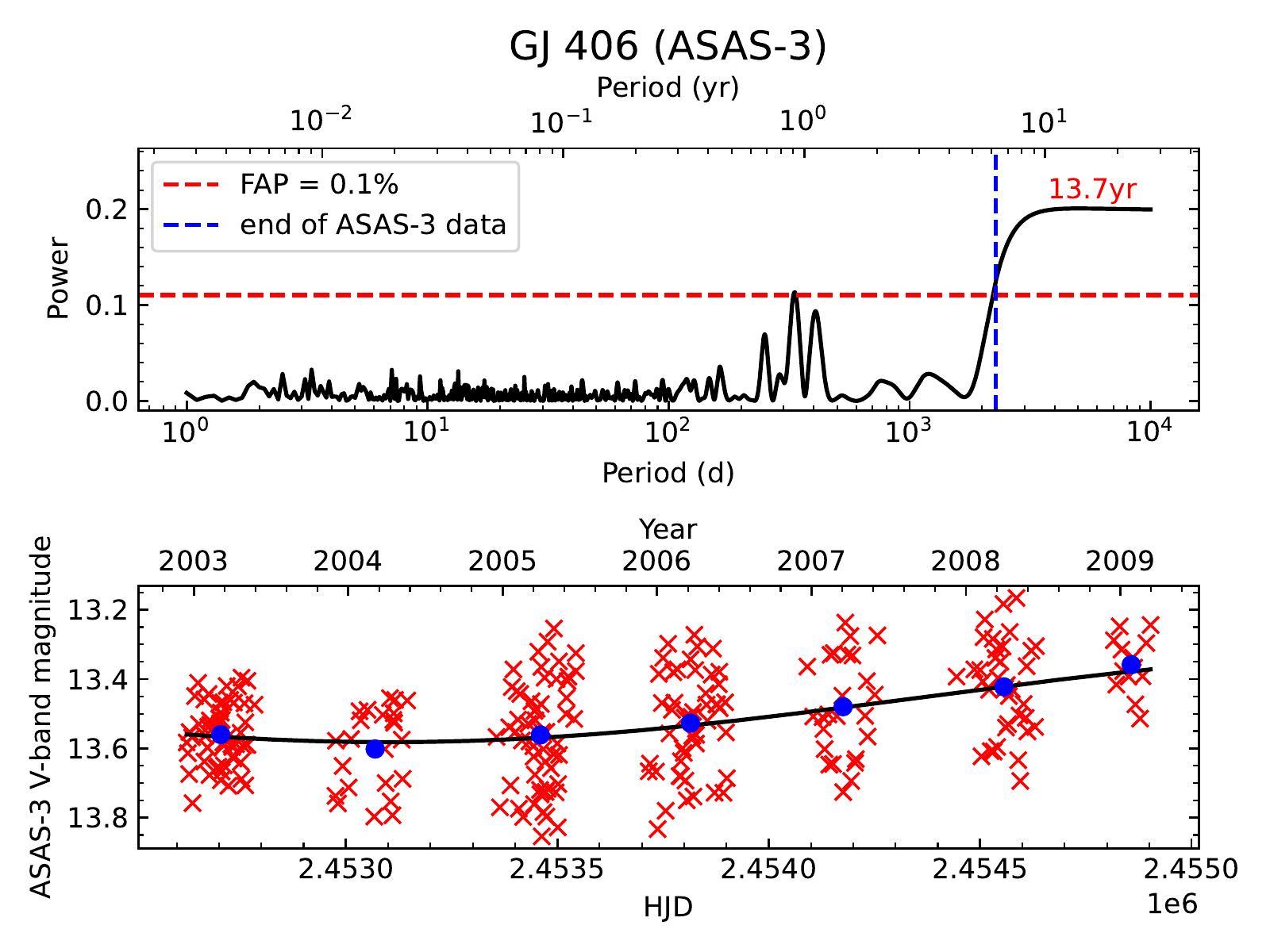}
\figsetgrpnote{Periodograms and ASAS-3/-4 light curves for our sample of M dwarfs.}
\figsetgrpend

\figsetgrpstart
\figsetgrpnum{1.7}
\figsetgrptitle{Periodogram and ASAS-3 data for GJ 447.}
\figsetplot{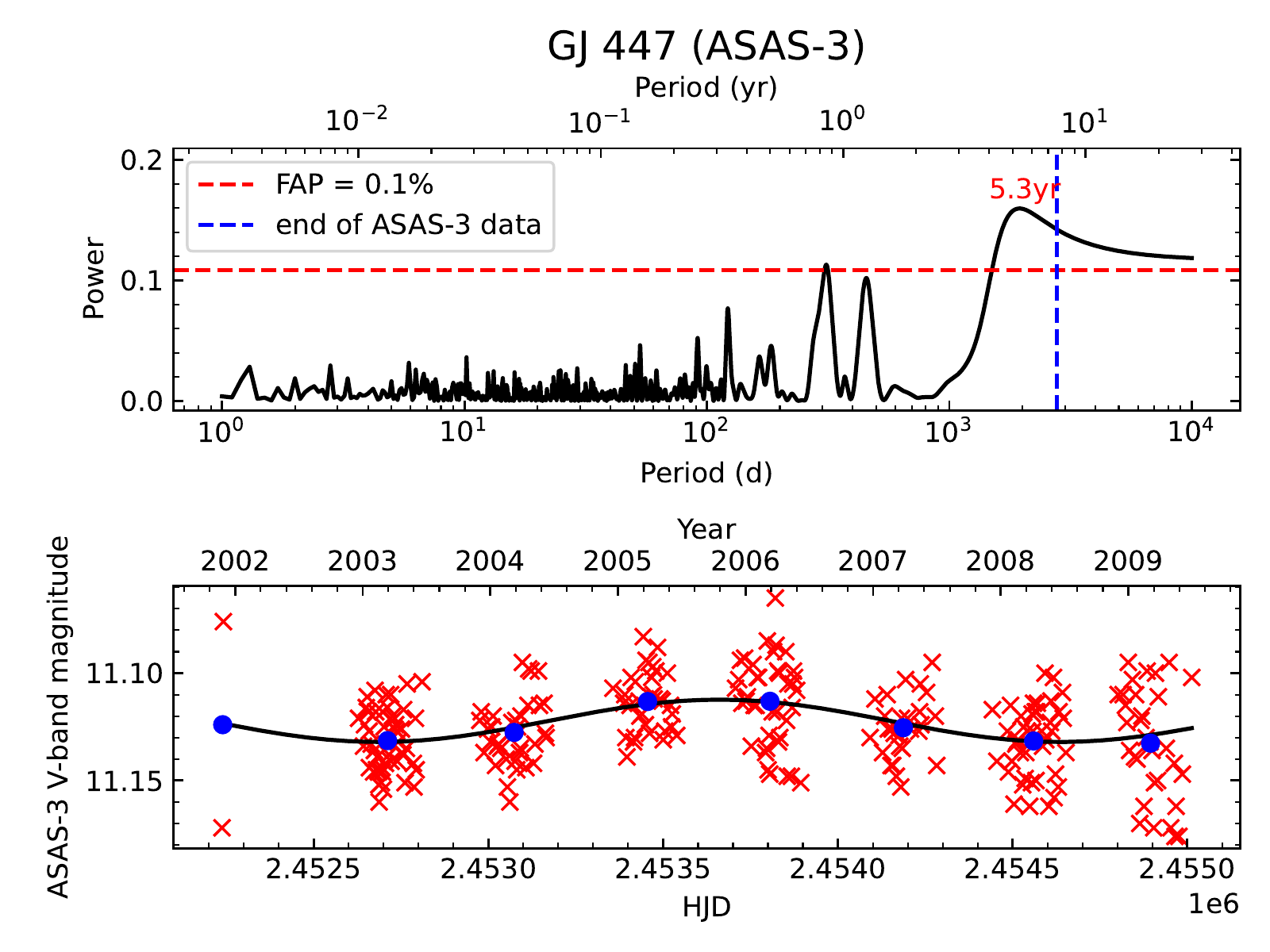}
\figsetgrpnote{Periodograms and ASAS-3/-4 light curves for our sample of M dwarfs.}
\figsetgrpend

\figsetgrpstart
\figsetgrpnum{1.8}
\figsetgrptitle{Periodogram and ASAS-3 data for GJ 54.1.}
\figsetplot{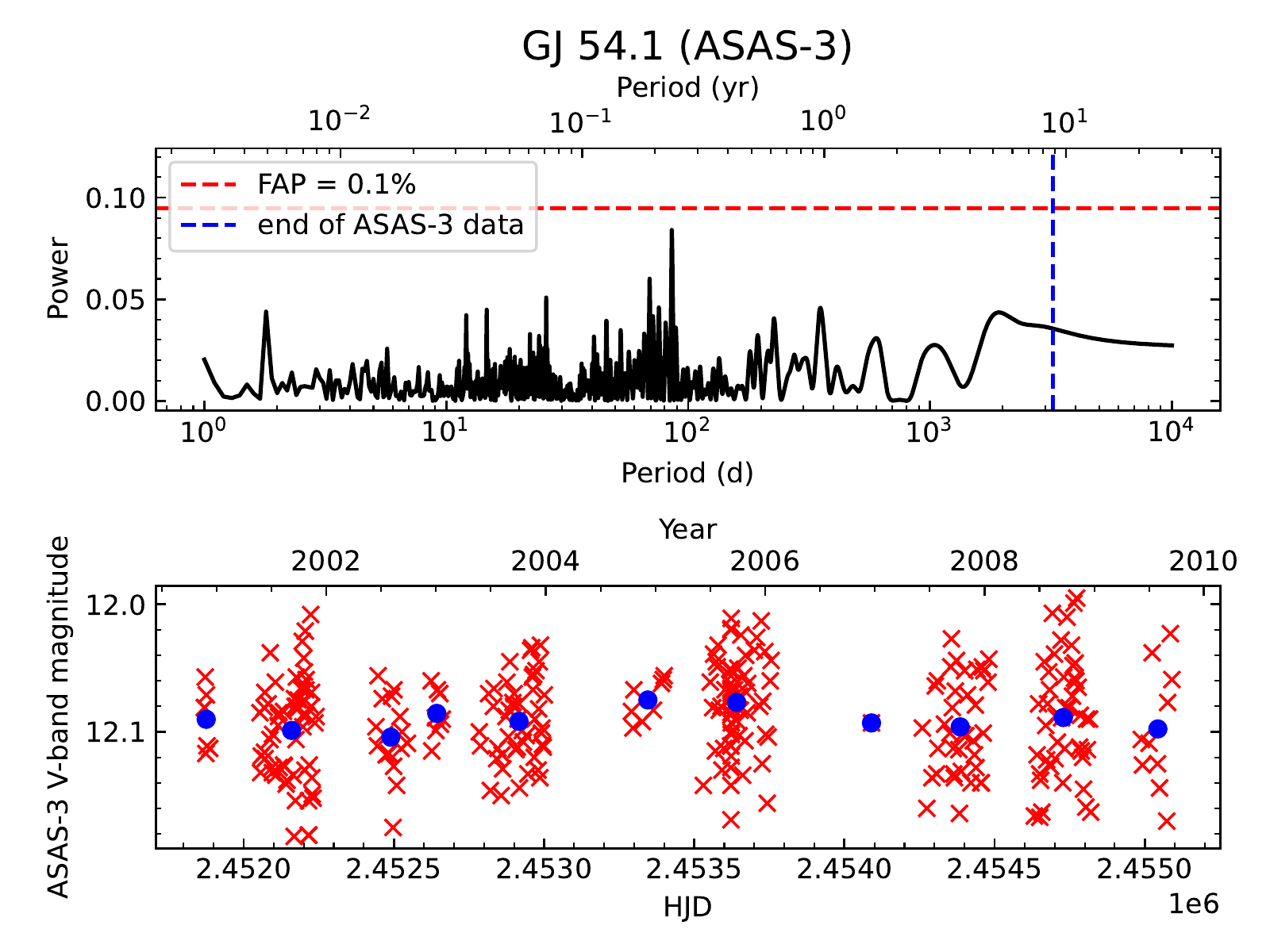}
\figsetgrpnote{Periodograms and ASAS-3/-4 light curves for our sample of M dwarfs.}
\figsetgrpend

\figsetgrpstart
\figsetgrpnum{1.9}
\figsetgrptitle{Periodogram and ASAS-3 data for GJ 581.}
\figsetplot{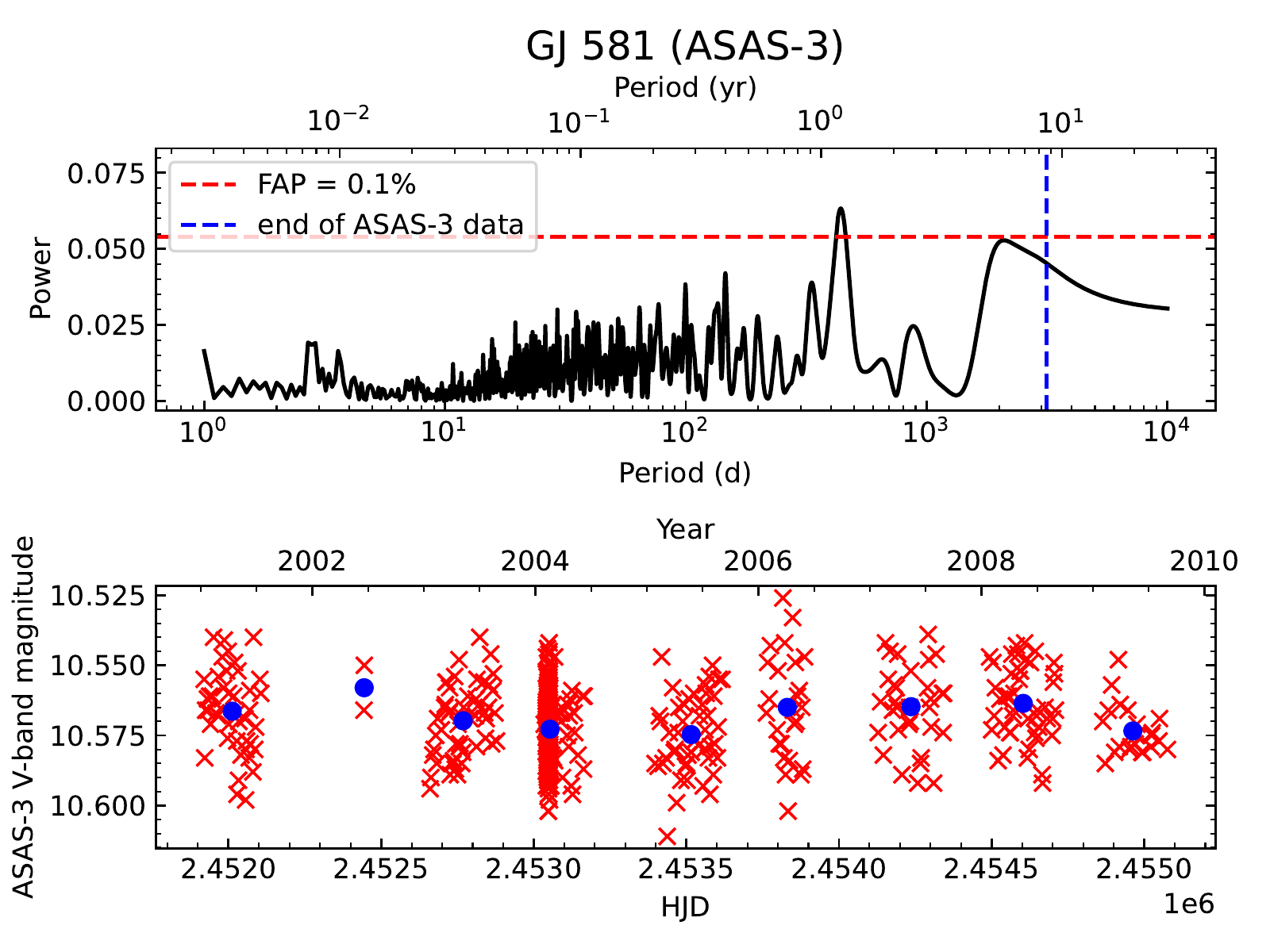}
\figsetgrpnote{Periodograms and ASAS-3/-4 light curves for our sample of M dwarfs.}
\figsetgrpend

\figsetgrpstart
\figsetgrpnum{1.10}
\figsetgrptitle{Periodogram and ASAS-3 data for GJ 628.}
\figsetplot{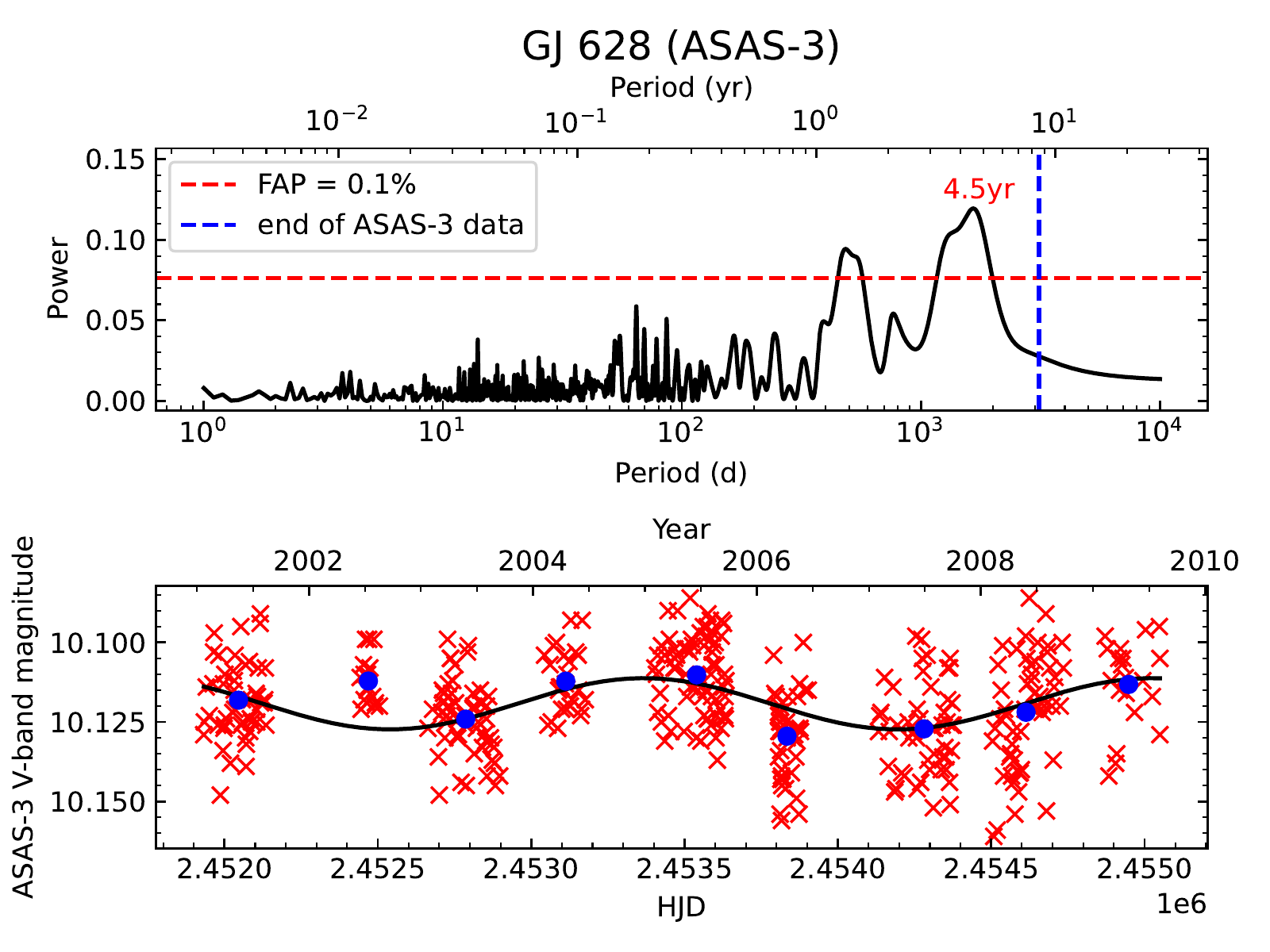}
\figsetgrpnote{Periodograms and ASAS-3/-4 light curves for our sample of M dwarfs.}
\figsetgrpend

\figsetgrpstart
\figsetgrpnum{1.11}
\figsetgrptitle{Periodogram and ASAS-3 data for GJ 729.}
\figsetplot{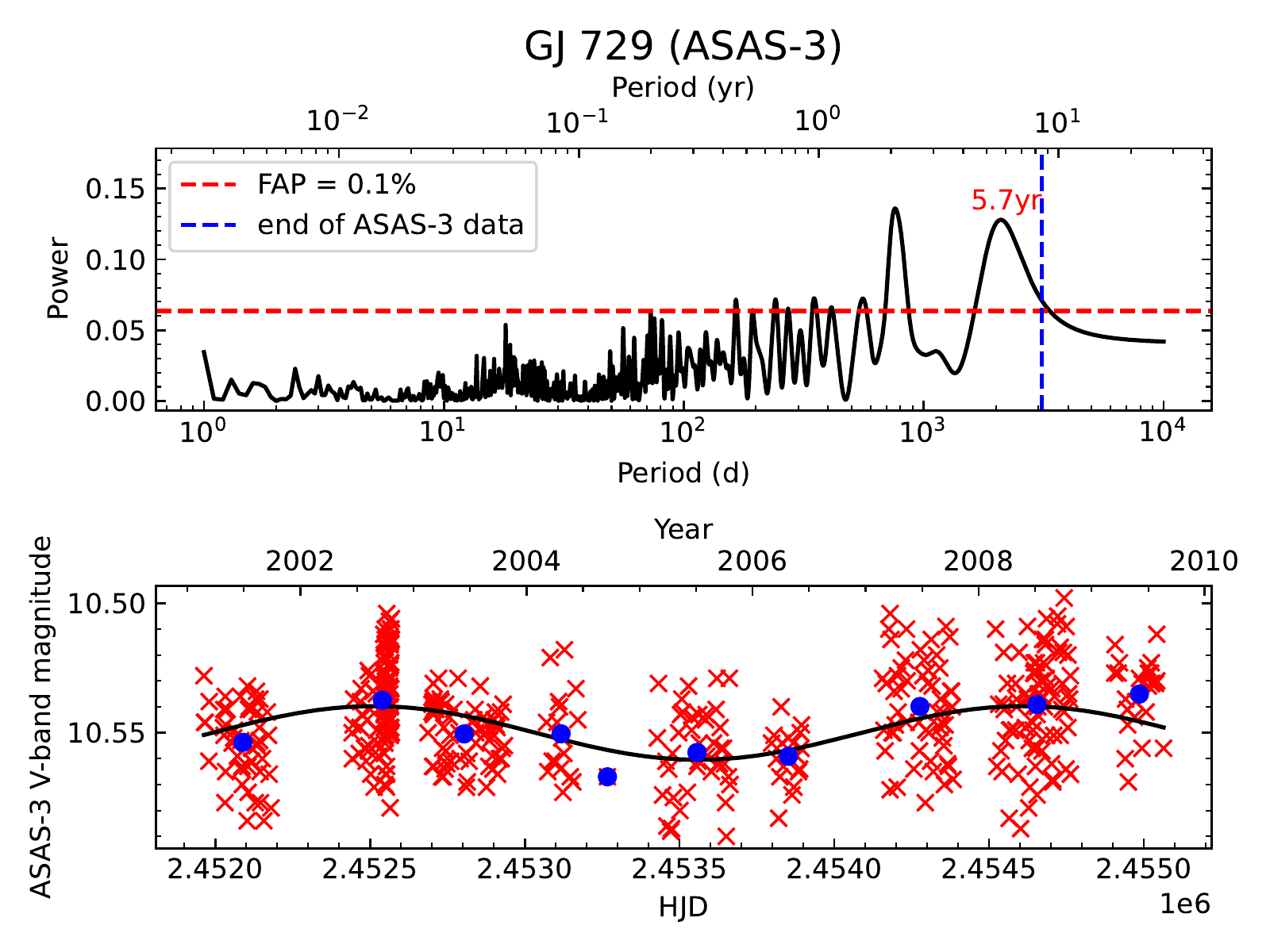}
\figsetgrpnote{Periodograms and ASAS-3/-4 light curves for our sample of M dwarfs.}
\figsetgrpend

\figsetgrpstart
\figsetgrpnum{1.12}
\figsetgrptitle{Periodogram and ASAS-3 data for GJ 849.}
\figsetplot{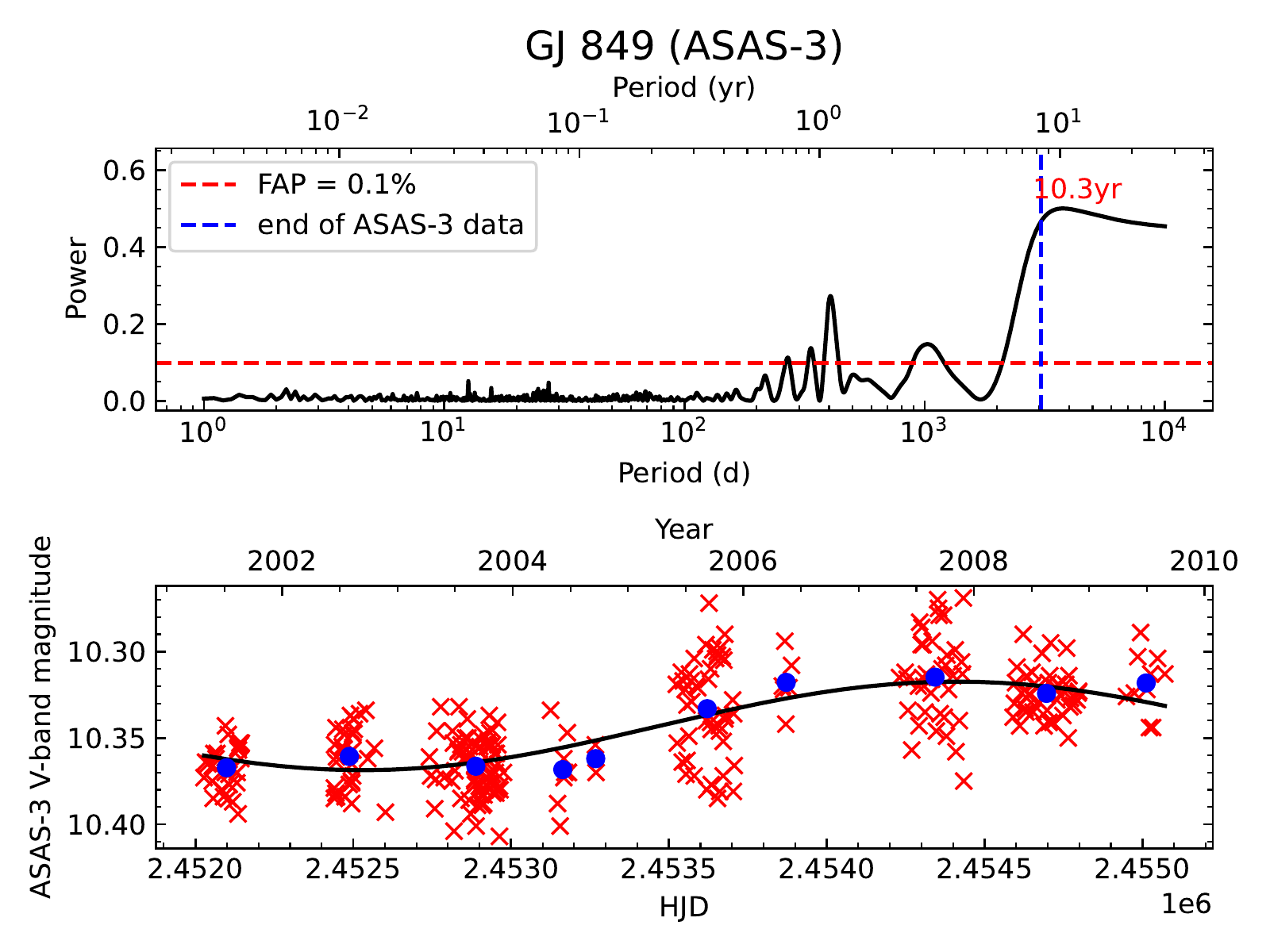}
\figsetgrpnote{Periodograms and ASAS-3/-4 light curves for our sample of M dwarfs.}
\figsetgrpend

\figsetgrpstart
\figsetgrpnum{1.13}
\figsetgrptitle{Periodogram and ASAS-3 data for GJ 896A.}
\figsetplot{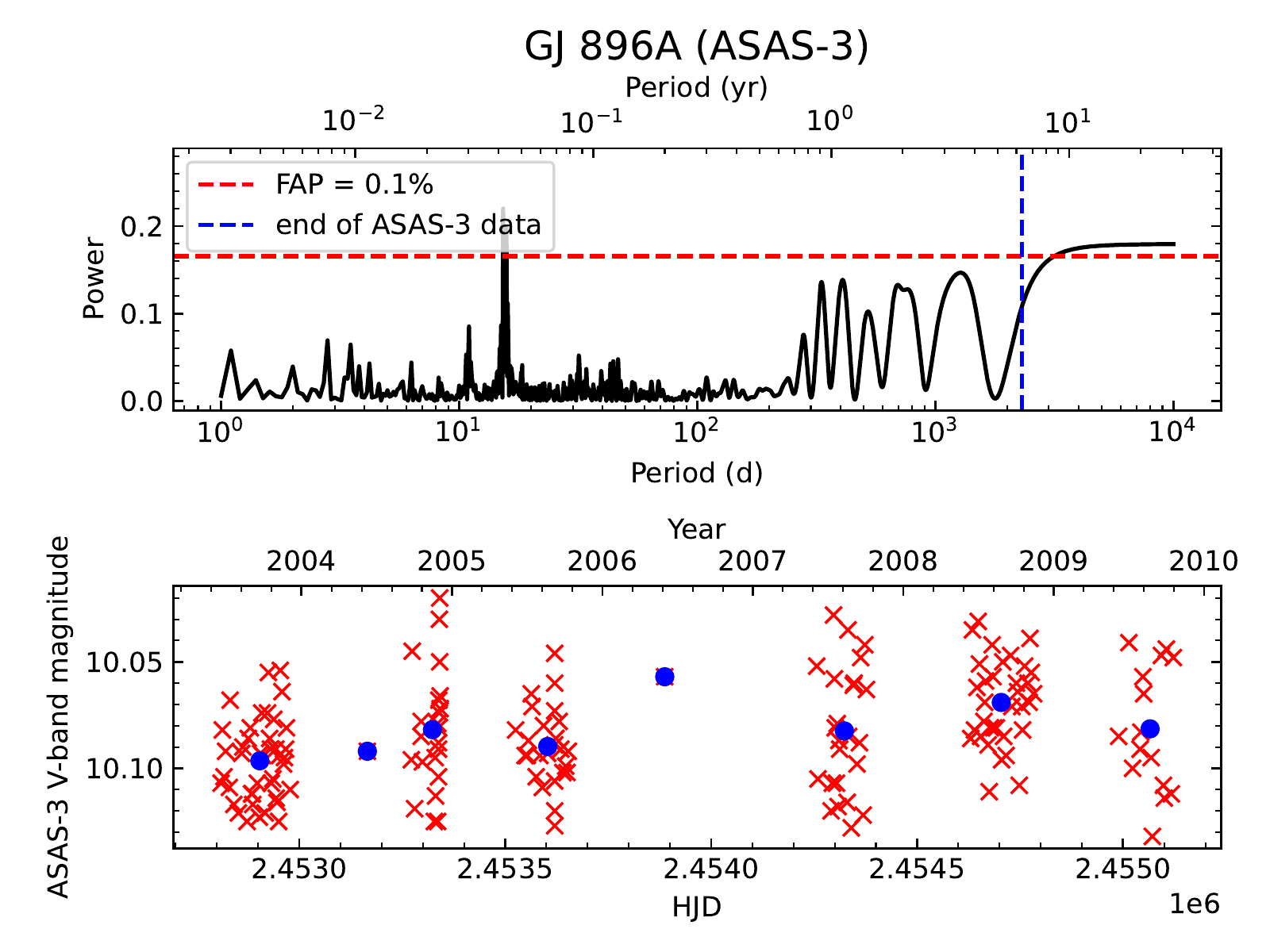}
\figsetgrpnote{Periodograms and ASAS-3/-4 light curves for our sample of M dwarfs.}
\figsetgrpend

\figsetgrpstart
\figsetgrpnum{1.14}
\figsetgrptitle{Periodogram and ASAS-3 data for GJ LP 816-60.}
\figsetplot{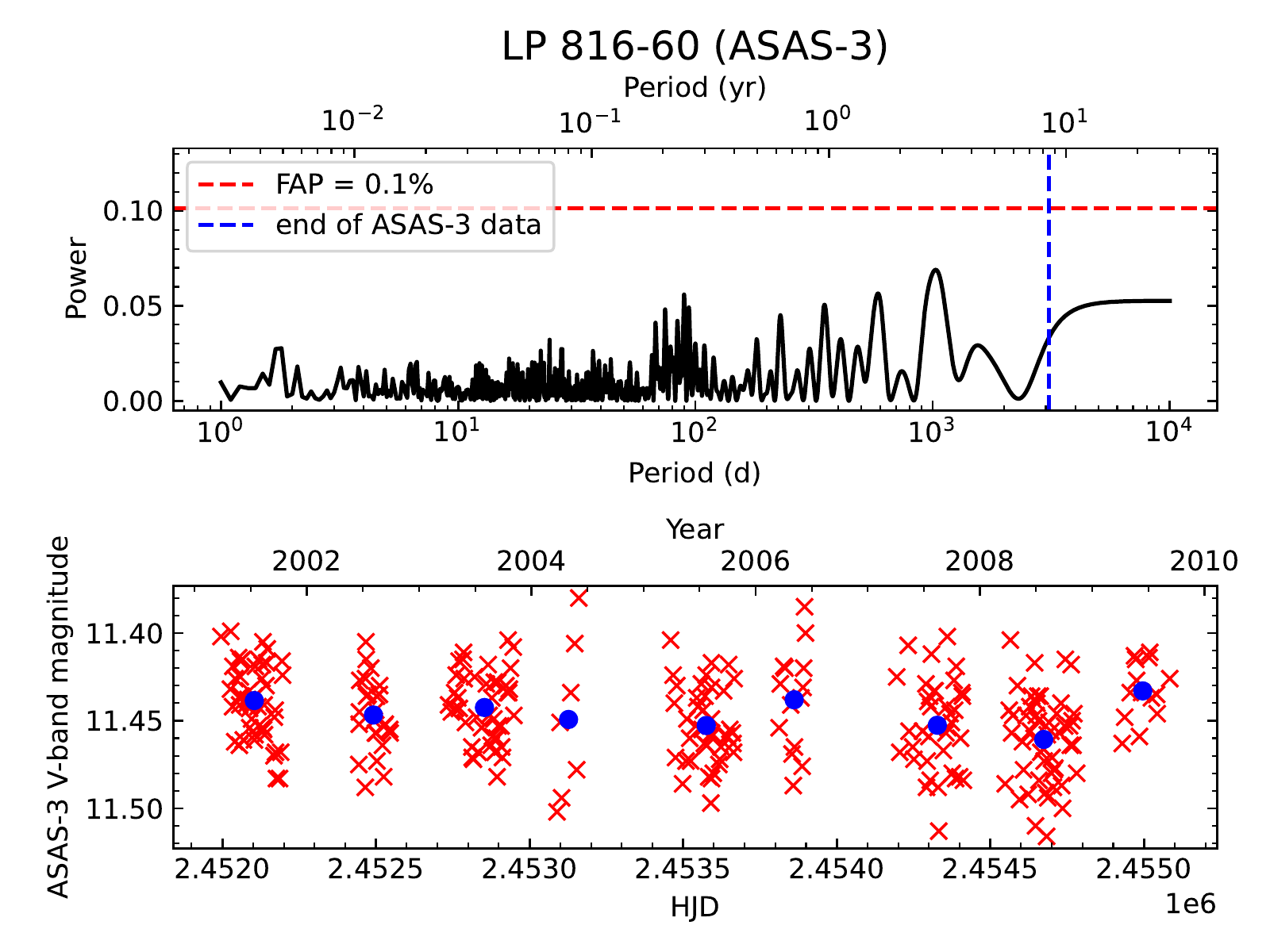}
\figsetgrpnote{Periodograms and ASAS-3/-4 light curves for our sample of M dwarfs.}
\figsetgrpend

\figsetgrpstart
\figsetgrpnum{1.15}
\figsetgrptitle{Periodogram and ASAS-3 data for Proxima.}
\figsetplot{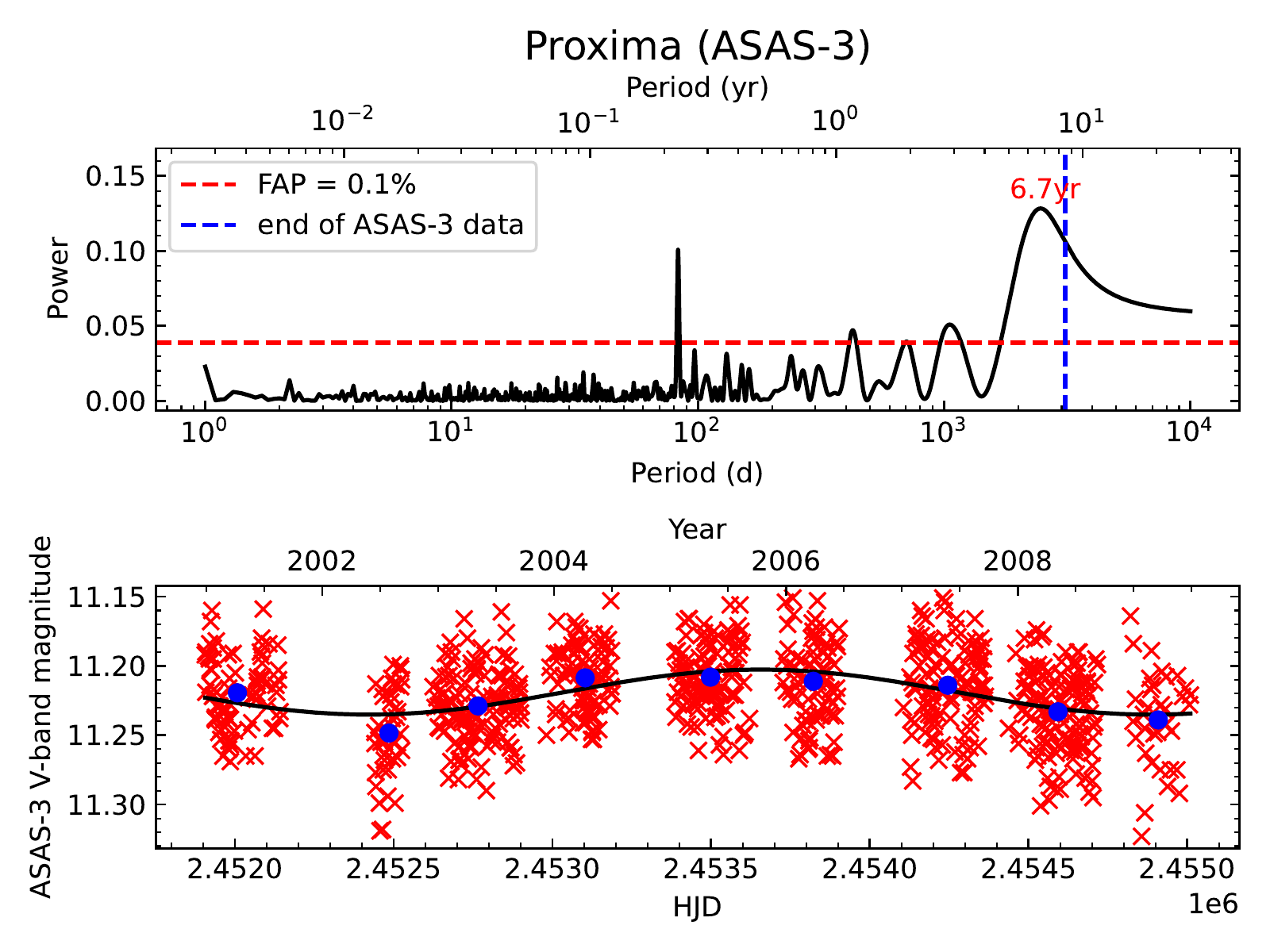}
\figsetgrpnote{Periodograms and ASAS-3/-4 light curves for our sample of M dwarfs.}
\figsetgrpend

\figsetgrpstart
\figsetgrpnum{1.16}
\figsetgrptitle{Periodogram and ASAS-4 data for Proxima.}
\figsetplot{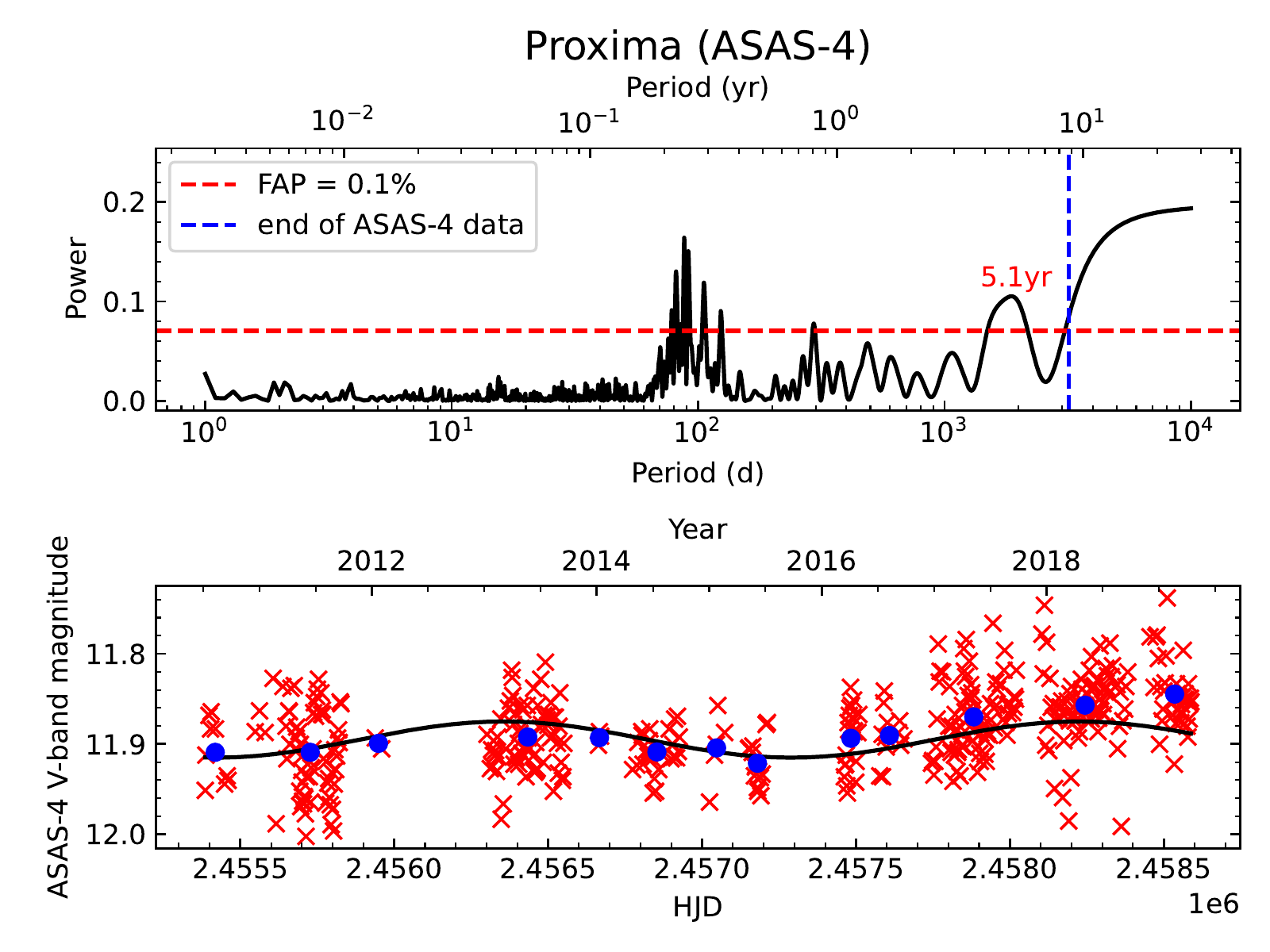}
\figsetgrpnote{Periodograms and ASAS-3/-4 light curves for our sample of M dwarfs.}
\figsetgrpend

\figsetend

If a periodogram contained multiple significant peaks, we fit sine functions using each peak and then combined these functions to create a superposition. If the fit of the superposition was worse than the fit of any single component, then the worst fitting components were removed until the fit of the superposition was better than the fit of any single component, or only the best fitting component was remaining. If the period(s) of the remaining component(s) was within the period of observation, then we deem this cycle to be ``well-defined" - and poorly constrained otherwise.

\figsetstart
\figsetnum{2}
\figsettitle{ASAS-SN}

\figsetgrpstart
\figsetgrpnum{2.1}
\figsetgrptitle{Periodogram and ASAS-SN data for GJ 234.}
\figsetplot{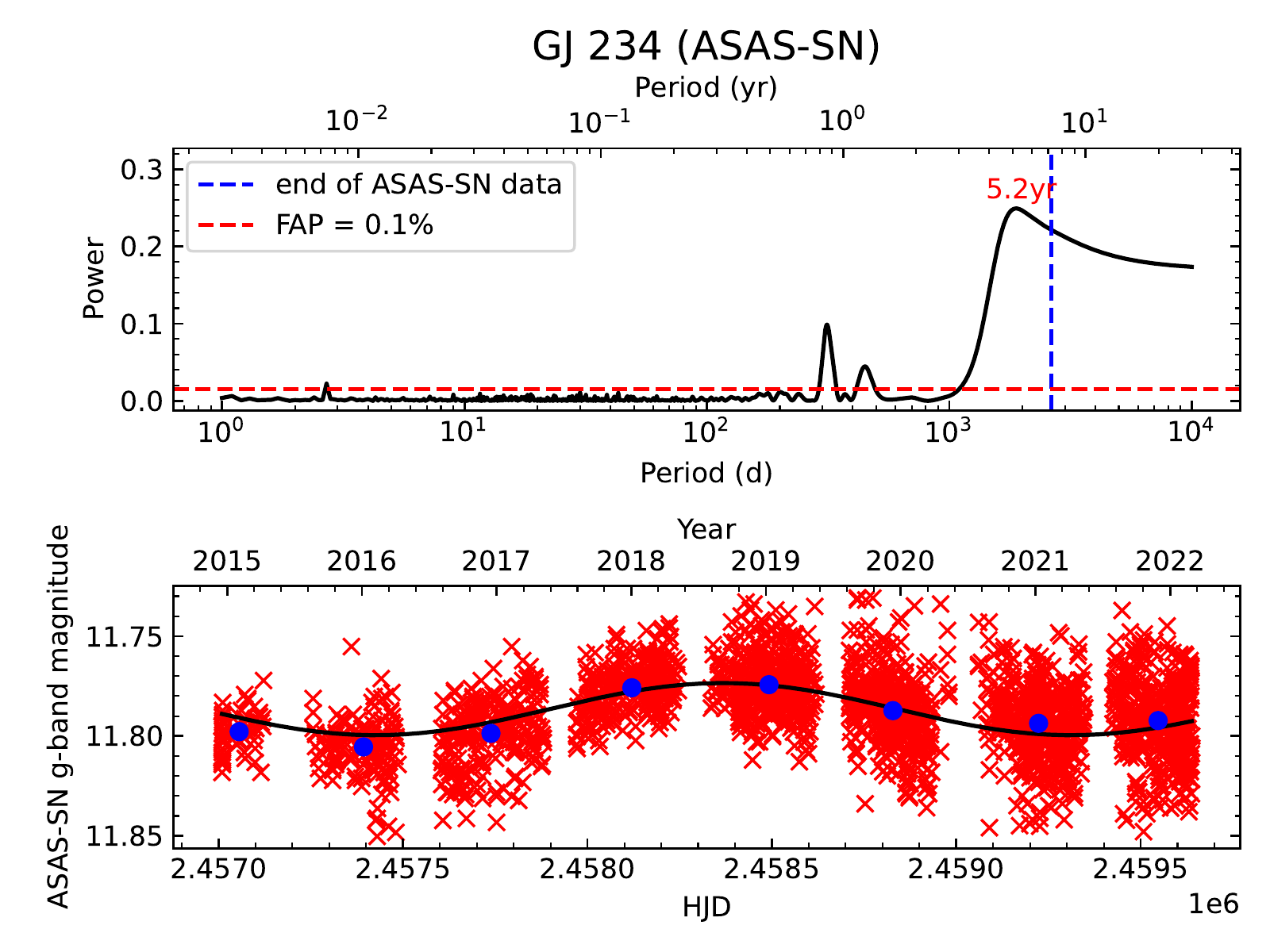}
\figsetgrpnote{Periodograms and ASAS-SN light curves for our sample of M dwarfs.}
\figsetgrpend

\figsetgrpstart
\figsetgrpnum{2.2}
\figsetgrptitle{Periodogram and ASAS-SN data for GJ 273.}
\figsetplot{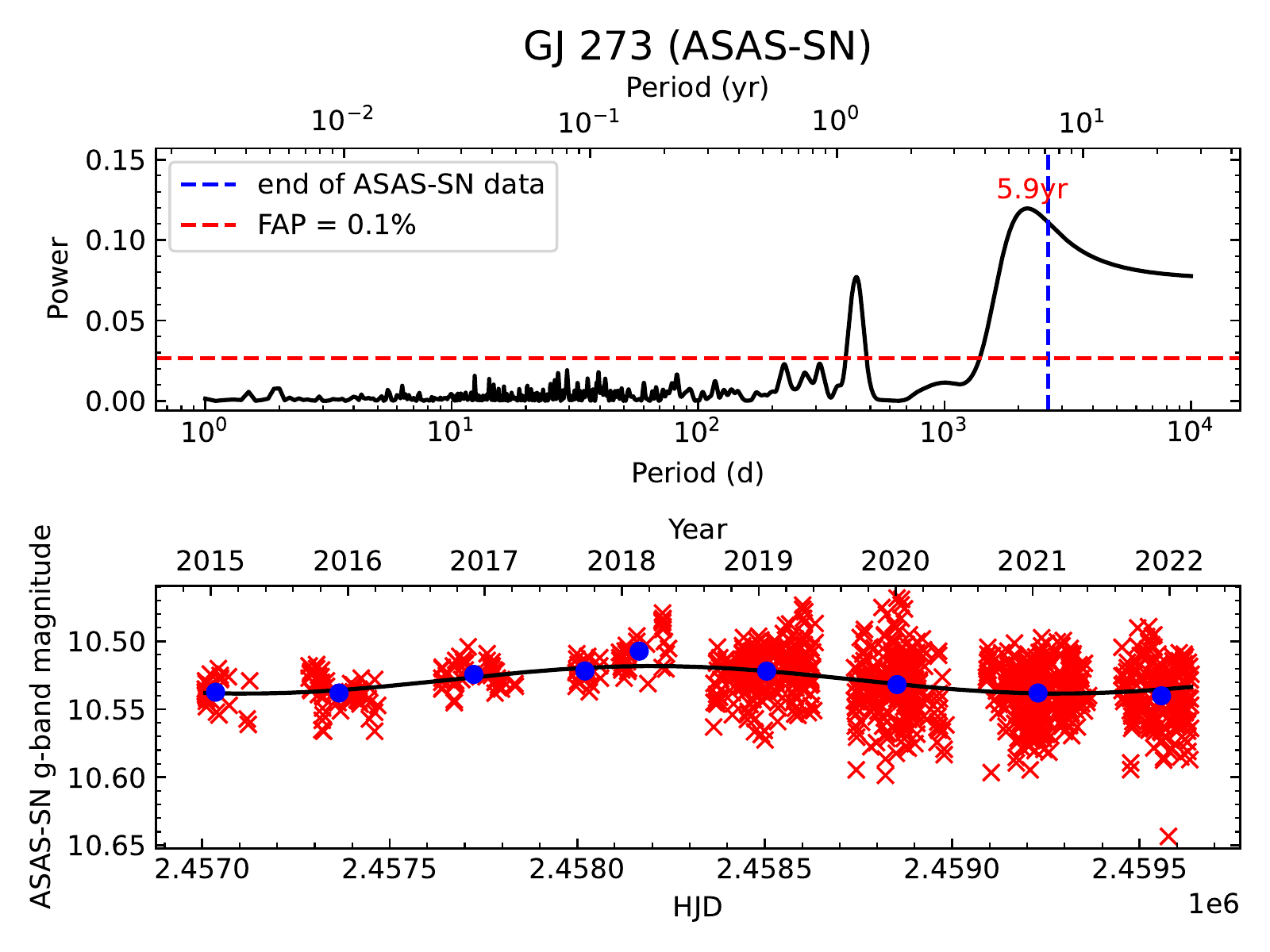}
\figsetgrpnote{Periodograms and ASAS-SN light curves for our sample of M dwarfs.}
\figsetgrpend

\figsetgrpstart
\figsetgrpnum{2.3}
\figsetgrptitle{Periodogram and ASAS-SN data for GJ 285.}
\figsetplot{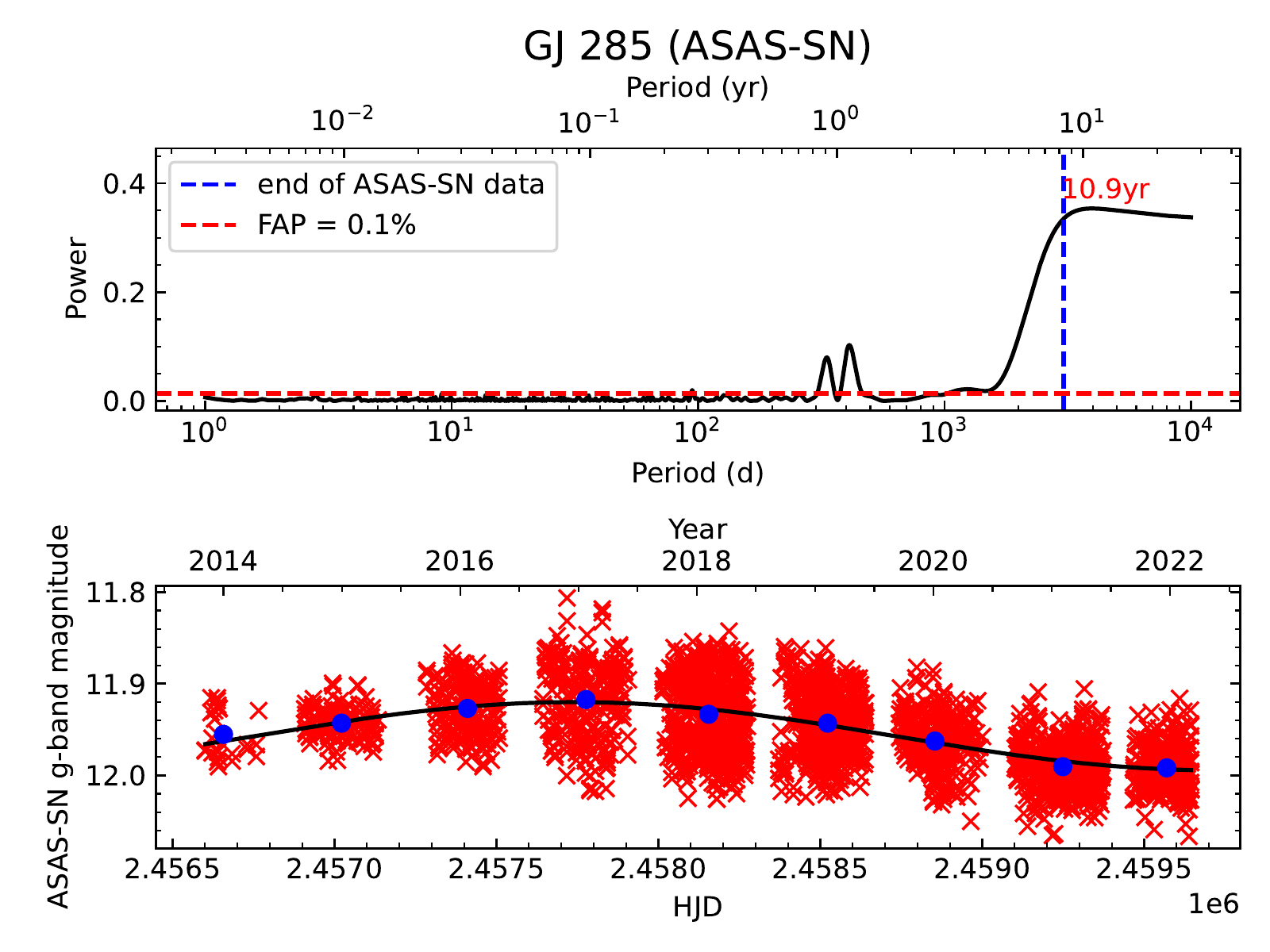}
\figsetgrpnote{Periodograms and ASAS-SN light curves for our sample of M dwarfs.}
\figsetgrpend

\figsetgrpstart
\figsetgrpnum{2.4}
\figsetgrptitle{Periodogram and ASAS-SN data for GJ 317.}
\figsetplot{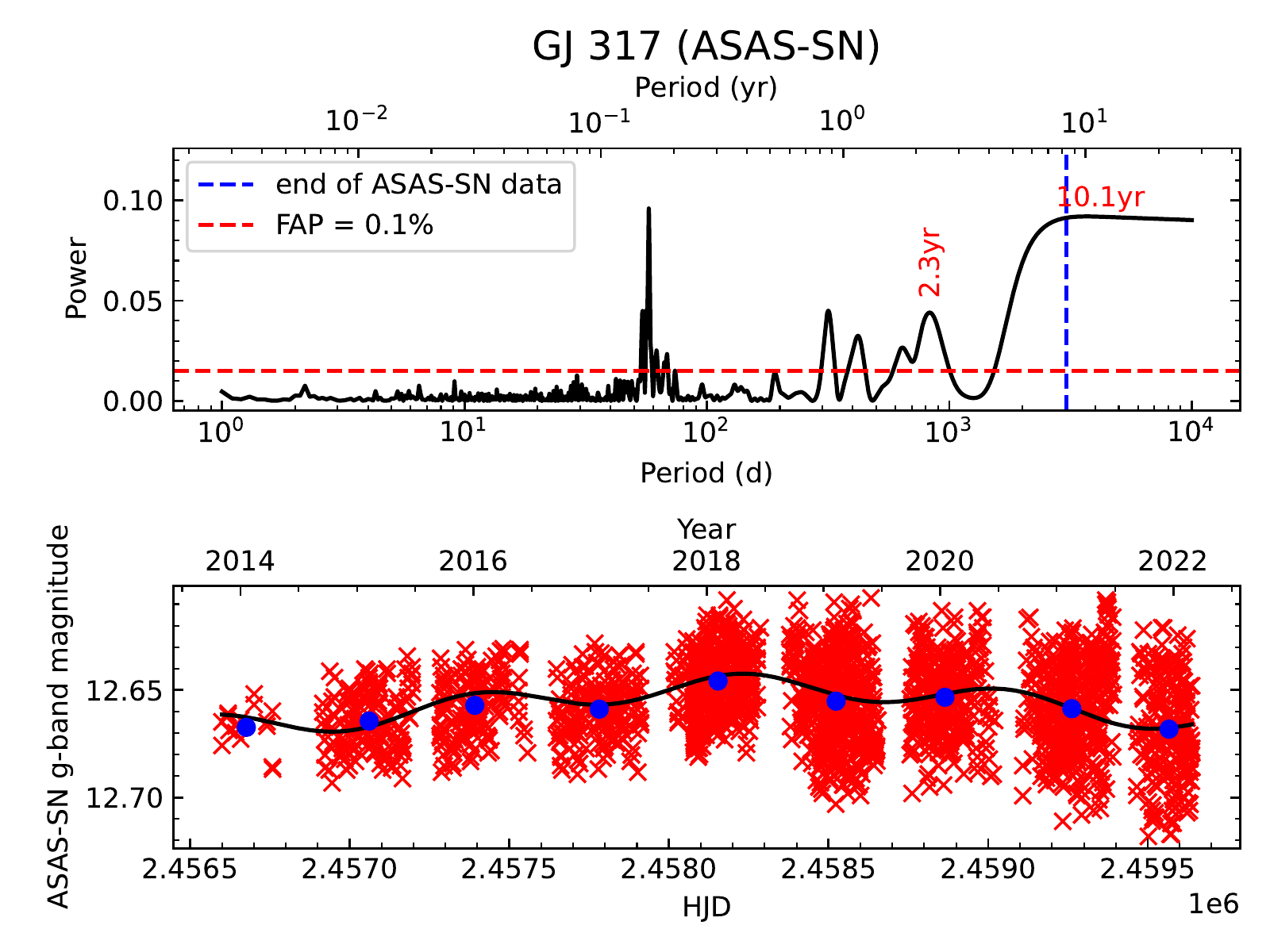}
\figsetgrpnote{Periodograms and ASAS-SN light curves for our sample of M dwarfs.}
\figsetgrpend

\figsetgrpstart
\figsetgrpnum{2.5}
\figsetgrptitle{Periodogram and ASAS-SN data for GJ 358.}
\figsetplot{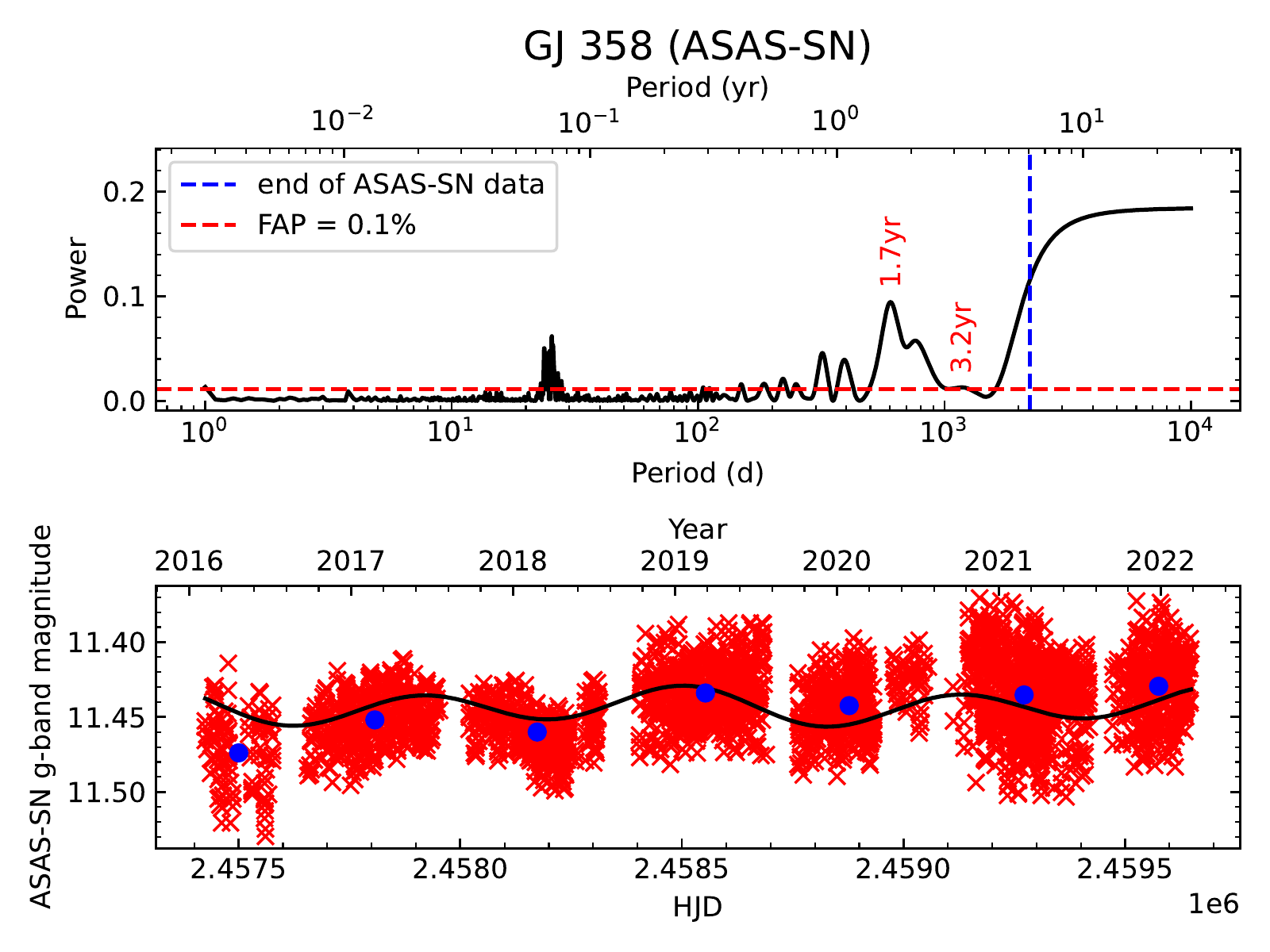}
\figsetgrpnote{Periodograms and ASAS-SN light curves for our sample of M dwarfs.}
\figsetgrpend

\figsetgrpstart
\figsetgrpnum{2.6}
\figsetgrptitle{Periodogram and ASAS-SN data for GJ 406.}
\figsetplot{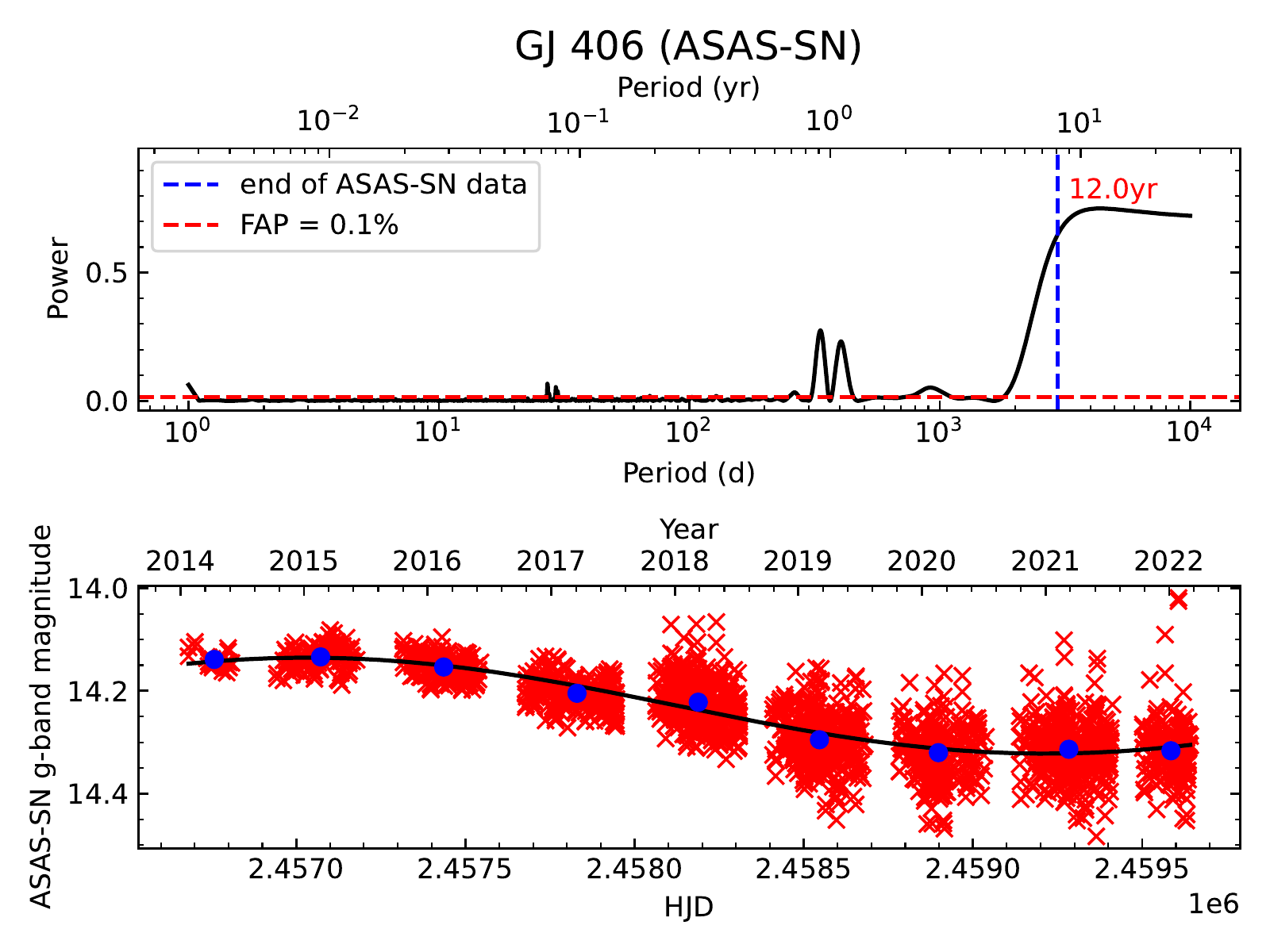}
\figsetgrpnote{Periodograms and ASAS-SN light curves for our sample of M dwarfs.}
\figsetgrpend

\figsetgrpstart
\figsetgrpnum{2.7}
\figsetgrptitle{Periodogram and ASAS-SN data for GJ 447.}
\figsetplot{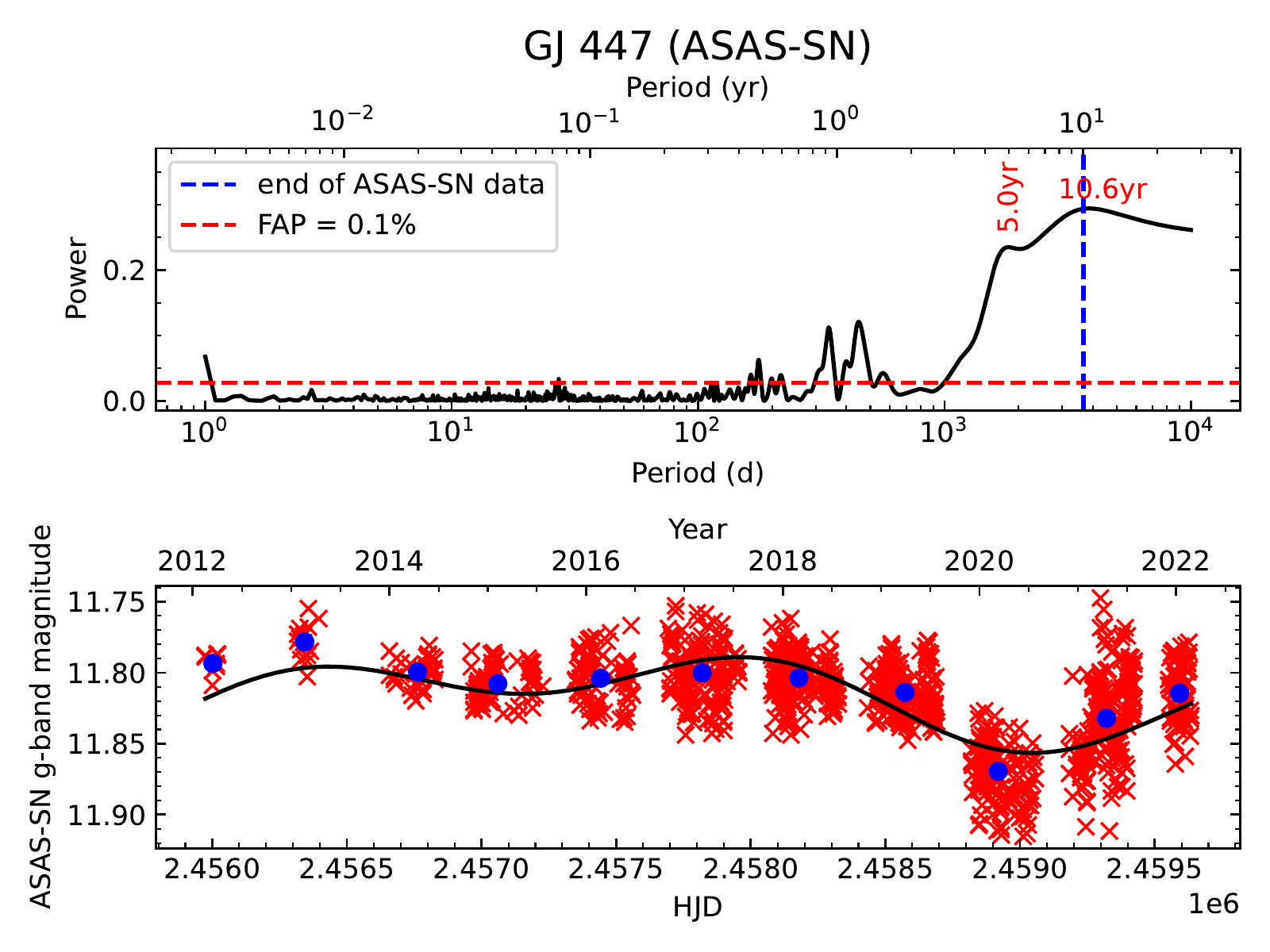}
\figsetgrpnote{Periodograms and ASAS-SN light curves for our sample of M dwarfs.}
\figsetgrpend

\figsetgrpstart
\figsetgrpnum{2.8}
\figsetgrptitle{Periodogram and ASAS-SN data for GJ 54.1.}
\figsetplot{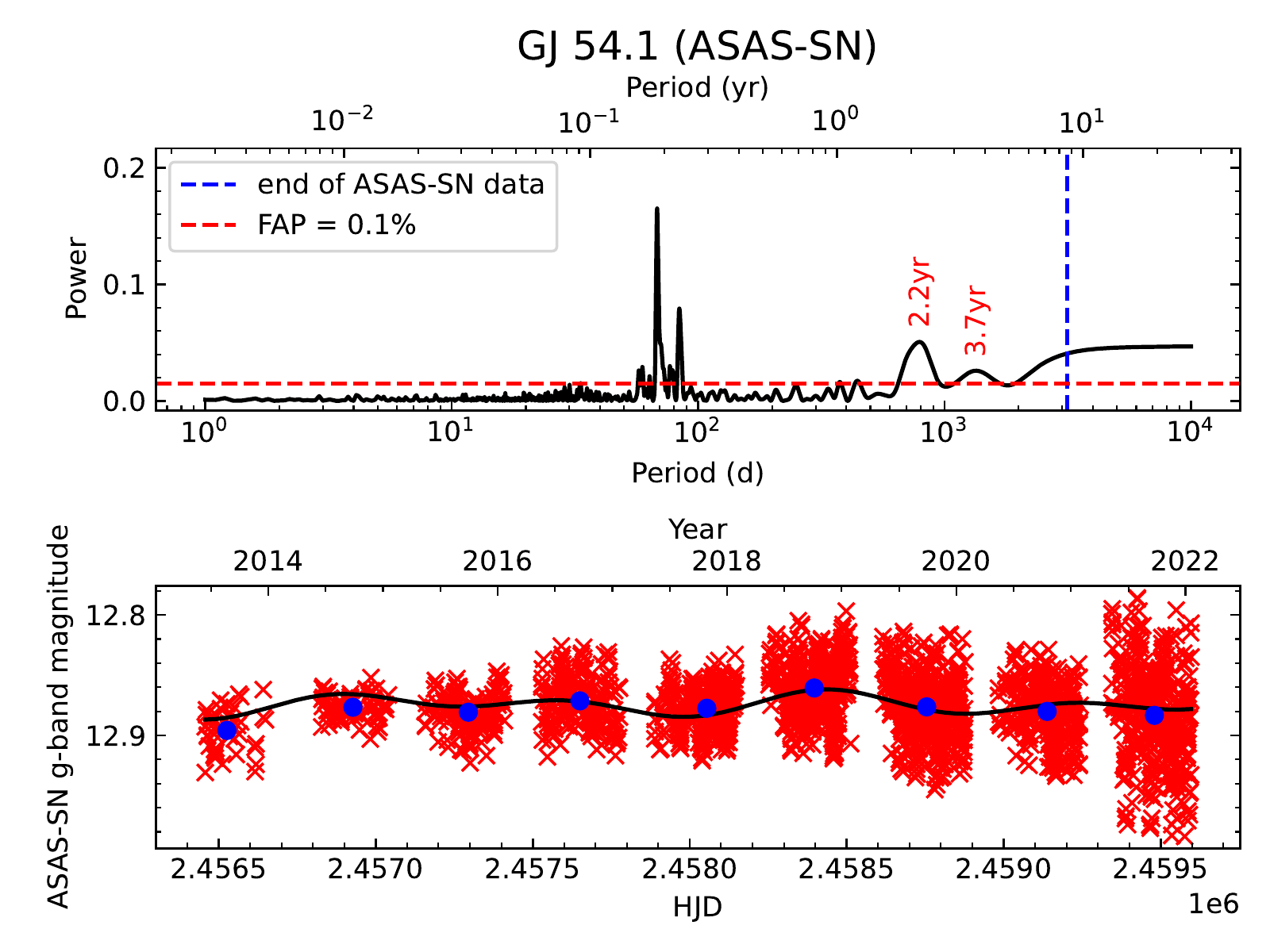}
\figsetgrpnote{Periodograms and ASAS-SN light curves for our sample of M dwarfs.}
\figsetgrpend

\figsetgrpstart
\figsetgrpnum{2.9}
\figsetgrptitle{Periodogram and ASAS-SN data for GJ 581.}
\figsetplot{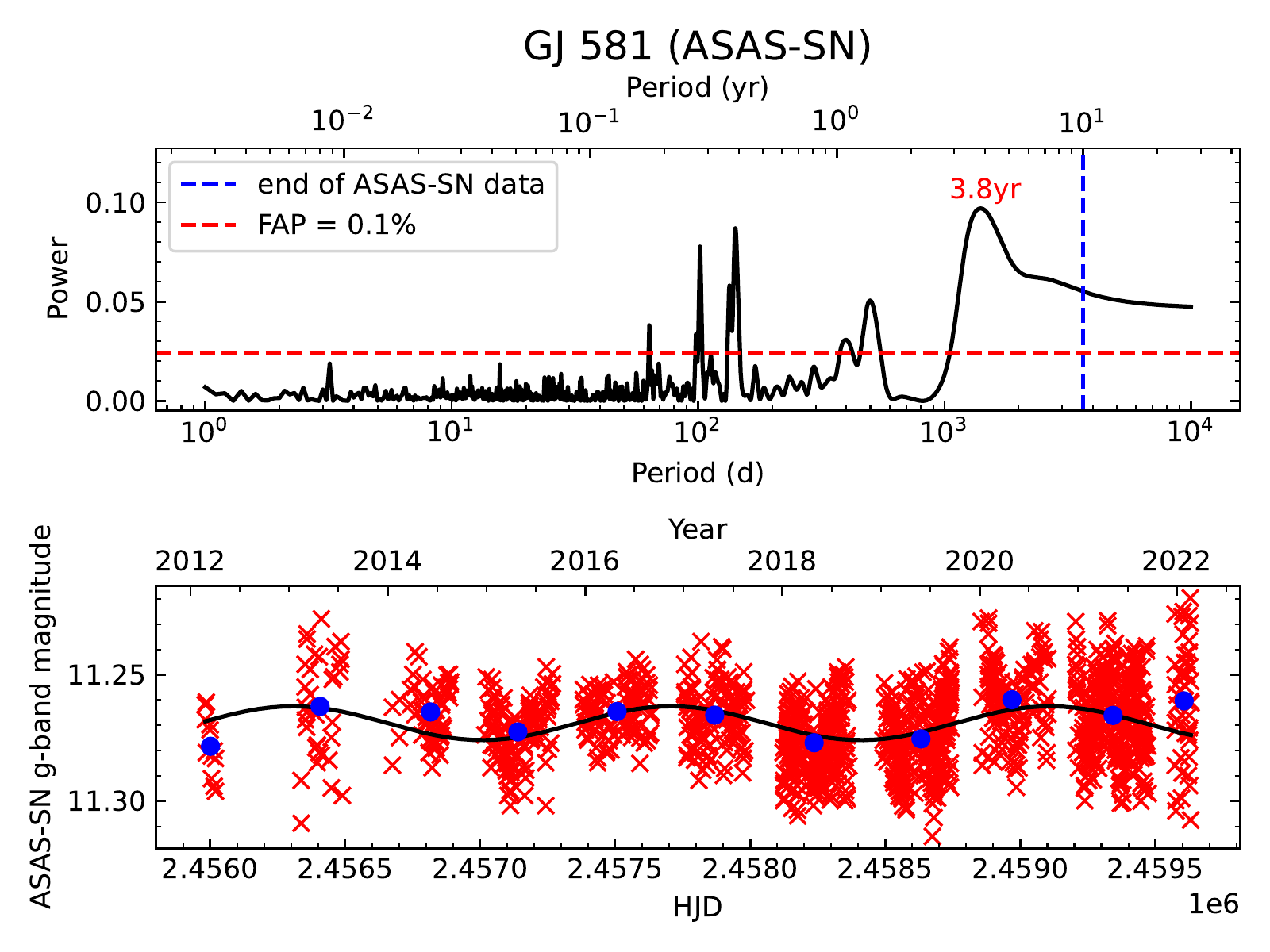}
\figsetgrpnote{Periodograms and ASAS-SN light curves for our sample of M dwarfs.}
\figsetgrpend

\figsetgrpstart
\figsetgrpnum{2.10}
\figsetgrptitle{Periodogram and ASAS-SN data for GJ 628.}
\figsetplot{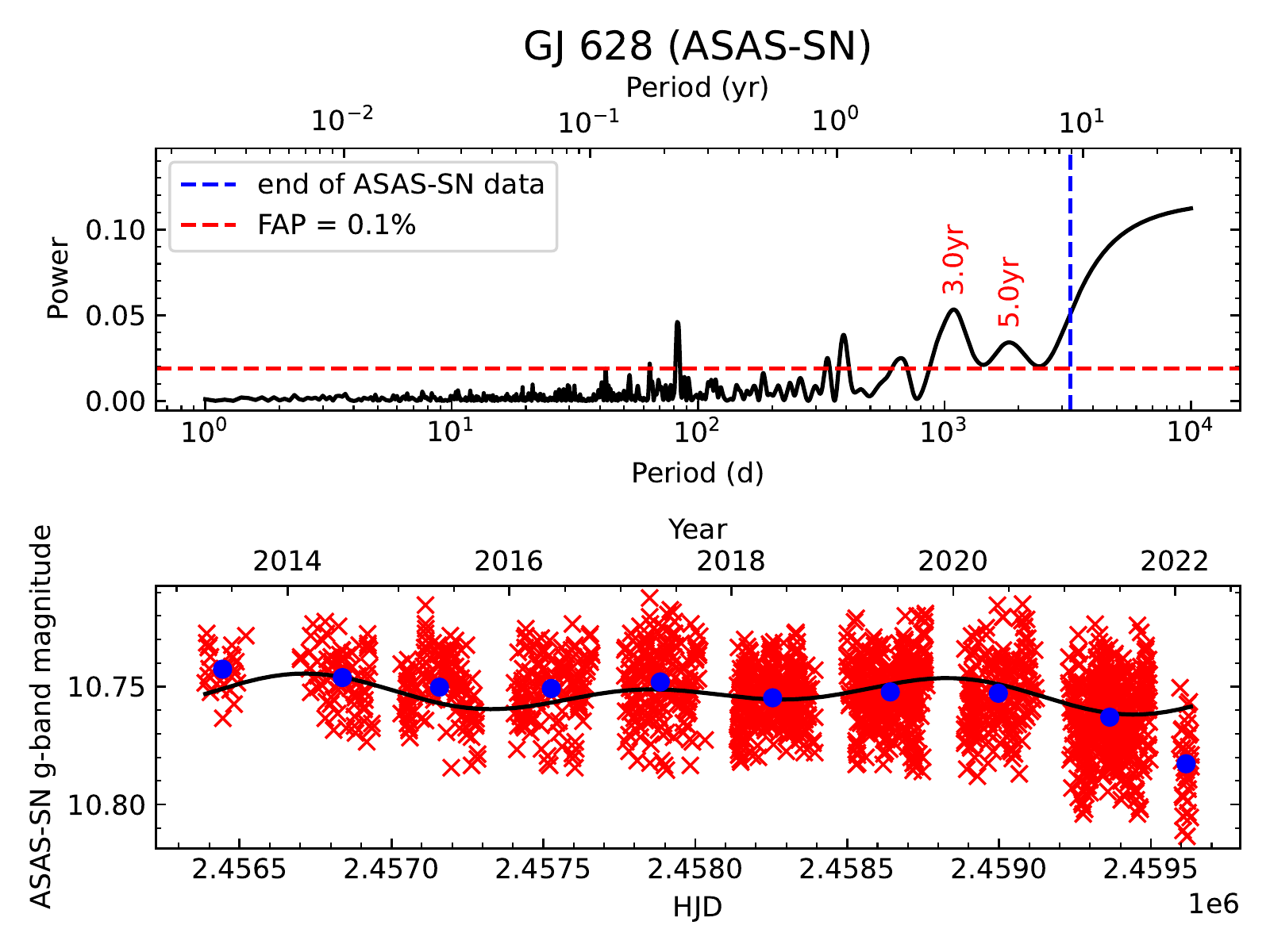}
\figsetgrpnote{Periodograms and ASAS-SN light curves for our sample of M dwarfs.}
\figsetgrpend

\figsetgrpstart
\figsetgrpnum{2.11}
\figsetgrptitle{Periodogram and ASAS-SN data for GJ 729.}
\figsetplot{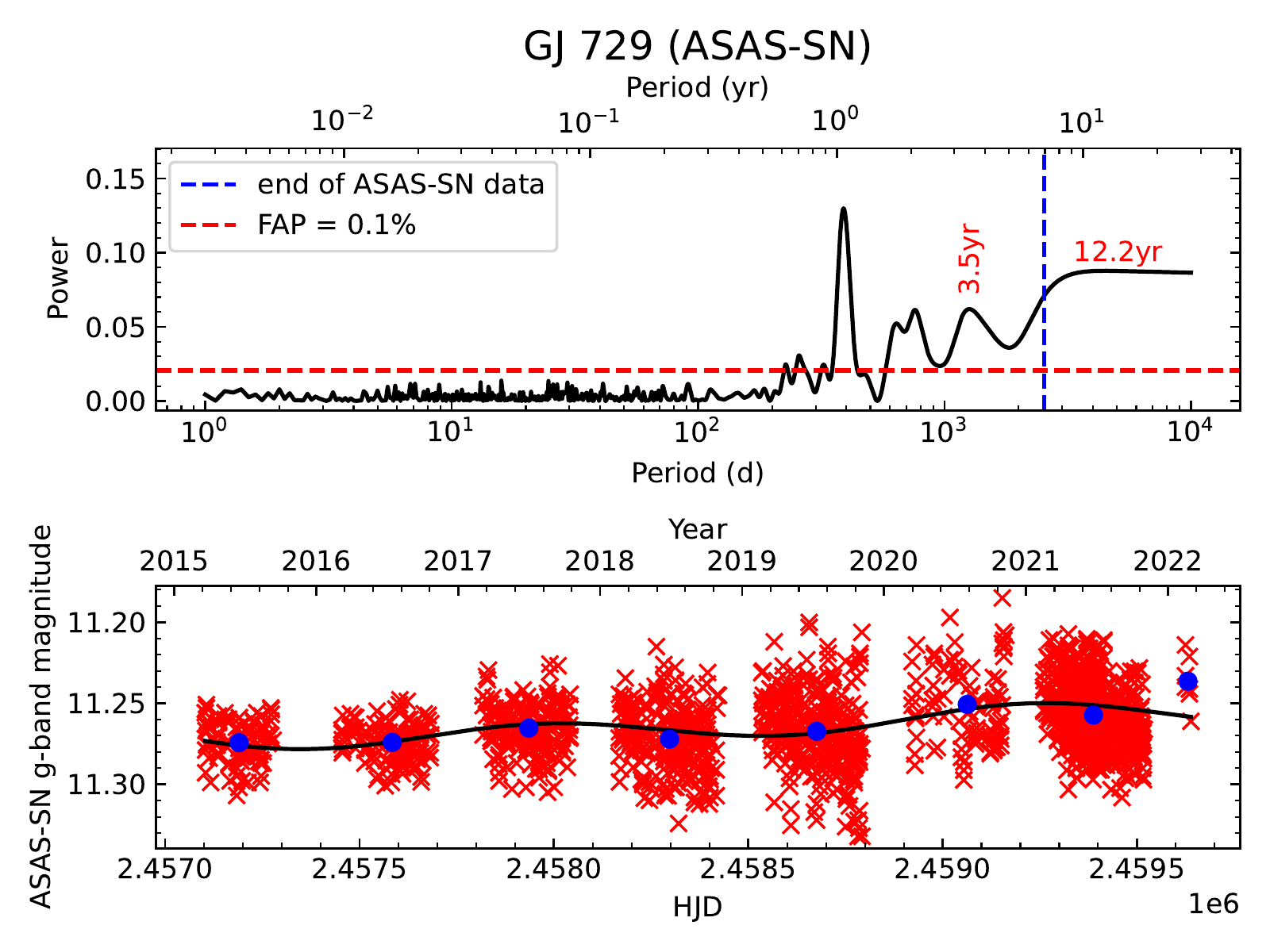}
\figsetgrpnote{Periodograms and ASAS-SN light curves for our sample of M dwarfs.}
\figsetgrpend

\figsetgrpstart
\figsetgrpnum{2.12}
\figsetgrptitle{Periodogram and ASAS-SN data for GJ 849.}
\figsetplot{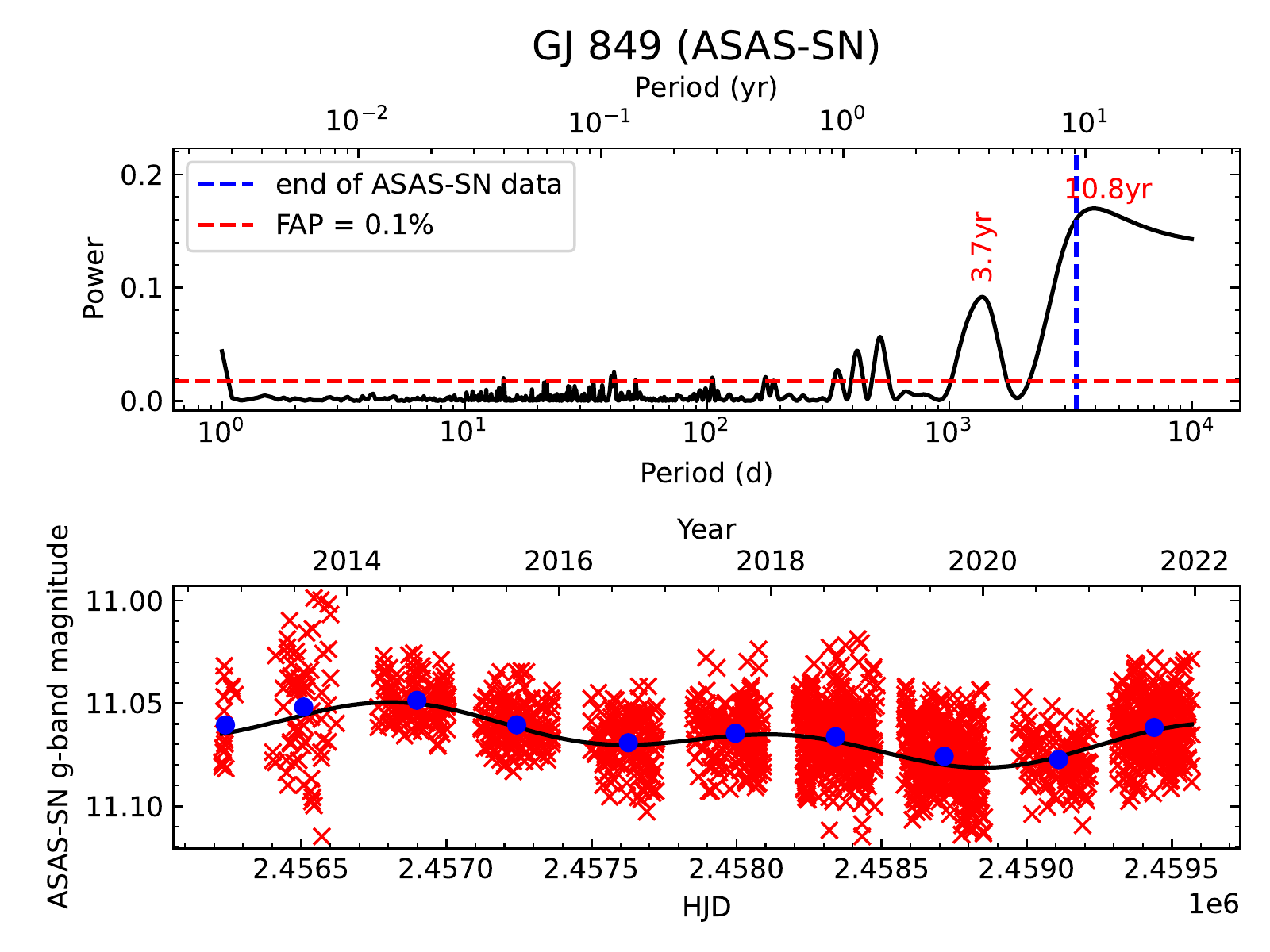}
\figsetgrpnote{Periodograms and ASAS-SN light curves for our sample of M dwarfs.}
\figsetgrpend

\figsetgrpstart
\figsetgrpnum{2.13}
\figsetgrptitle{Periodogram and ASAS-SN data for GJ 896A.}
\figsetplot{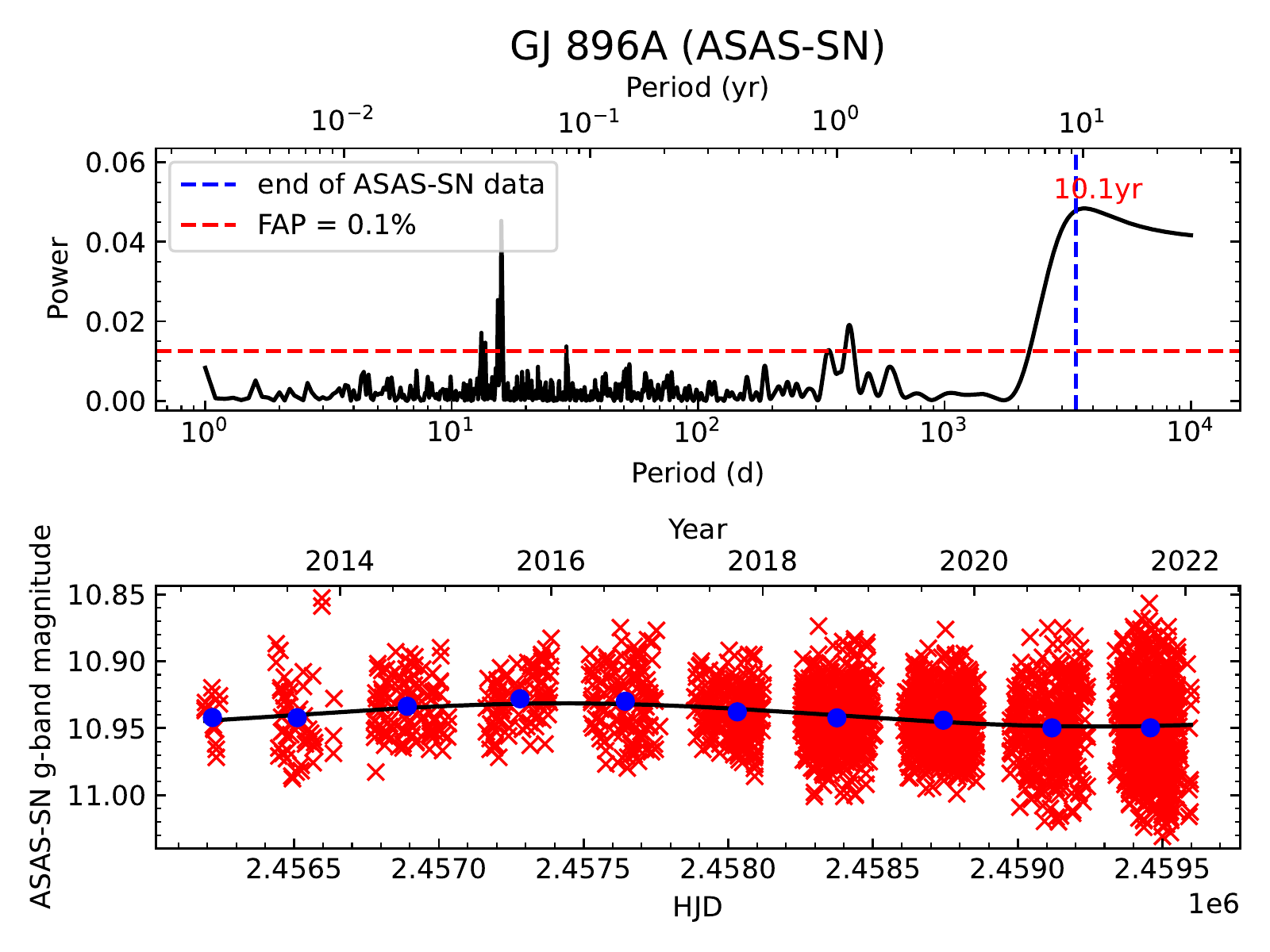}
\figsetgrpnote{Periodograms and ASAS-SN light curves for our sample of M dwarfs.}
\figsetgrpend

\figsetgrpstart
\figsetgrpnum{2.14}
\figsetgrptitle{Periodogram and ASAS-SN data for GJ LP 816-60.}
\figsetplot{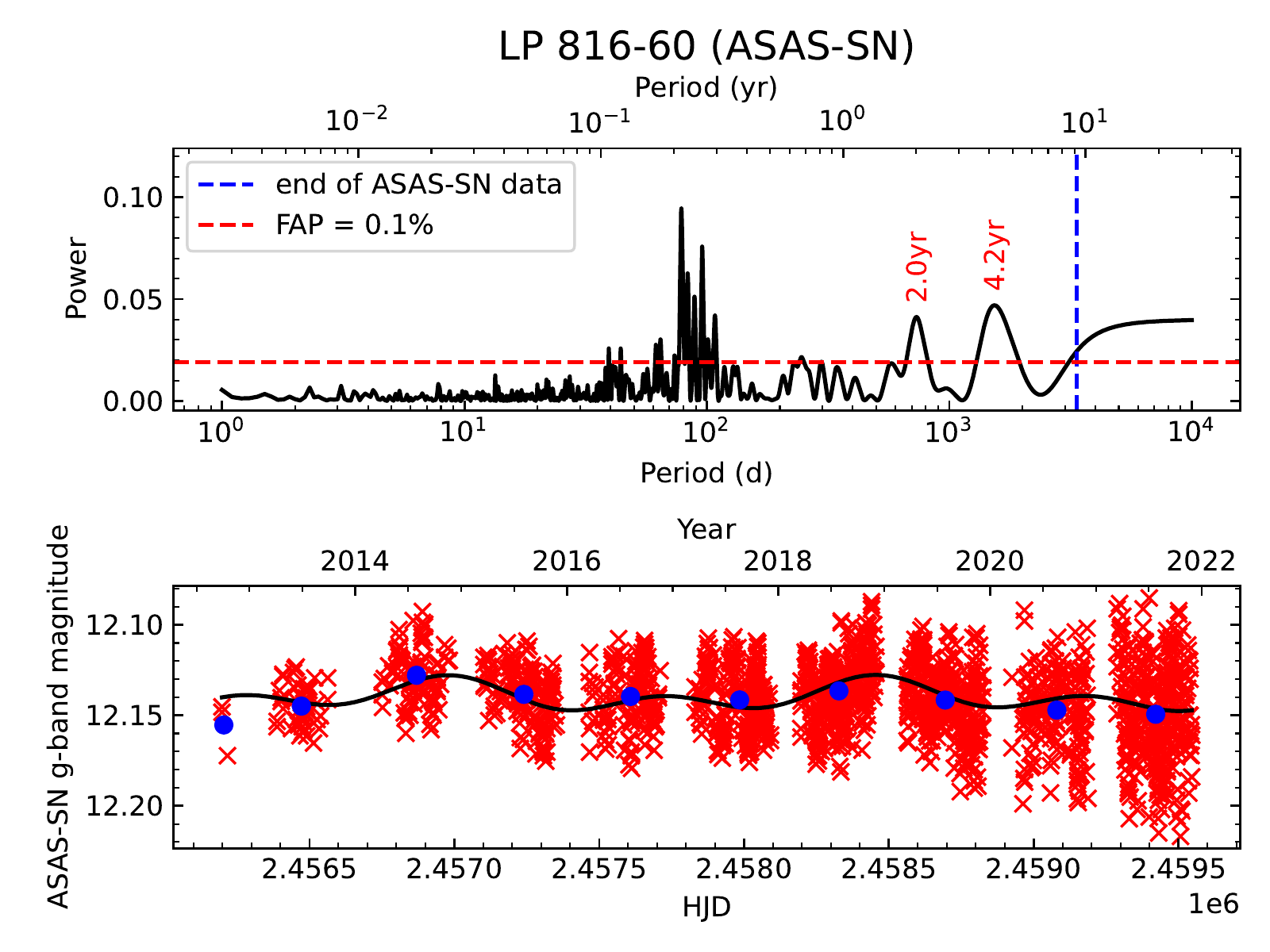}
\figsetgrpnote{Periodograms and ASAS-SN light curves for our sample of M dwarfs.}
\figsetgrpend

\figsetgrpstart
\figsetgrpnum{2.15}
\figsetgrptitle{Periodogram and ASAS-SN data for Proxima.}
\figsetplot{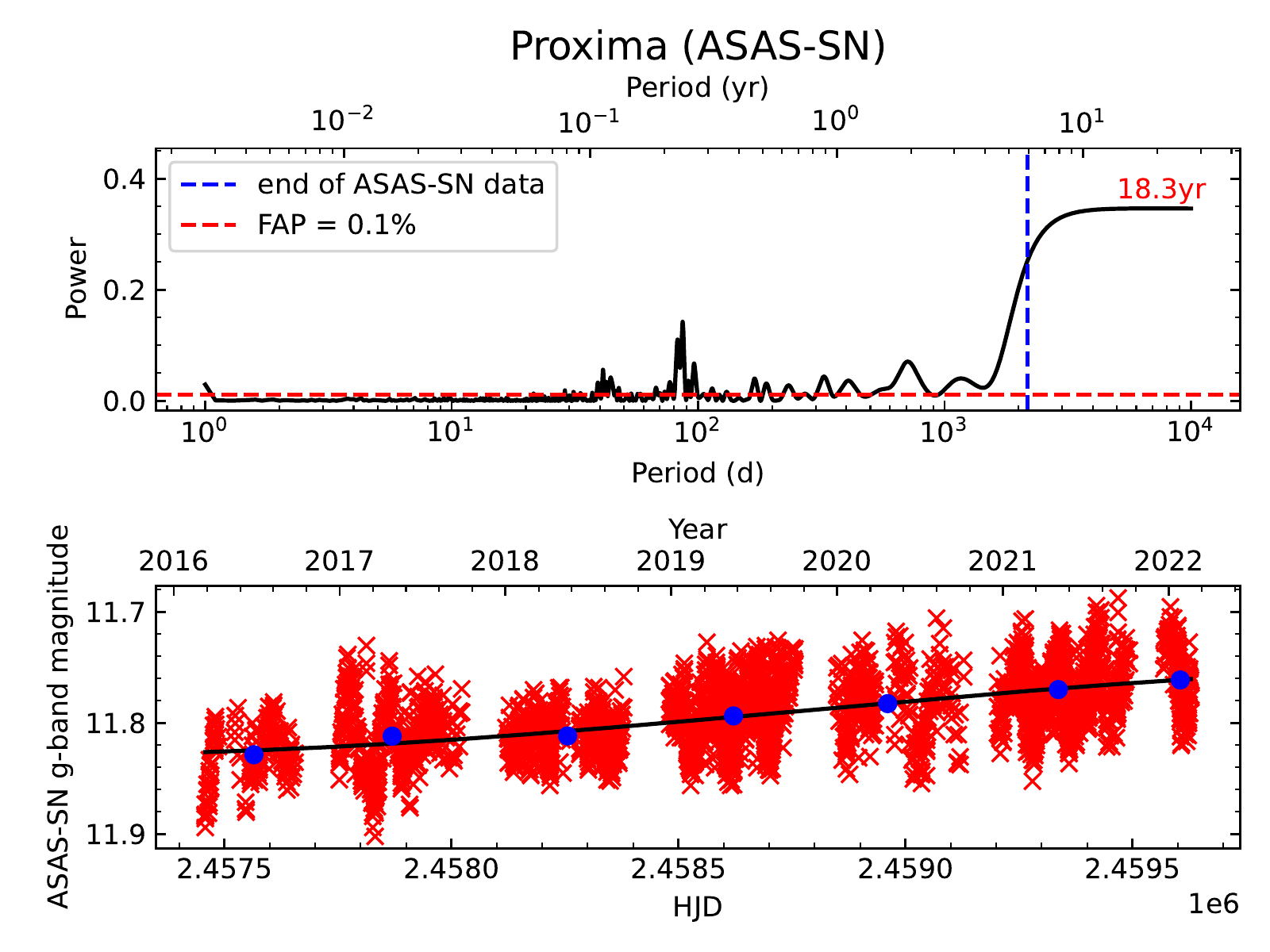}
\figsetgrpnote{Periodograms and ASAS-SN light curves for our sample of M dwarfs.}
\figsetgrpend

\figsetend

\section{Results}\label{sec: results}

Rotation and cycle periods are presented in Table \ref{tab: optical cycle periods}. All results from this work have FAPs $\leq 0.1\%$.

\begin{sidewaystable*}
    \centering
    .
    \\[250pt]
    \caption{Cycle Periods Inferred From Photometric Time Series.}
    \begin{tabular}{ccccccccccccc}
        \hline
        \hline
         &  &  &  &  &  & \multicolumn{2}{c}{ASAS-3} & \multicolumn{2}{c}{Previous ASAS-3 Results} & \multicolumn{2}{c}{ASAS-SN} \\
        Star & Sp. Type & V--K$_{\rm S}$ & $P_{\rm rot}$ & $P_{\rm rot}$ Lit. & Ref. & $P_{\rm cyc}$ & $A_{\rm cyc}$ & $P_{\rm cyc}$ & $A_{\rm cyc}$ & $P_{\rm cyc}$ & $A_{\rm cyc}$ & Phase \\
         &  &  & (d) & (d) &  & (yr) & (mmag) & (yr) & (mmag) & (yr) & (mmag) & Match \\
         \hline
        GJ 358$^{\rm *}$ & M3 & 4.638 & $25.2 \pm 1.1$ & $25.2 \pm 0.1$ & 2 & \bf 4.7 $\pm$ 0.9 & \bf 21.9 $\pm$ 1.9 & $4.9 \pm 0.1$ & $20.5 \pm 0.9$ & \bf 1.7 $\pm$ 0.2 & \bf 10.7 $\pm$ 0.6 & poor \\
        \nodata & \nodata & \nodata & \nodata & \nodata & \nodata & \nodata & \nodata & \nodata & \nodata & {\bf 3.2 $\pm$ 0.5} & {\bf 4.0 $\pm$ 0.6} & \nodata \\
        GJ 581 & M3 & 4.723 & $141.6 \pm 3.5$ & $132.5 \pm 6.3^{\rm c}$ & 1 & \nodata & \nodata & $6.2 \pm 0.9$ & 3.8 $\pm$ 0.9 & \bf 3.8 $\pm$ 0.6 & \bf 6.7 $\pm$ 0.6 & \nodata \\ 
        GJ 628 & M3 & 4.997 & $79.6 \pm 2.6$ & $119.3 \pm 0.5$ & 2 & \bf 4.5 $\pm$ 1.3 & \bf 8.1 $\pm$ 1.2 & $4.4 \pm 0.2$ & 8.3 $\pm$ 1.1 & \bf 3.0 $\pm$ 0.5 & \bf 5.0 $\pm$ 0.5 & good \\
        \nodata & \nodata & \nodata & \nodata & \nodata & \nodata & \nodata & \nodata & \nodata & \nodata & \bf 5.0 $\pm$ 1.2 & \bf 3.8 $\pm$ 0.5 & \nodata \\ 
        \nodata & \nodata & \nodata & \nodata & \nodata & \nodata & \nodata & \nodata & \nodata & \nodata & \it long & \nodata & \nodata \\ 
        GJ 849$^{\rm *}$ & M3.5 & 4.772 & $41.4 \pm 1.3$ & $39.2 \pm 6.3^{\rm c}$ & 1 & \it 10.3 $\pm$ 3.0 & \it 25.6 $\pm$ 1.5 & $10.2 \pm 0.9$ & 12.4 $\pm$ 1.0 & \bf 3.7 $\pm$ 0.6 & \bf 6.7 $\pm$ 0.5 & poor \\ 
        \nodata & \nodata & \nodata & \nodata & \nodata & \nodata & \nodata & \nodata & \nodata & \nodata & \it 10.8 $\pm$ 3.0 & \it 9.5 $\pm$ 0.5 & \nodata \\
        GJ 896A$^{\rm *}$ & M3.5 & 4.847 & $15.58 \pm 0.06$ & $15.3 \pm 0.1$ & 2 & {\it long} & \nodata & \nodata & \nodata & \it 10.1 $\pm$ 2.7 & \it 8.7 $\pm$ 0.7 & \nodata \\
        GJ 317$^{\rm *}$ & M3.5 & 4.947 & $57.5 \pm 1.7$ & \nodata & \nodata & \it long & \nodata & $5.2 \pm 0.3$ & 12.4 $\pm$ 1.1 & \bf 2.3 $\pm$ 0.3 & \bf 5.7 $\pm$ 0.6 & \nodata \\ 
        \nodata & \nodata & \nodata & \nodata & \nodata & \nodata & \nodata & \nodata & \nodata & \nodata & \it 10.1 $\pm$ 4.4 & \it 9.6 $\pm$ 0.6 & \nodata \\ 
        GJ 273 & M3.5 & 5.015 & \nodata & $115.9 \pm 19.4^{\rm c}$ & 1 & \bf 5.8 $\pm$ 1.3 & \bf 9.4 $\pm$ 1.2 & $6.6 \pm 1.3$ & 7.7 $\pm$ 1.5 & \bf 5.9 $\pm$ 1.4 & \bf 10.1 $\pm$ 0.8 & good \\
        GJ 729 & M3.5 & 5.06 & $2.9 \pm 0.1$ & $2.9 \pm 0.1$ & 2 & \bf 5.7 $\pm$ 1.1 & \bf 10.4 $\pm$ 1.3 & $7.1 \pm 0.1$ & 5.3 $\pm$ 0.7 & \bf 3.5 $\pm$ 1.1 & \bf 6.8 $\pm$ 0.7 & fair \\
        \nodata & \nodata & \nodata & \nodata & \nodata & \nodata & \nodata & \nodata & \nodata & \nodata & \it 12.2 $\pm$ 6.7 & \it 8.1 $\pm$ 0.7 & \nodata \\ 
        LP 816-60 & M4 & 5.259 & $86.3 \pm 02$ & $67.6 \pm 0.1$ & 2 & \nodata & \nodata & 10.6 $\pm$ 1.7$^{\rm f}$ & 8.2 $\pm$ 1.2$^{\rm f}$ & \bf 2.0 $\pm$ 0.2 & \bf 6.0 $\pm$ 0.8 & \nodata \\ 
        \nodata & \nodata & \nodata & \nodata & \nodata & \nodata & \nodata & \nodata & \nodata & \nodata & \bf 4.2 $\pm$ 0.7 & \bf 5.9 $\pm$ 0.6 & \nodata \\
        GJ 447 & M4 & 5.499 & $175.9 \pm 3.4$ & $165.1 \pm 0.8^{\rm e}$ & 2 & \bf 5.3 $\pm$ 1.3 & \bf 9.8 $\pm$ 1.4 & $4.1 \pm 0.3^{\rm g}$ & 7.1 $\pm$ 1.1$^{\rm g}$ & \bf 5.0 $\pm$ 1.1 & \bf 19.6 $\pm$ 1.0 & poor \\
        \nodata & \nodata & \nodata & \nodata & \nodata & \nodata & \nodata & \nodata & \nodata & \nodata & \it 10.6 $\pm$ 6.9 & \it 21.5 $\pm$ 1.0 & \nodata \\ 
        GJ 285 & M4 & 5.527 & $2.775 \pm 0.001$ & $2.8 \pm 0.1$ & 2 & \it 10.0 $\pm$ 4.1 & \it 39.1 $\pm$ 2.2 & $10.6 \pm 0.4$ & 37.2 $\pm$ 1.1 & \it 10.9 $\pm$ 4.0 & \it 37.2 $\pm$ 1.1 & good \\ 
        GJ 54.1 & M4 & 5.654 & $71.4 \pm 2.4$ & $69.2 \pm 0.1$ & 2 & \nodata & \nodata & \nodata & \nodata & \bf 2.2 $\pm$ 0.3 & \bf 6.8 $\pm$ 0.6 & \nodata \\ 
        \nodata & \nodata & \nodata & \nodata & \nodata & \nodata & \nodata & \nodata & \nodata & \nodata & \bf 3.7 $\pm$ 0.8 & \bf 5.8 $\pm$ 0.7 & \nodata \\ 
        \nodata & \nodata & \nodata & \nodata & \nodata & \nodata & \nodata & \nodata & \nodata & \nodata & \it long & \nodata & \nodata \\
        GJ 234$^{\rm a}$ & M4.5 & 5.585 & $1.582 \pm 0.002$ & $8.1 \pm 0.1^{\rm d}$ & 2 & \bf 5.6 $\pm$ 1.5 & \bf 10.9 $\pm$ 1.4 & $5.9 \pm 0.5$ & 10.1 $\pm$ 1.4 & \bf 5.2 $\pm$ 1.0 & \bf 13.1 $\pm$ 0.5 & fair \\
        GJ 551$^{\rm b}$ & M5.5 & 6.75 & $85.1 \pm 1.2$ & $83.2 \pm 0.1^{\rm c}$ & 2 & \bf 6.7 $\pm$ 1.4 & \bf 16.3 $\pm$ 1.5 & $6.8 \pm 0.3$ & 15.5 $\pm$ 0.9 & \it 18.3 $\pm$ 14.2 & \it 39.5 $\pm$ 1.1 & poor \\
        \nodata & \nodata & \nodata & \nodata & \nodata & \nodata & 5.1 $\pm$ 1.0$^{\rm h}$ & 20.0 $\pm$ 3.0$^{\rm h}$ & \nodata & \nodata & \nodata & \nodata & \nodata \\
        \nodata & \nodata & \nodata & \nodata & \nodata & \nodata & \it long$^{\rm h}$ & \nodata & \nodata & \nodata & \nodata & \nodata & \nodata \\
        GJ 406$^{\rm b}$ & M6 & 7.423 & $2.6 \pm 0.2$ & \nodata & \nodata & \it 13.7 $\pm$ 7.3 & \it 128.4 $\pm$ 22.4 & $8.9 \pm 0.2$ & 64.8 $\pm$ 1.2 & \it 12.0 $\pm$ 4.4 & \it 93.3 $\pm$ 1.1 & fair \\
        \hline
    \end{tabular}
    \\[0pt]
    \justifying
    \tablecomments{$^{\rm *}$ partially convective (see Section \ref{sec: convectiveness}); $^{\rm a}$ known binary system; $^{\rm b}$ photometric monitoring may include significant contamination from physically unassociated nearby objects on the sky (see Section \ref{sec: contamination}); $^{\rm c}$ FAP $\leq$ 0.3\%; $^{\rm d}$ FAP = 0.3\%; $^{\rm e}$ FAP $\leq$ 0.2\%; $^{\rm f}$ FAP = 3.0\%; $^{\rm g}$ FAP = 1.7\%; $^{\rm h}$ from ASAS-4 data; ``{\it long}" denotes high amplitude at large $P_{\textrm{cyc}}$ in power spectrum without well defined peak. All cycle periods reported from this work have FAPs $\leq 0.1 \%$; likewise for previous results unless noted. Cycles in bold font are deemed well-defined, while poorly constrained cycles are italicized. The ``Phase Match" column gives a subjective measure of how well the phases (and amplitudes) of the cycles found in this work agree where they overlap. References: 1) \protect\cite{S-M2015}; 2) \protect\cite{S-M2016}. Previous ASAS-3 results were taken from Reference 2.}
    \label{tab: optical cycle periods}
\end{sidewaystable*}

\subsection{Rotation Periods}\label{sec: optical rotation periods}

\begin{figure}
    \centering
    \includegraphics[width=\columnwidth]{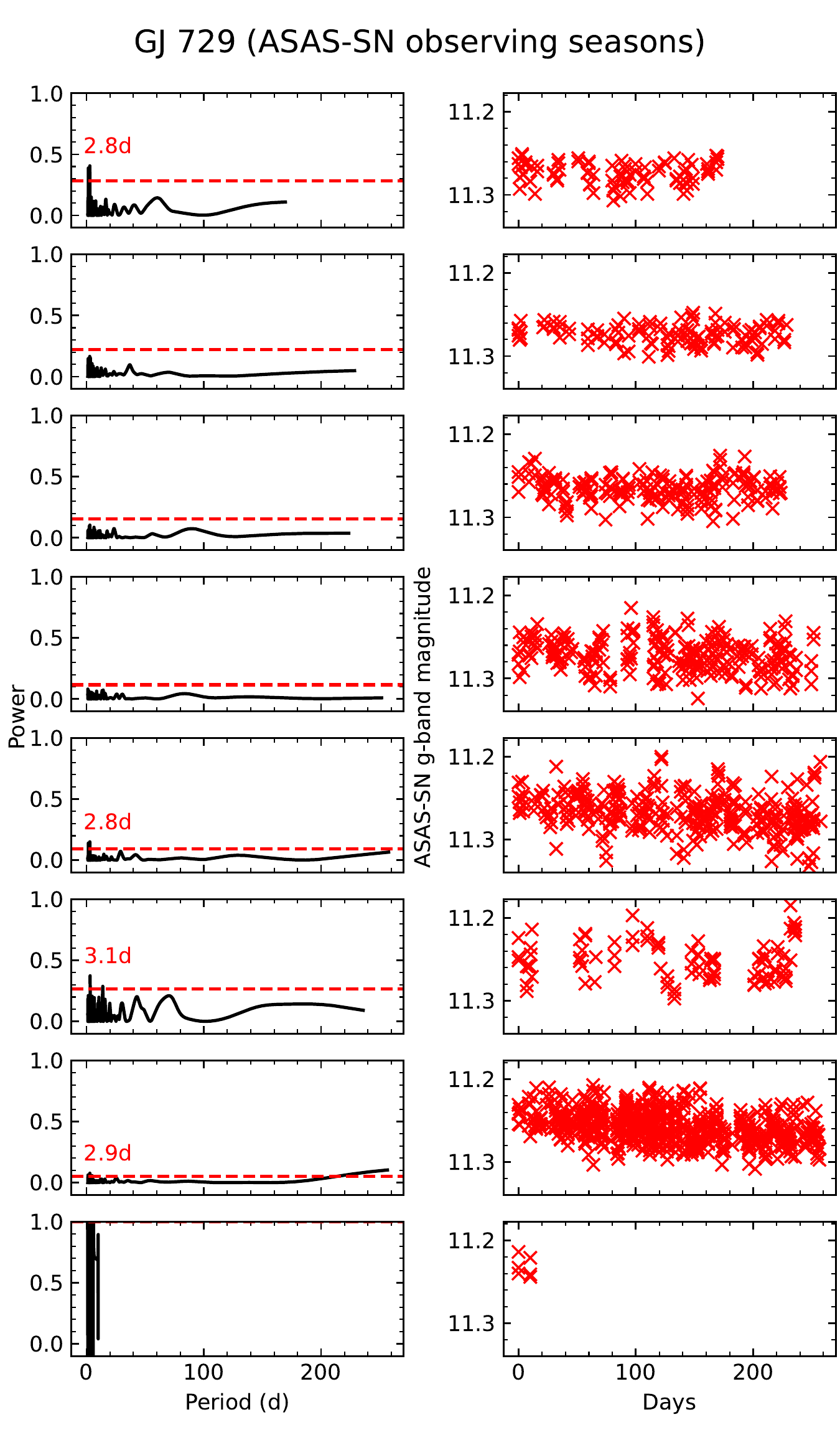}
    \caption{L--S periodograms (left) computed from ASAS-SN observing seasons (right) on GJ 729. Due to the short inferred rotation periods, no sine fits are shown as the modulations would be unresolvable.}
    \label{fig: GJ 729 ASAS-SN obs seasons}
\end{figure}

Using the analysis procedures described in Section \ref{sec: optical rotation method}, high-confidence rotation periods were determined for all but two of our stars. For one exception, GJ 729, $P_{\rm rot}$ is less than 15d but TESS data were not available for our work. We therefore used the same method as for stars with longer rotation periods, computing the mean and standard deviation of similar significant peaks in each ASAS-SN observing season's periodogram; these periodograms are shown in Figure \ref{fig: GJ 729 ASAS-SN obs seasons}. This resulted in a rotation period of $2.9 \pm 0.1$d, in agreement with \cite{S-M2016, Ibanez2020}, who found a rotation periods of $2.9 \pm 0.1$d (from photometric monitoring) and $2.848 \pm 0.001$d (from chromospheric indicators), respectively.

\begin{figure}
    \centering
    \includegraphics[width=\columnwidth]{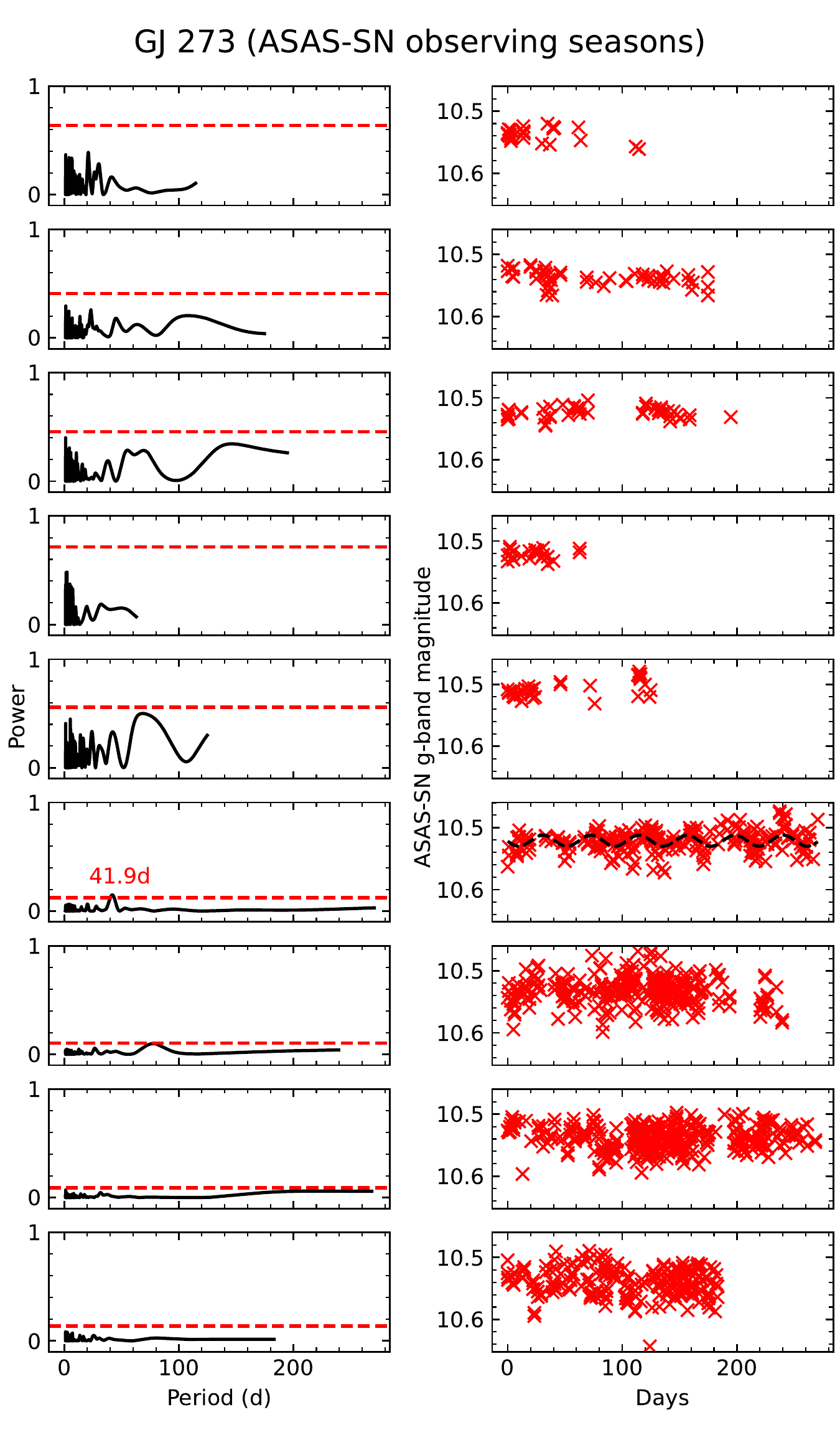}
    \caption{L--S periodograms (left) computed from ASAS-SN observing seasons (right) on GJ 273.}
    \label{fig: GJ 273 ASAS-SN obs seasons}
\end{figure}

\begin{figure}
    \centering
    \includegraphics[width=\columnwidth]{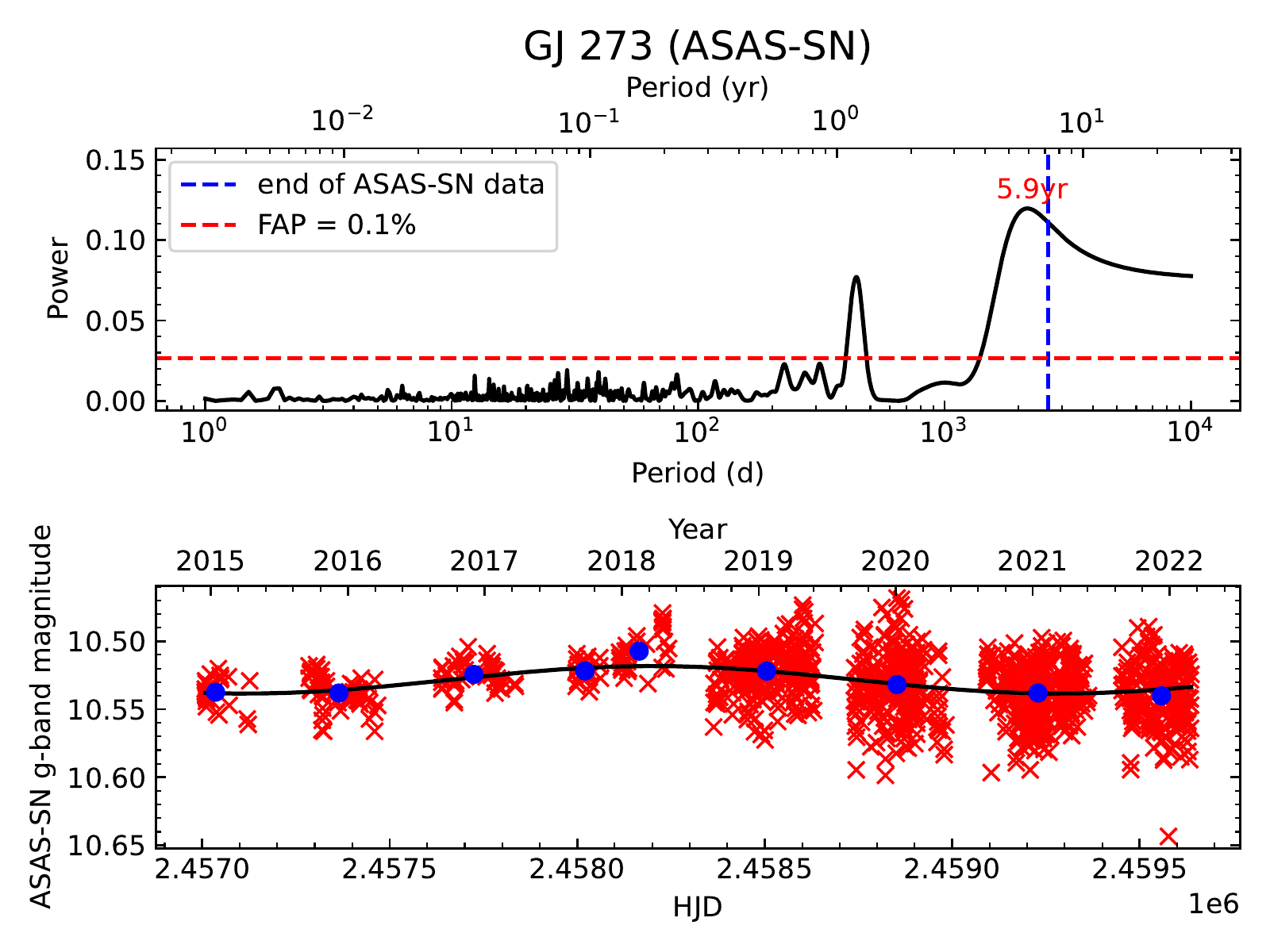}
    \caption{Periodogram and ASAS-SN data for GJ 273.}
    \label{fig: GJ 273 ASAS-SN L-S periodogram}
\end{figure}

For the second exception, GJ 273, we could not determine a rotation period from our data sources. Breaking ASAS-SN data down by observing season, as described in Section \ref{sec: optical rotation method}, and shown in Figure \ref{fig: GJ 273 ASAS-SN obs seasons}, produced only one significant peak at $\sim$42d. However, no other observing seasons showed any significant peaks at all. Computing periodograms from the entire ASAS-SN light curve, as shown in Figure \ref{fig: GJ 273 ASAS-SN L-S periodogram}, also produced no significant peaks with a period of less than one year. Figures \ref{fig: GJ 273 ASAS-SN obs seasons} and \ref{fig: GJ 273 ASAS-SN L-S periodogram} show that this star has fewer ASAS-SN observations than many of the other stars in our sample, especially in the early seasons (which use V-band filters). The sparsity of these observations undoubtedly contributed to our inability to identify rotational modulations within this light curve. For these reasons, we adopted the $P_{\rm rot}$ of $115.9 \pm 19.4$d measured by \cite{S-M2015} using HARPS {Ca\,{\sc ii}} H and K and H$\alpha$ data.

Given analysis uncertainties and the inherent variation in rotational measurements caused by starspot evolution, migration in latitude and longitude, and differential rotation, our results are generally in good agreement with previous measurements. Where there are significant disagreements (GJ 234, GJ 628, and LP 816-60) we believe our measurements are more trustworthy given the higher quality and cadence of TESS and ASAS-SN data compared to the ASAS-3 data used by \cite{S-M2016}. Note that two of our stars, GJ 317 and GJ 406, have $P_{\rm rot}$ values reported for the first time.

\subsection{Cycle Periods}\label{sec: optical cycle periods}

L--S analysis yields FAPs of less than 0.1\% using ASAS-3/-4 and/or ASAS-SN data for 12 stars, not including a few cases where the significant peak corresponds to a cycle period exceeding the period of observation. Below, we comment on three stars of particular note. For stars not mentioned below, light curves and associated L--S periodograms are provided in Figure Sets 1 (ASAS-3/-4) and 2 (ASAS-SN).

\subsubsection{LP 816-60}\label{sec: LP 816-60 optical cycle}

Figure \ref{fig: LP 816-60 ASAS-SN sine fit} provides strong evidence for at least one activity cycle in LP 816-60. From this figure, it is clear that there are two strong periodic signals in LP 816-60's ASAS-SN light curve, with periods of $2.0$- and $4.2$-yr. Upon first inspection, one of these peaks could be interpreted as a harmonic of the other. Indeed, 4.2 is very close to double 2.0, and taking into account the uncertainties on these periods, one could well be a harmonic of the other. 

Evidence presented by do Nascimento et al. (in preparation) suggests that it is not uncommon to measure cycles with $P_{\rm cyc}$ and $P_{\rm cyc}/2$. In that paper, they interpret such results as the separate Hale (magnetic polarity) and Schwabe (activity, starspots) cycles having different amplitudes due to (for reasons which are currently unclear) one polarity being stronger. Given the two clear cycles visible in LP 816-60's ASAS-SN data, this star may therefore provide evidence for M dwarfs also possessing (or, at least, appearing to have) asymmetric polarity strengths. Another possibility is that both peaks are the result of a single quasi-periodic cycle, as discussed in \cite{Olspert2018b}. 

Alternatively, since these cycles have been inferred using optical photometry data, they could be the result of an observing effect. Assuming a 4.2-yr cycle period, it could be the case that for one half of this cycle, starspots are preferentially generated out of our line of sight (i.e., sin(i) $\ll$ 1) - although the clear rotational modulations make this less likely. Without Zeeman-Doppler imaging (ZDI) \citep[e.g.,][]{Hebrard2016}, or X-ray observations \citep[e.g.,][]{Wargelin2017}, however, it is difficult to determine whether an observed cycle is a result of magnetic activity (specifically a Schwabe cycle) or some other phenomenon.

\subsubsection{GJ 358}\label{sec: GJ 358 cycle}

\begin{figure}
    \centering
    \includegraphics[width=\columnwidth]{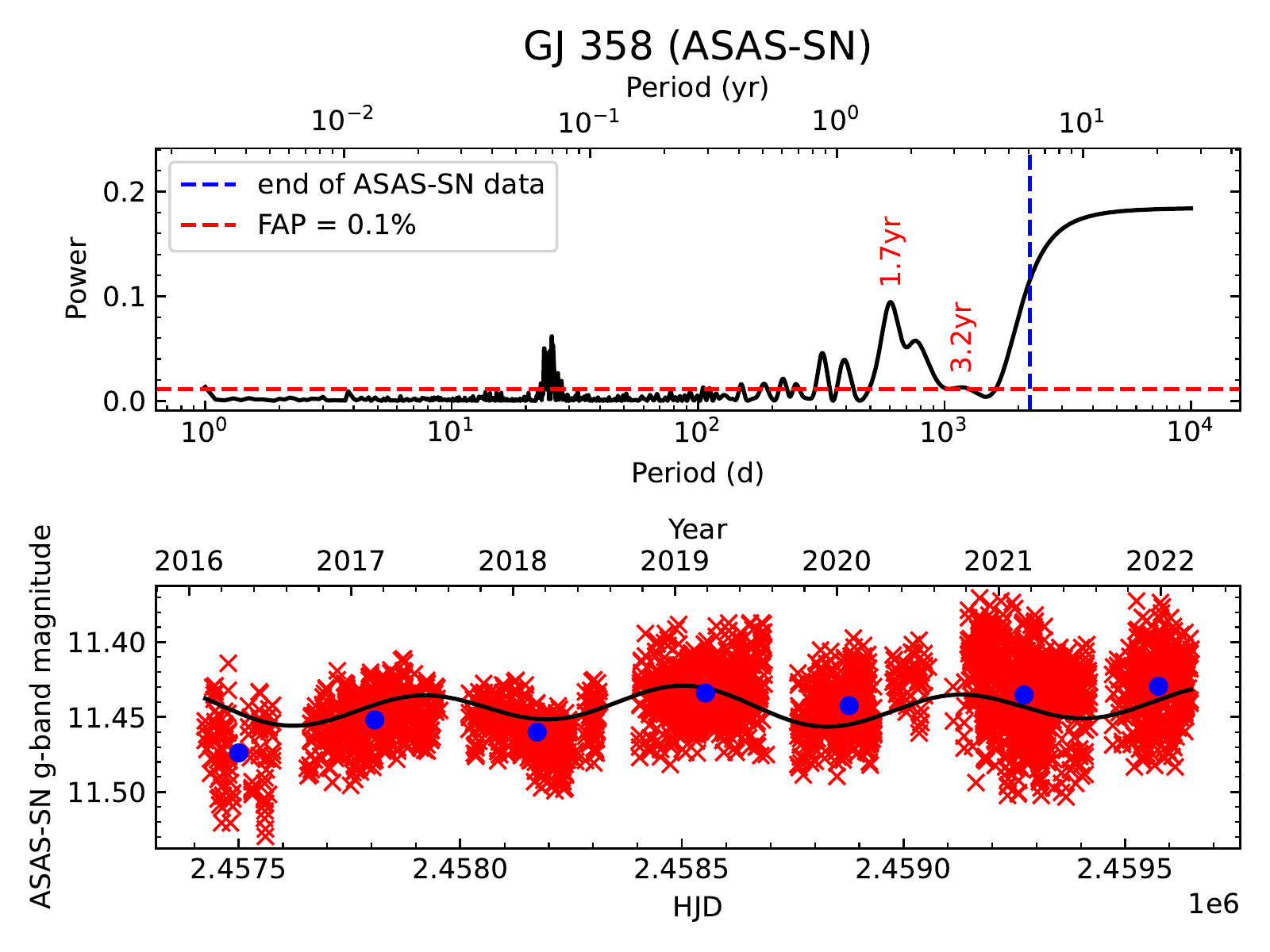}
    \caption{Periodogram and ASAS-SN data for GJ 358.}
    \label{fig: GJ 358 ASAS-SN}
\end{figure}

\begin{figure}
    \centering
    \includegraphics[width=\columnwidth]{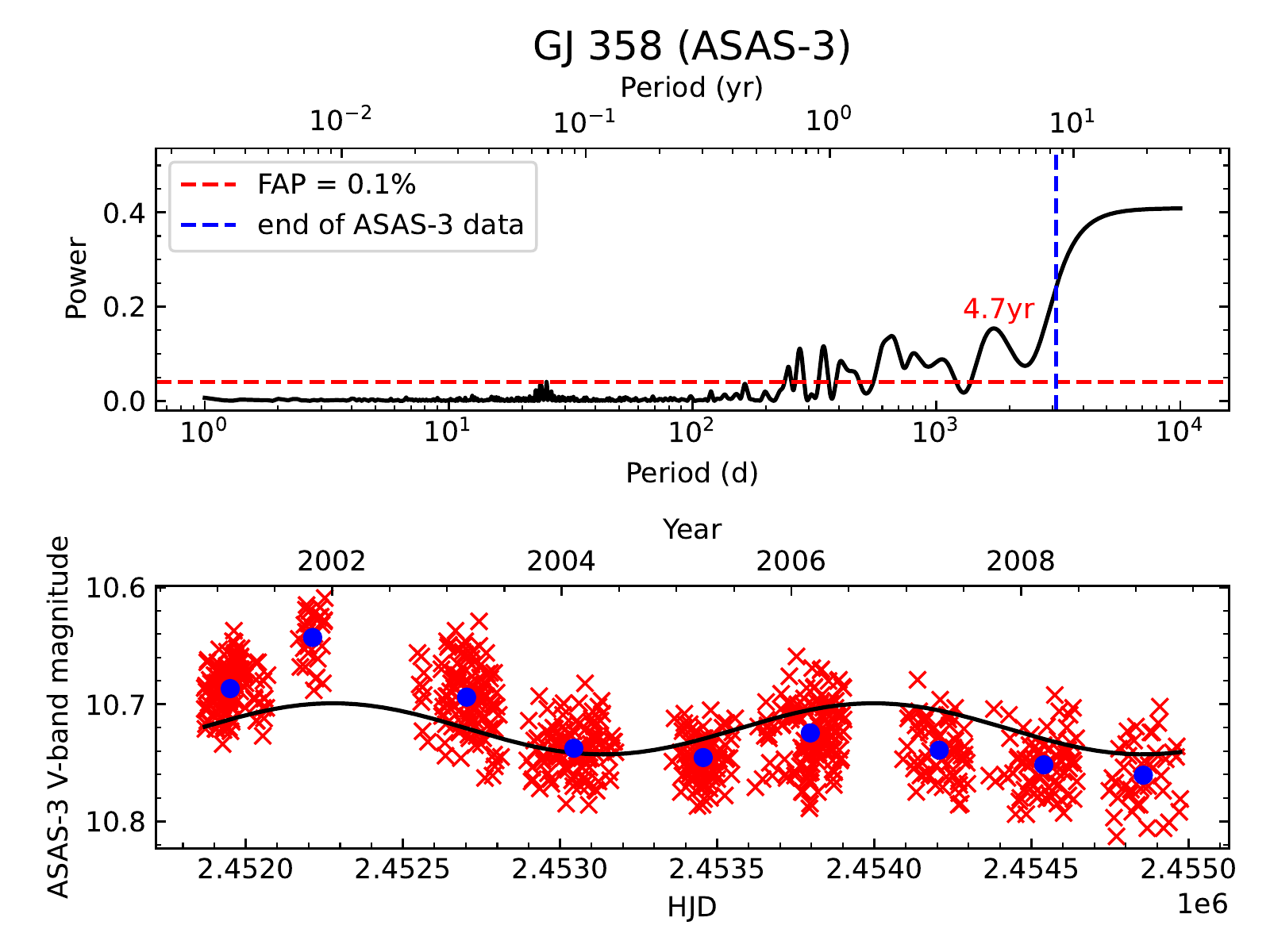}
    \caption{Periodogram and ASAS-3 data for GJ 358.}
    \label{fig: GJ 358 ASAS-3}
\end{figure}

GJ 358's ASAS-SN light curve (Figure \ref{fig: GJ 358 ASAS-SN}) shows that the cyclic modulations appear to be reasonably well described by two sinusoidal functions, following the $P_{\rm cyc}$ and $P_{\rm cyc}/2$ rule suggested by do Nascimento et al. (in preparation). Interestingly, both the ASAS-3 and ASAS-SN light curves (Figures \ref{fig: GJ 358 ASAS-3} and \ref{fig: GJ 358 ASAS-SN}) show long-term ``linear" trends whose gradients are of opposite sign; this may suggest that GJ 358 exhibits a longer-term, Gleissberg-esque cycle, although many more years of data will be needed to confirm this.

\begin{figure}
    \centering
    \includegraphics[width=\columnwidth]{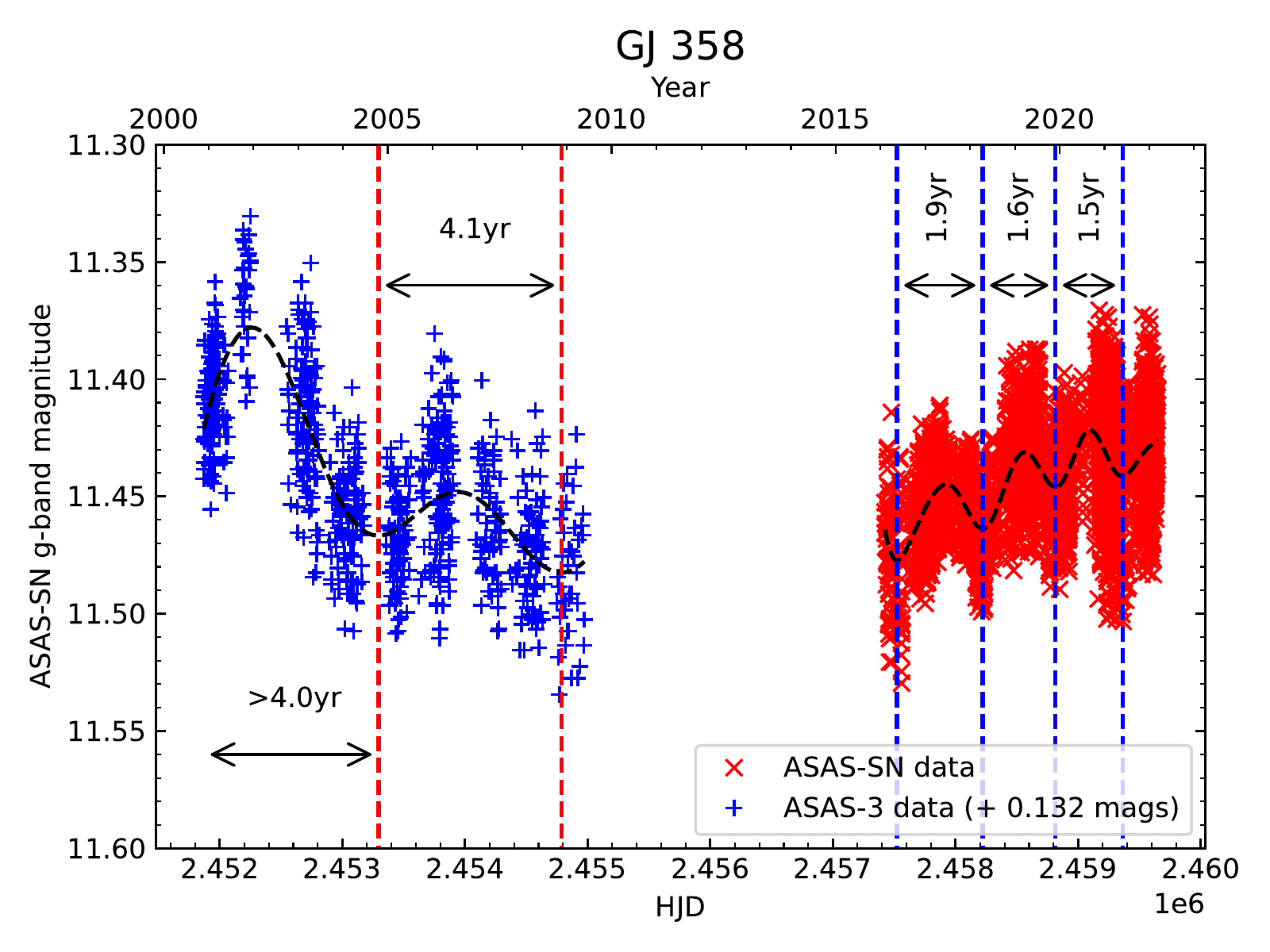}
    \caption{GP fits (dashed black line) to ASAS-3 and ASAS-SN data (blue and red markers, respectively) on GJ 358. ASAS-3 data were color corrected using Equation \ref{eq: colour correction}, and then further offset, to align vertically with ASAS-SN data. Local minima in these GP fits are marked by vertical dashed lines and the periods between successive minima are labeled.}
    \label{fig: GJ 358 GP fit}
\end{figure}

Further study of the lightcurves from ASAS-3 and ASAS-SN, both of which are shown in a single plot in Figure \ref{fig: GJ 358 GP fit}, reveals that GJ 358's cycle period has been steadily decreasing for at least the past 20 years. ASAS-3 data show cyclic modulations with periods greater than four years, while the most recent modulation seen in ASAS-SN data has a period of just 1.5 years. This star therefore appears to be quite unusual. Unfortunately, there is a $\sim$6-yr gap between ASAS-3 and ASAS-SN data on this object. This gap could be filled with ASAS-4 data, however these data are not publicly available. Analysis of these data would prove very interesting and reveal, for example, whether the cycle period has changed monotonically, and when and how the longer-term trends seen in ASAS-3 and ASAS-SN data intersected.

At present, it is also unclear whether GJ 358's cycle is approaching some kind of (quasi-)steady-state, or if this cycle period will reach a local minimum before increasing again. It is also unclear if this changing period is related to the long-term cycle suggested by the ``linear" trends seen in GJ 358's ASAS-3 and ASAS-SN data. Future observations on this object are therefore needed to better understand this seemingly exceptional cyclic activity.

\subsubsection{GJ 551}\label{sec: GJ 551}

GJ 551 (Proxima Centauri) is an interesting case for several reasons. It is the most well-studied example of a fully convective M star exhibiting cyclic activity, and is the only fully convective star (in fact, the only M star) to have undergone long-term X-ray monitoring. \cite{S-M2016} found a 7-yr cycle in ASAS-3 data, and \cite{Wargelin2017} included an additional $\sim$5 years of ASAS-4
data, plus UV and X-ray observations to further support that conclusion.

\begin{figure}
    \centering
    \includegraphics[width=\columnwidth]{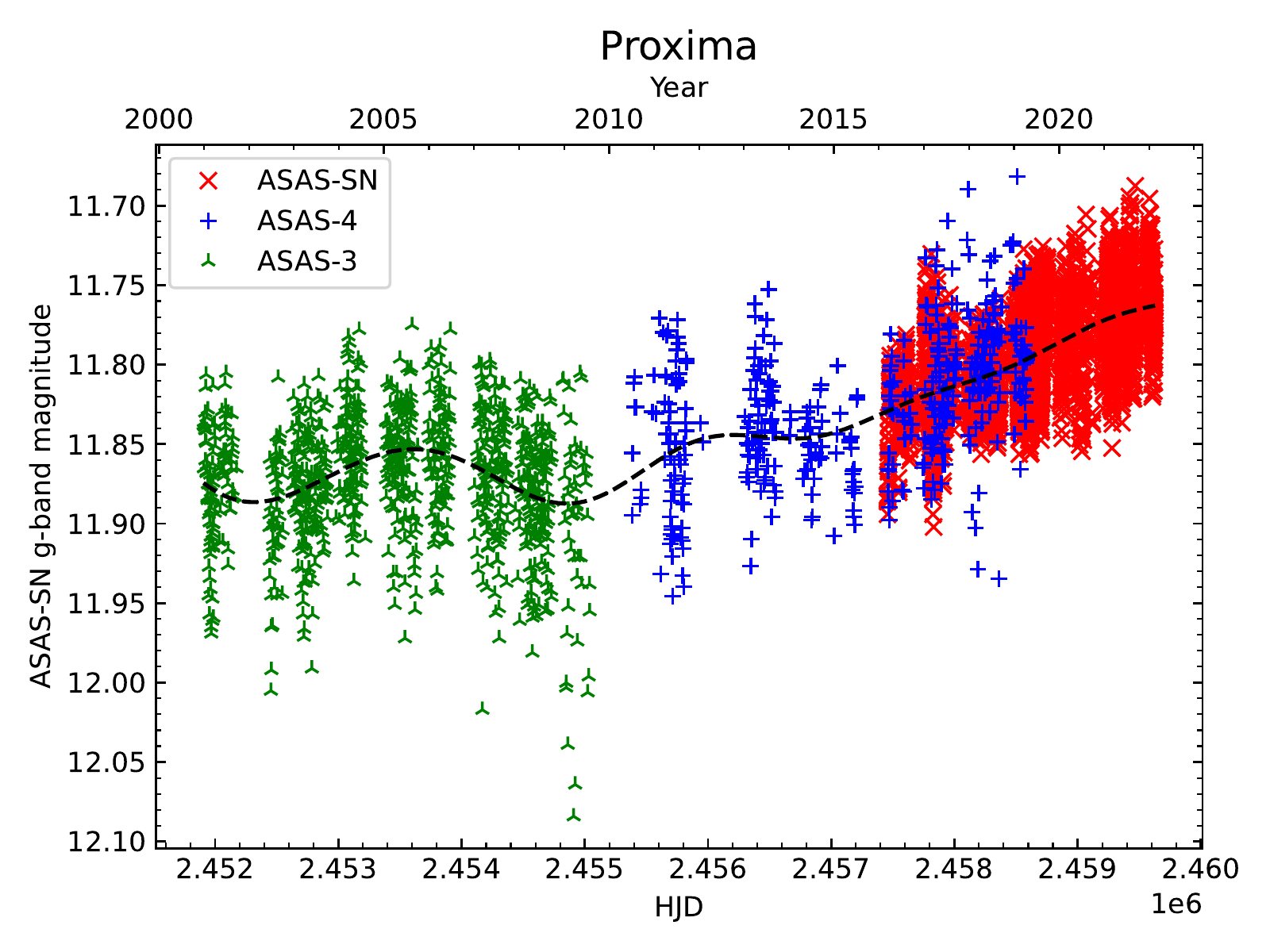}
    \caption{GP fit (dashed black line) to Proxima's ASAS-3, ASAS-4, and ASAS-SN data.}
    \label{fig: Prox Cen all data GP fit}
\end{figure}

More recent optical monitoring data, however, are less clear, as reflected in the ASAS-SN results listed in Table \ref{tab: optical cycle periods}, and shown in Figure \ref{fig: Prox Cen all data GP fit}, which includes ASAS-4 data extending into 2019 that were presented in \cite{Damasso2020}. ASAS-3 and ASAS-4 data were cross calibrated (private communication, G.~Pojma\'{n}ski), and then calibrated versus ASAS-SN data following the ``overlapping data" procedures described in Section \ref{sec: ASAS-SN prep}. As seen in Figure \ref{fig: Prox Cen all data GP fit}, periodic behavior is quite apparent at the beginning, but steadily weakens while average brightness increases. We suspect that the increasing stellar contamination noted in Section \ref{sec: contamination} is significantly affecting the data, but attempting to model and correct the contamination is beyond the scope of this paper. In contrast, X-ray data from Swift, now spanning more than 12 years (with some gaps), indicate well behaved cyclic behavior, currently with a period of $\sim$9 years (Wargelin, et al., in preparation).

\subsection{Periodicities and Magnetic Cycles}\label{sec: false cycles}

It should be noted that some of the cycles presented in Table \ref{tab: optical cycle periods} may not be related to a star's global magnetic field, but could be the result of other physical phenomena (or, if they are, may not be Schwabe cycles; do Nascimento et al. in preparation). For example, Rossby waves (horizontally flowing eddies in the convective zone/photosphere), as seen in the Sun, can act as dynamos \citep[e.g.,][]{Gilman1969}, leading to perceived cyclic activity which is not necessarily related to the star's global magnetic field. 

In the case of our Sun, there is also the Rieger cycle \citep{Rieger1984}, where hard solar flares are observed to occur in groups with a mean spacing of $\sim$154d. Similar cycles surely exist in other stars; while flares are a result of magnetic activity, they are local effects and it is not well understood how they relate to a star's global magnetic field - even in the case of our Sun \citep[e.g.,][]{Toriumi2022}.

\subsection{Phase Matching}\label{sec: phase checking}

One way to assess the validity of a star's apparent cycles is to see if they are consistent through ASAS-3 and ASAS-SN data. If the best fitting cyclic models from both data sets show a reasonable level of agreement (in terms of both phase and amplitude) where they overlap, then these cycles are more likely to be real. Note, however, that this method assumes a roughly constant cycle period, which, as discussed in Section \ref{sec: intro}, may not always be the case.

\begin{figure}
    \centering
    \includegraphics[width=\columnwidth]{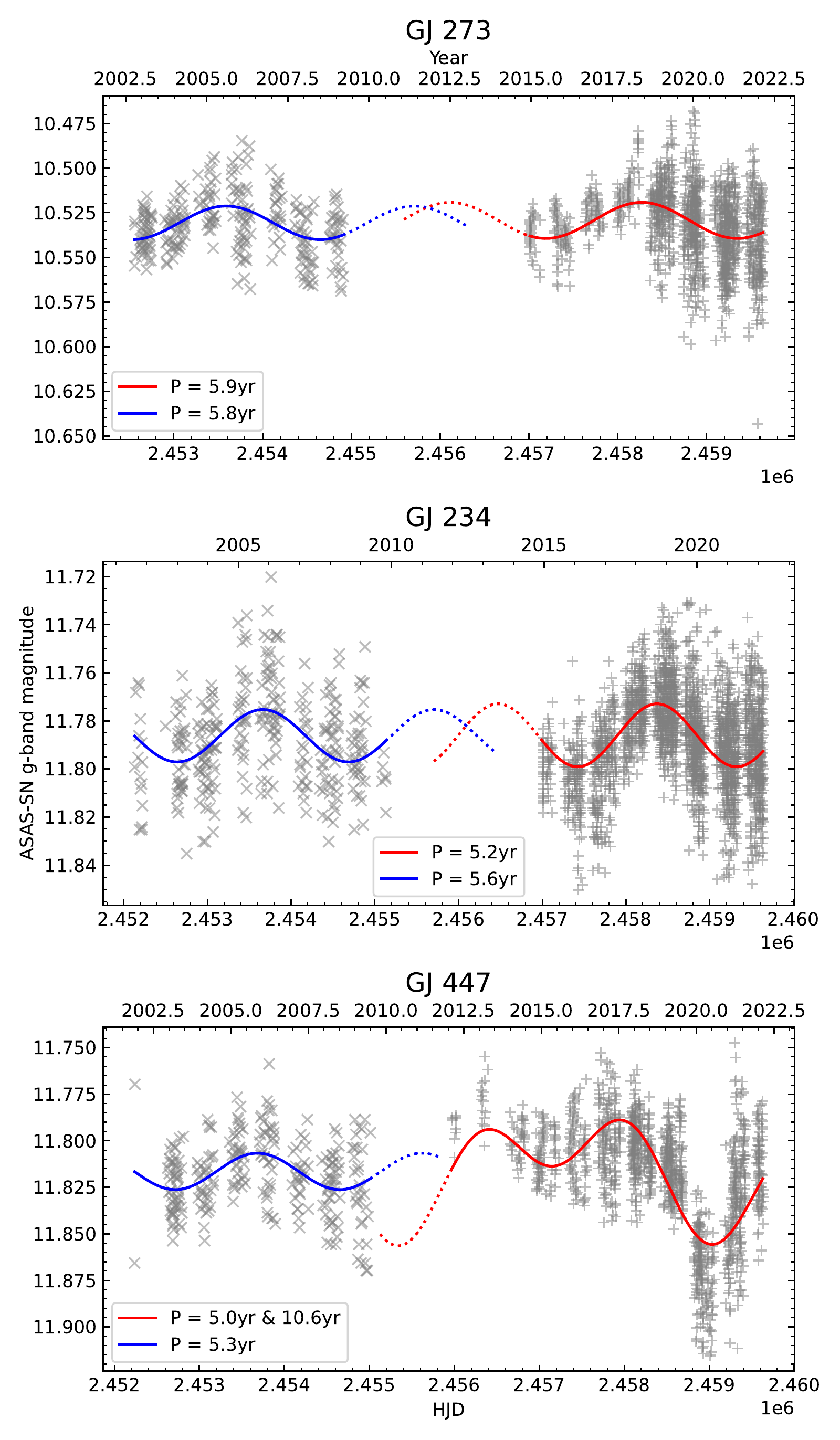}
    \caption{Plots of the cyclic models found from ASAS-3 (blue line) and ASAS-SN (red line) data for GJ 273, GJ 234, and GJ 447 from Table \ref{tab: optical cycle periods}, showing good, fair, and poor agreement, respectively. The complete figure set (10 images) is available in the online journal (Figure Set 3).}
    \label{fig: phase check}
\end{figure}

To investigate the amplitude and phase match-ups, we extend the cycles identified in Table \ref{tab: optical cycle periods} such that the ASAS-3 and ASAS-SN cyclic models for each star overlap by one year. In addition to this, we also offset the ASAS-3 data such that the mean magnitude of this light curve is the same as the mean magnitude of the ASAS-SN light curve, thus allowing easy comparison of the models' phases and amplitudes. Example plots are shown in Figure \ref{fig: phase check}, showing the results for GJ 273, GJ 234, and GJ 447. All phase match plots (10 images) are provided in Figure Set 3. Note, however, that we place little emphasis on these plots; indeed, they are primarily included so that the inferred cycles from both data sets can be more easily compared.

\figsetstart
\figsetnum{3}
\figsettitle{Phase Check}

\figsetgrpstart
\figsetgrpnum{3.1}
\figsetgrptitle{Phase check for GJ 234's well-defined cycles.}
\figsetplot{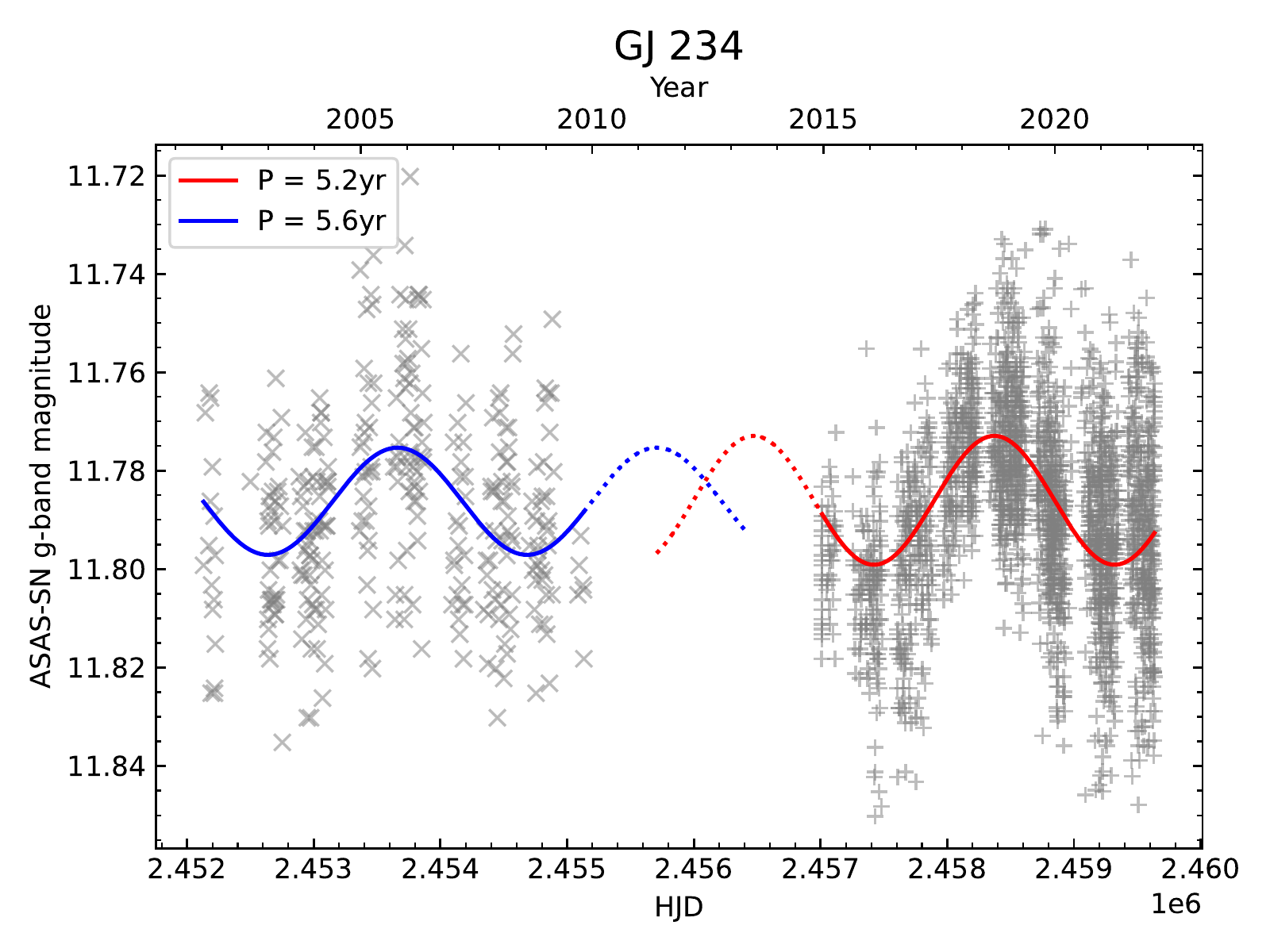}
\figsetgrpnote{Phase checks between well-defined ASAS-3 and ASAS-SN cycles.}
\figsetgrpend

\figsetgrpstart
\figsetgrpnum{3.2}
\figsetgrptitle{Phase check for GJ 273's well-defined cycles.}
\figsetplot{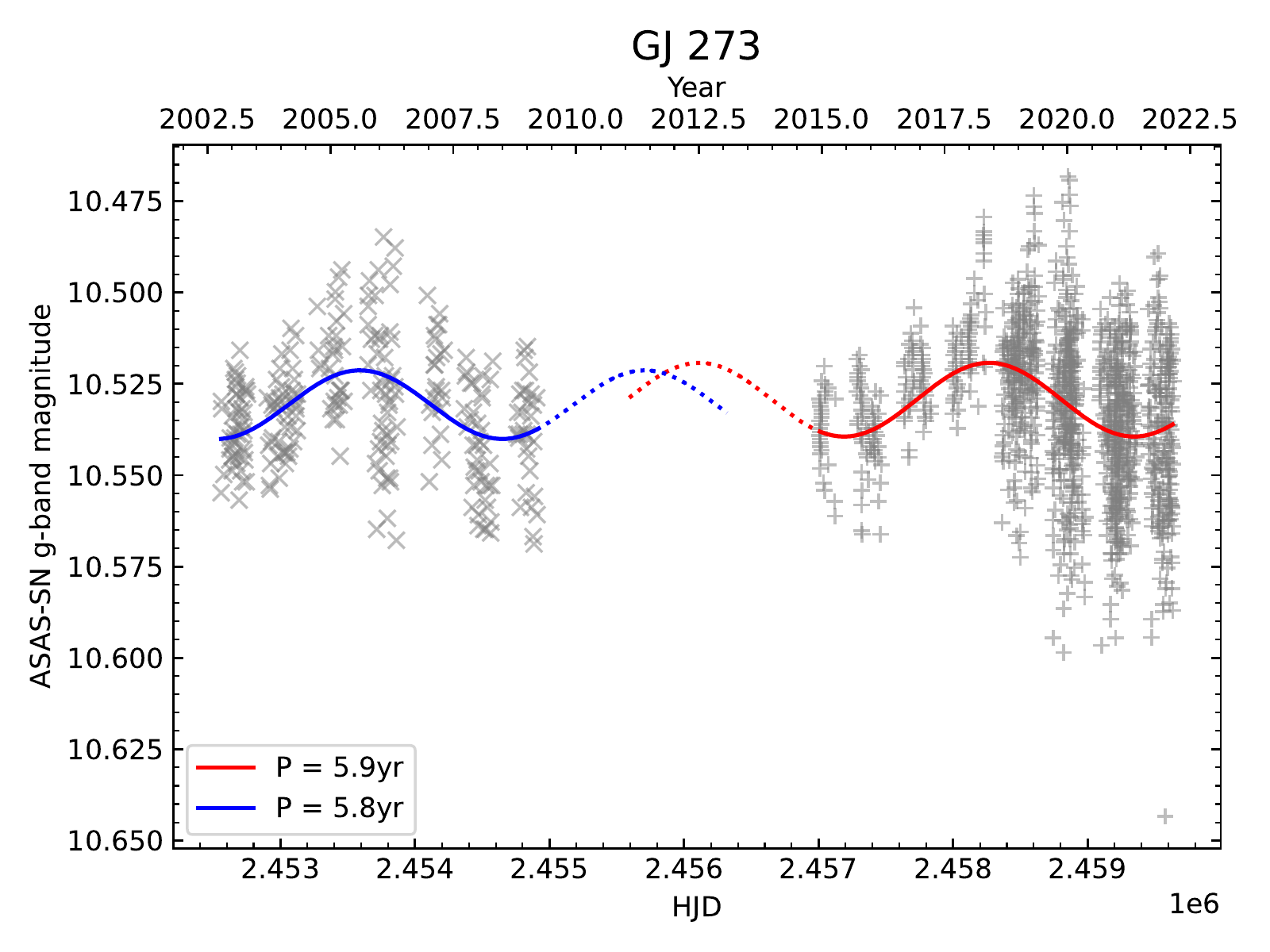}
\figsetgrpnote{Phase checks between well-defined ASAS-3 and ASAS-SN cycles.}
\figsetgrpend

\figsetgrpstart
\figsetgrpnum{3.3}
\figsetgrptitle{Phase check for GJ 285's well-defined cycles.}
\figsetplot{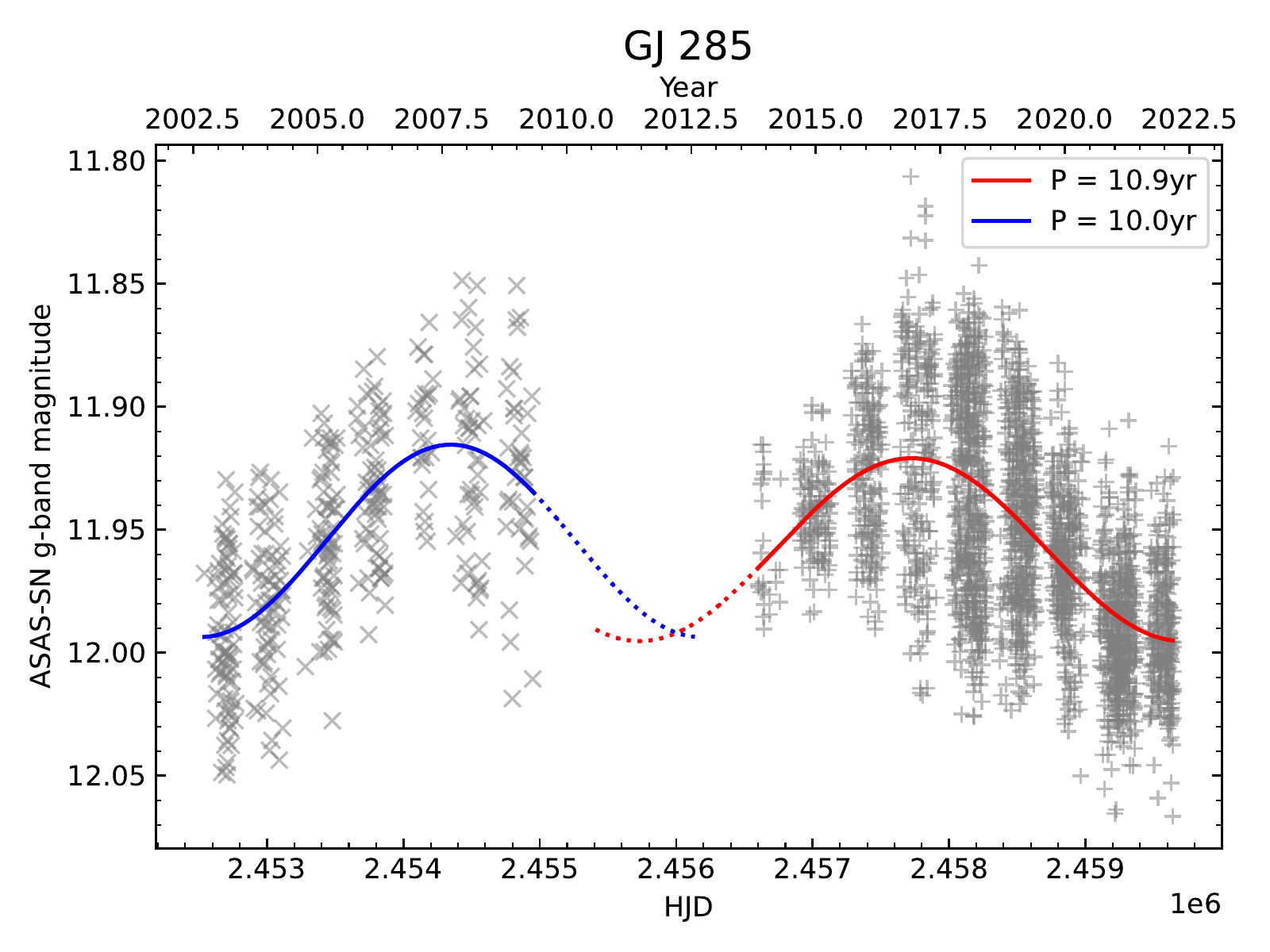}
\figsetgrpnote{Phase checks between well-defined ASAS-3 and ASAS-SN cycles.}
\figsetgrpend

\figsetgrpstart
\figsetgrpnum{3.4}
\figsetgrptitle{Phase check for GJ 358's well-defined cycles.}
\figsetplot{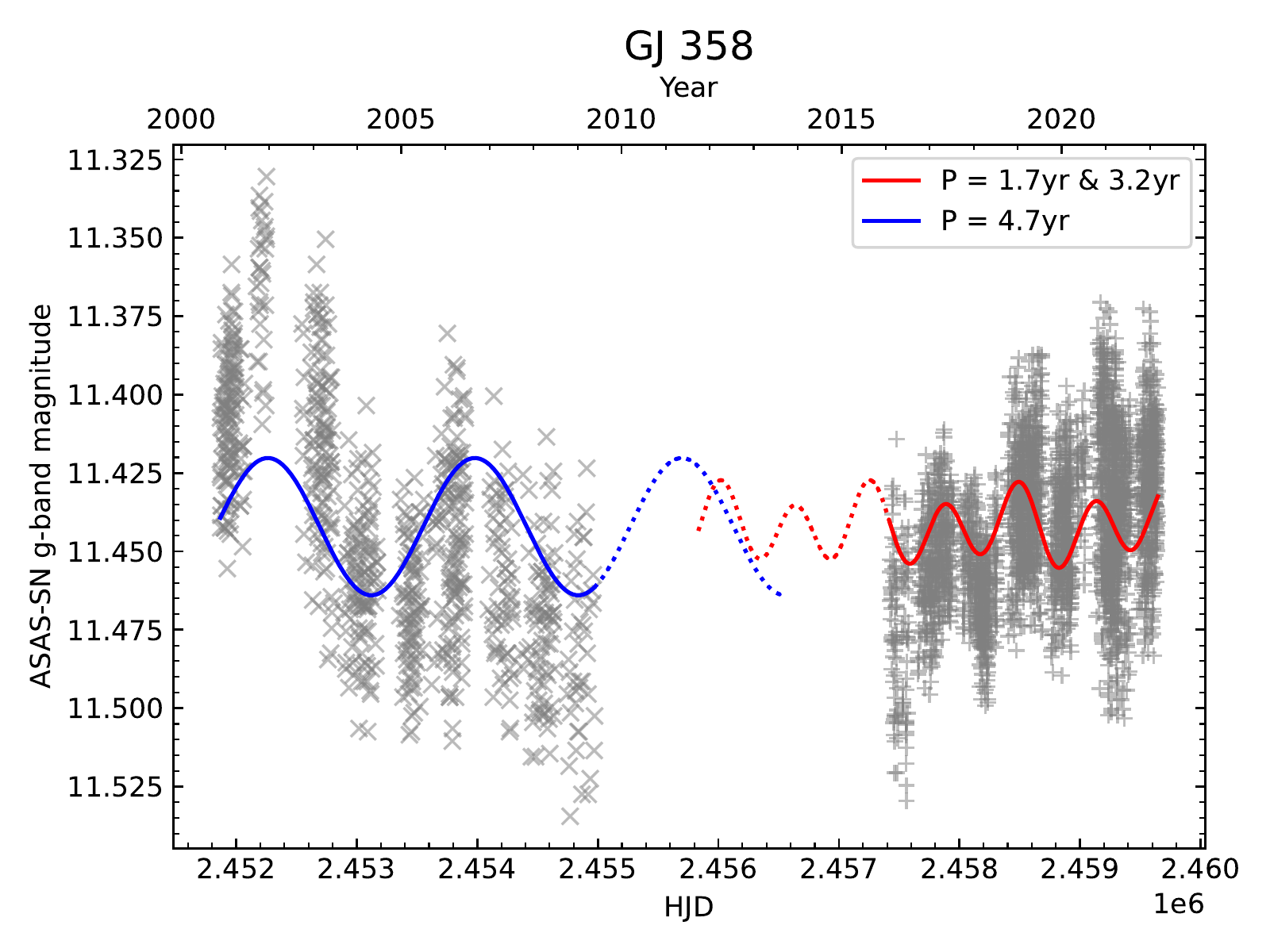}
\figsetgrpnote{Phase checks between well-defined ASAS-3 and ASAS-SN cycles.}
\figsetgrpend

\figsetgrpstart
\figsetgrpnum{3.5}
\figsetgrptitle{Phase check for GJ 406's well-defined cycles.}
\figsetplot{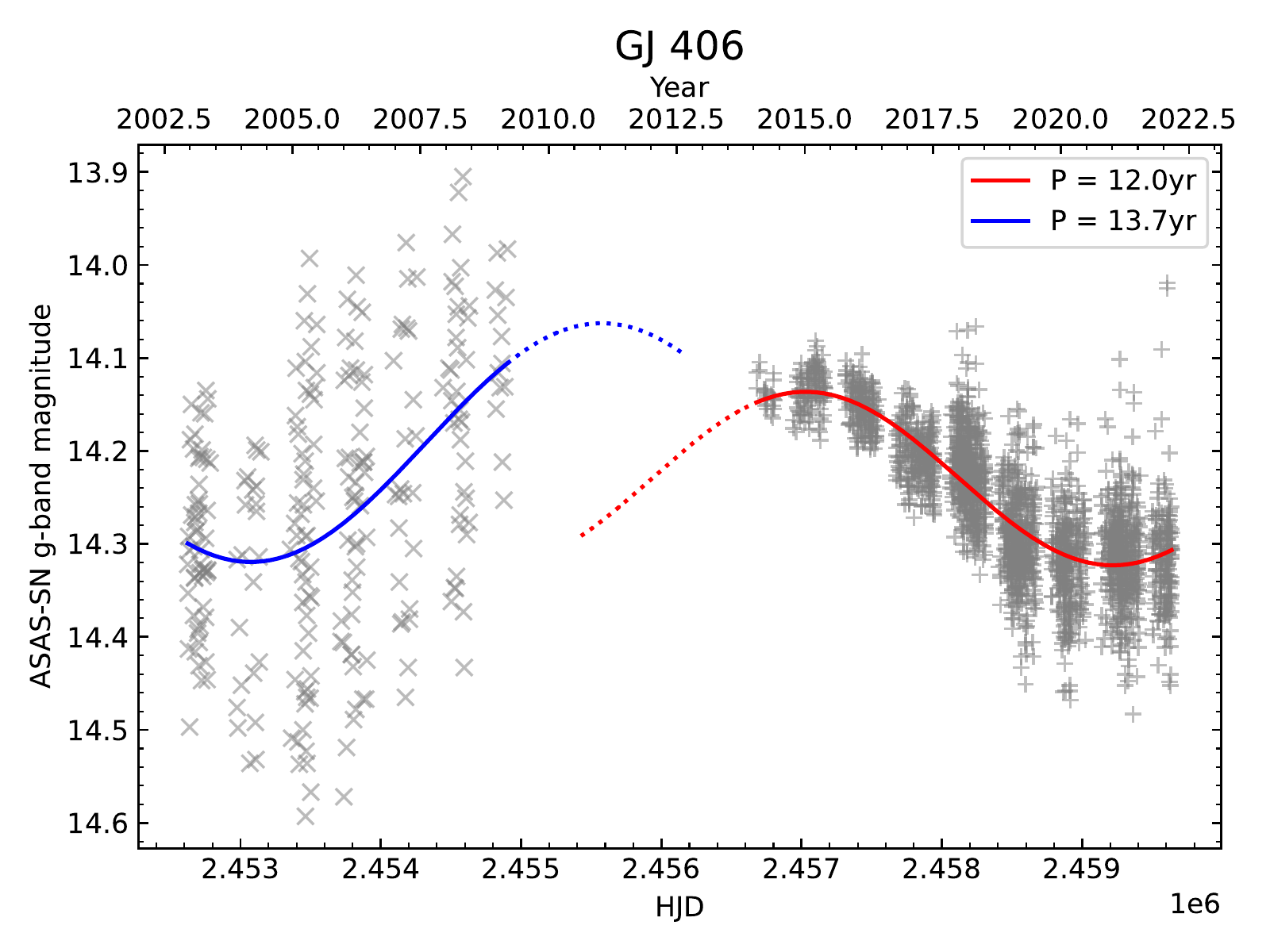}
\figsetgrpnote{Phase checks between well-defined ASAS-3 and ASAS-SN cycles.}
\figsetgrpend

\figsetgrpstart
\figsetgrpnum{3.6}
\figsetgrptitle{Phase check for GJ 447's well-defined cycles.}
\figsetplot{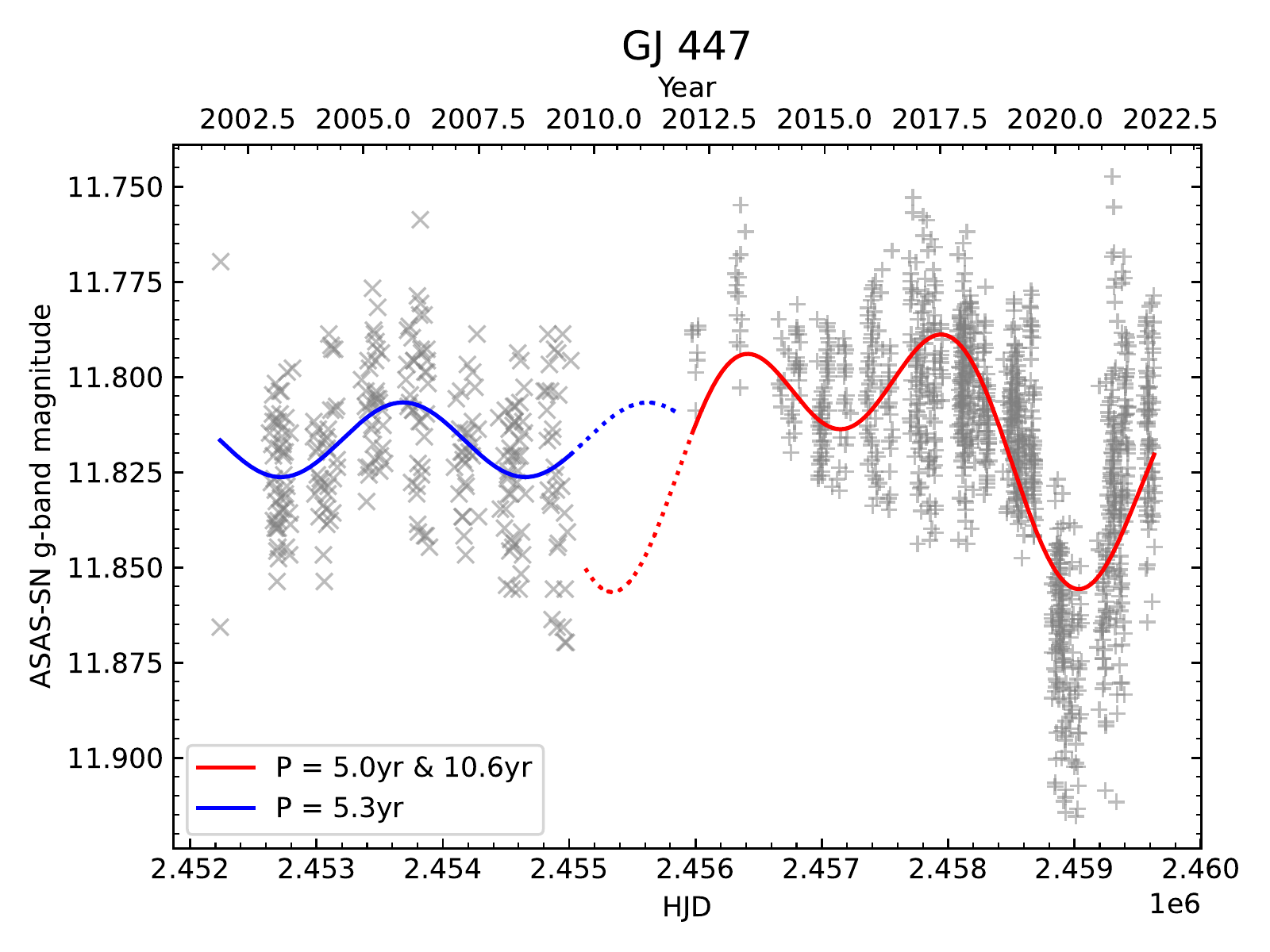}
\figsetgrpnote{Phase checks between well-defined ASAS-3 and ASAS-SN cycles.}
\figsetgrpend

\figsetgrpstart
\figsetgrpnum{3.7}
\figsetgrptitle{Phase check for GJ 628's well-defined cycles.}
\figsetplot{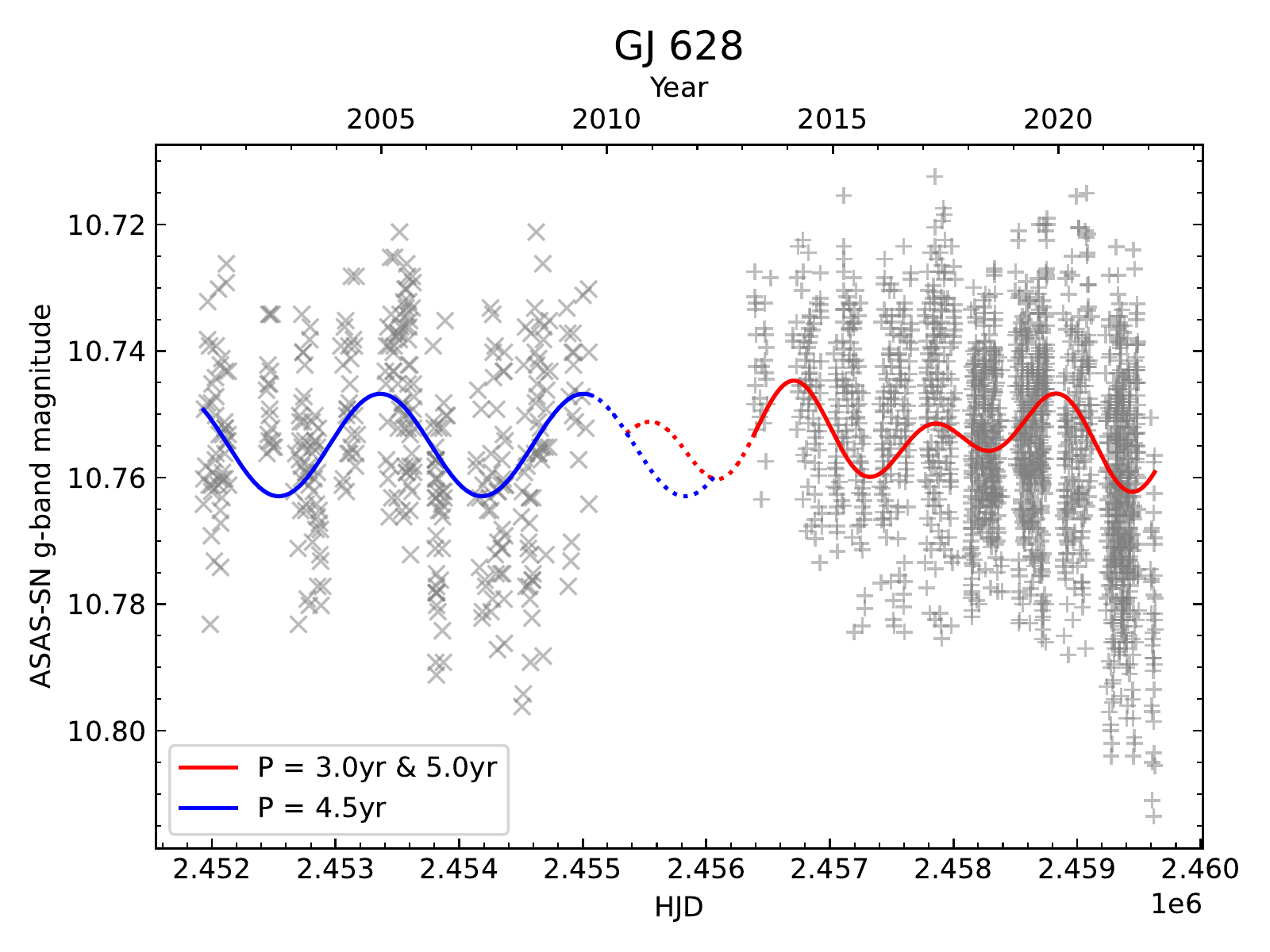}
\figsetgrpnote{Phase checks between well-defined ASAS-3 and ASAS-SN cycles.}
\figsetgrpend

\figsetgrpstart
\figsetgrpnum{3.8}
\figsetgrptitle{Phase check for GJ 729's well-defined cycles.}
\figsetplot{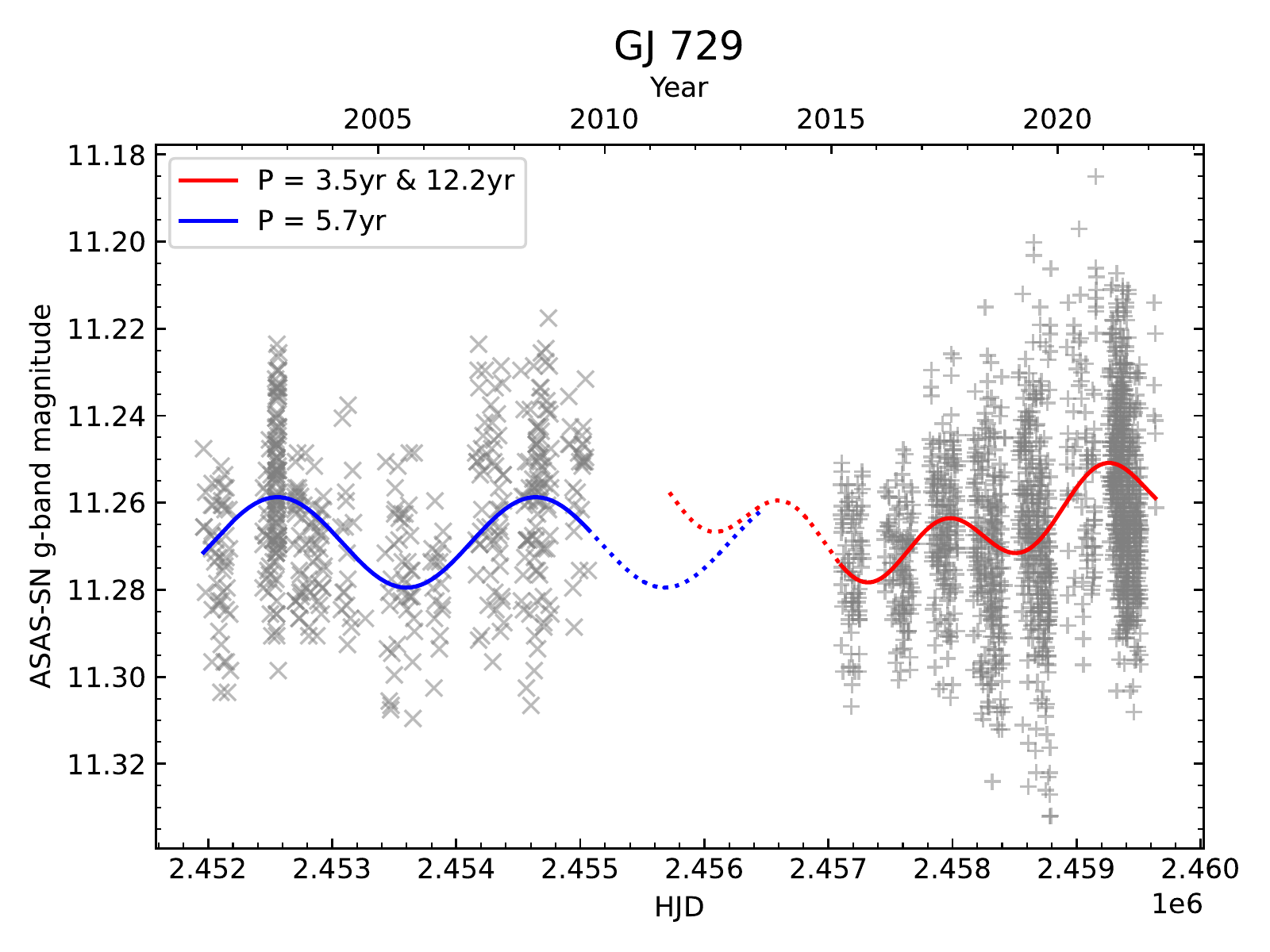}
\figsetgrpnote{Phase checks between well-defined ASAS-3 and ASAS-SN cycles.}
\figsetgrpend

\figsetgrpstart
\figsetgrpnum{3.9}
\figsetgrptitle{Phase check for GJ 849's well-defined cycles.}
\figsetplot{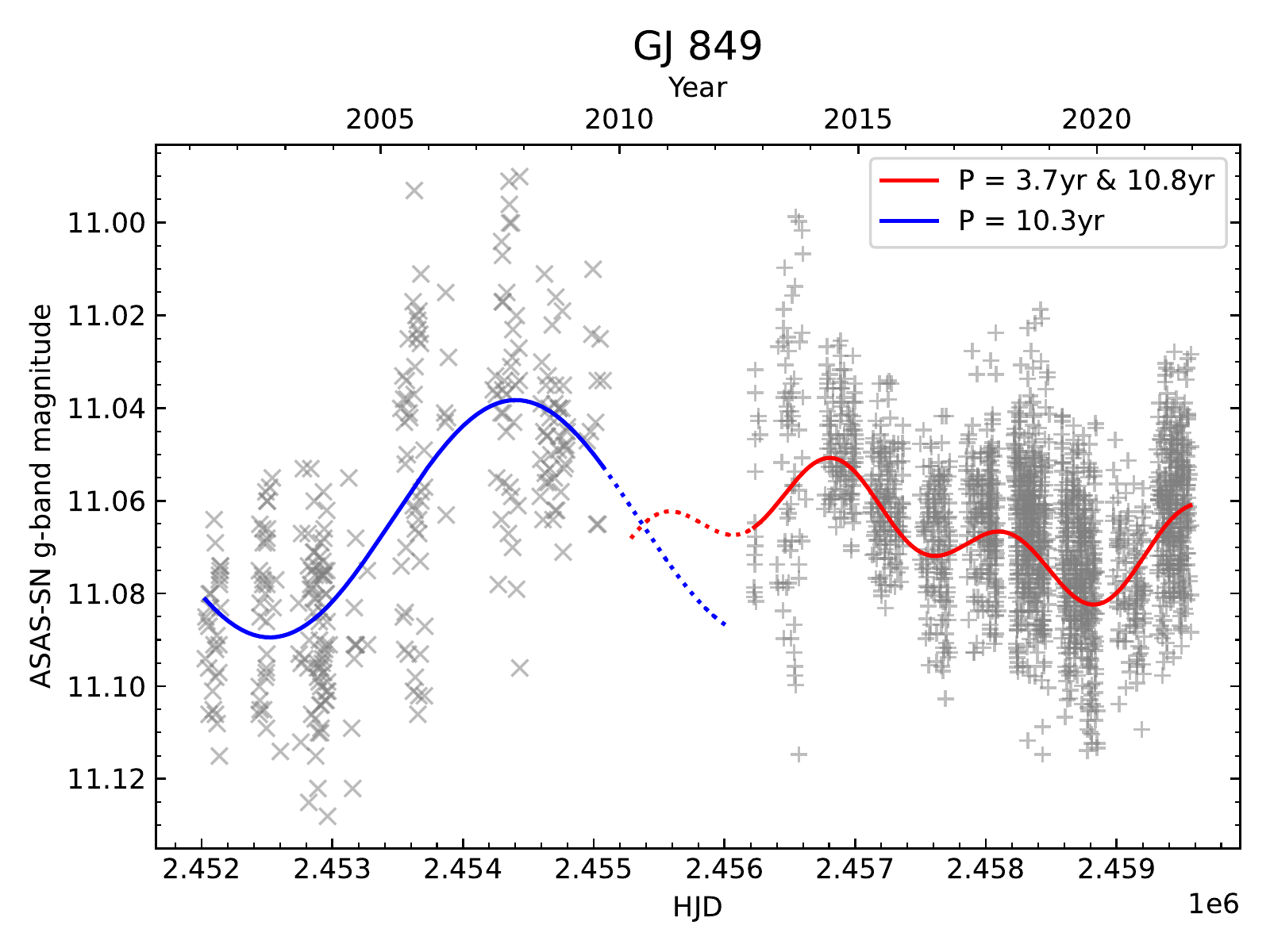}
\figsetgrpnote{Phase checks between well-defined ASAS-3 and ASAS-SN cycles.}
\figsetgrpend

\figsetgrpstart
\figsetgrpnum{3.10}
\figsetgrptitle{Phase check for Proxima's well-defined cycles.}
\figsetplot{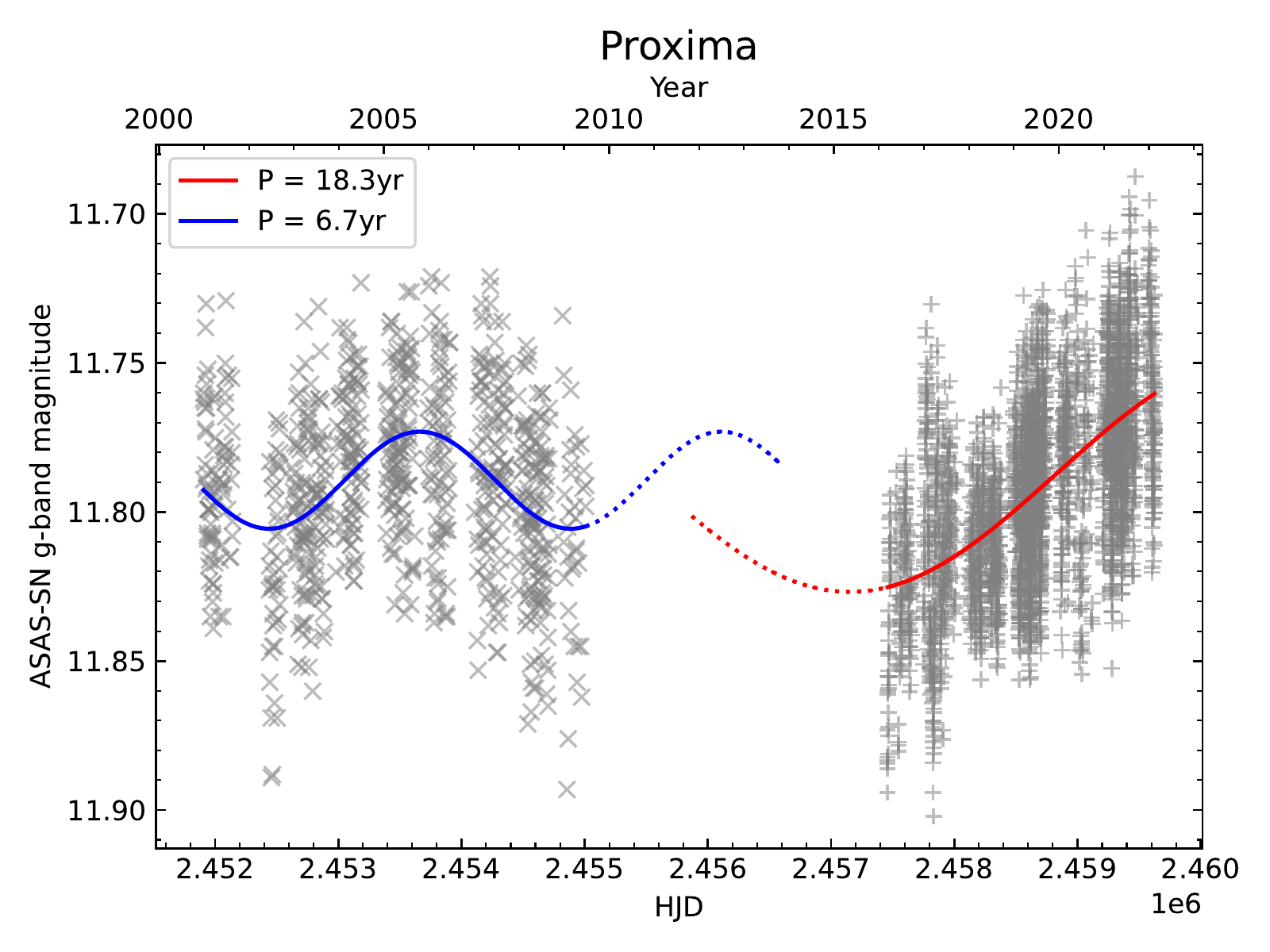}
\figsetgrpnote{Phase checks between well-defined ASAS-3 and ASAS-SN cycles.}
\figsetgrpend

\figsetend

As can be seen in the top panel of Figure \ref{fig: phase check}, both the ASAS-3 and ASAS-SN cyclic models on GJ 273 agree very well, in terms of both phase and amplitude, where they overlap. We interpret this as enhanced evidence for these cycles being real. On the other hand, GJ 234 (middle panel of Figure \ref{fig: phase check}) shows good amplitude agreement, but poorer phase agreement, while GJ 447 (bottom panel of Figure \ref{fig: phase check}) shows poor agreement in both cases. In the final column of Table \ref{tab: optical cycle periods}, we assign each star's ASAS-3 and ASAS-SN cycles a subjective ``phase match" grade of either good, fair, or poor. For reference, we judge that GJ 273, GJ 234, GJ 447 show good, fair, and poor phase agreement, respectively. The remaining plots are provided in Figure Set 3.

\section{Discussion}\label{sec: discussion}

With rotation and cycle parameters in hand, we can investigate correlations among various physical quantities. First, however, we must determine a largely theoretical quantity for each star: its convective turnover time, $\tau$.

\subsection{Determining Convective Turnover Times and Rossby Numbers}\label{sec: turnover time}

The convective turnover time is a measure of how long it takes for convection to move material from the bottom of the convective zone to the top, or vice versa, and is related to the ratio of the thickness of the convective envelope (thin in F stars, fully convective by $\sim$M3.5) and the mean convective velocity (which increases with temperature). To compare stars with different sized convective zones, temperatures, and rotation periods, a dimensionless quantity known as the Rossby number, Ro, is commonly used. Another dimensionless quantity is the linear $\alpha \Omega$ dynamo number, D, a measure of the dynamo's strength, which is proportional to $\textrm{Ro}^{-2}$. The Rossby number of a star is defined as:
\begin{equation}
    \textrm{Ro} = \frac{P_{\rm rot}}{\tau},
\end{equation}
where $P_{\rm rot}$ is the rotation period and $\tau$ is the convective turnover time. Note, however, that there are two definitions for the convective turnover time: a local and a global definition. The local convective turnover time, $\tau_L$, applies to the bottom of the convective zone (in the tachocline, where the main dynamo amplification takes place in some models), while the global convective turnover time, $\tau_G$, is an average over the entire convective zone. The global convective turnover time is therefore a better quantity for M stars, where the tachocline is negligible or absent.

The choice of a convective turnover time is important. We need both a local (for tachocline-based dynamos) and a global (for full convection zone dynamos) $\tau$, which we denote $\tau_{\rm L}$ and $\tau_{\rm G}$, respectively. Since we are interested in the operation of dynamos, a ``$\tau$" derived from any indirect activity diagnostic (e.g., Ca\,{\sc ii} HK or X-ray emission) is of dubious utility. This is because such empirical $\tau$s implicitly {\it assume} that the function $f(M,\log g, [Fe/H])$ which best connects the given emission to rotation can be {\it defined} to be ``$\tau$," ignoring the complex physics connecting magnetic flux to the heating which produces the emission in question \citep[e.g.,][]{Cuntz1999}. The hidden inclusion of this additional heating physics makes these empirical ``$\tau$s" inappropriate for understanding how dynamos themselves operate. Note that a $\tau$ derived by best fitting rotation with unsigned magnetic flux measurements would be more appropriate. This has not yet been attempted to our knowledge, likely because data on unsigned fluxes are sparse and generally come with large errors \citep[e.g.,][]{Reiners2012, Saar1994}. Calculating such a $\tau$ is outside the scope of the current work; however new flux measurement methods may make it feasible in the near future \citep[e.g.,][]{Lehmann2015, Mortier2016}.

\begin{figure}
    \centering
    \includegraphics[width=\columnwidth]{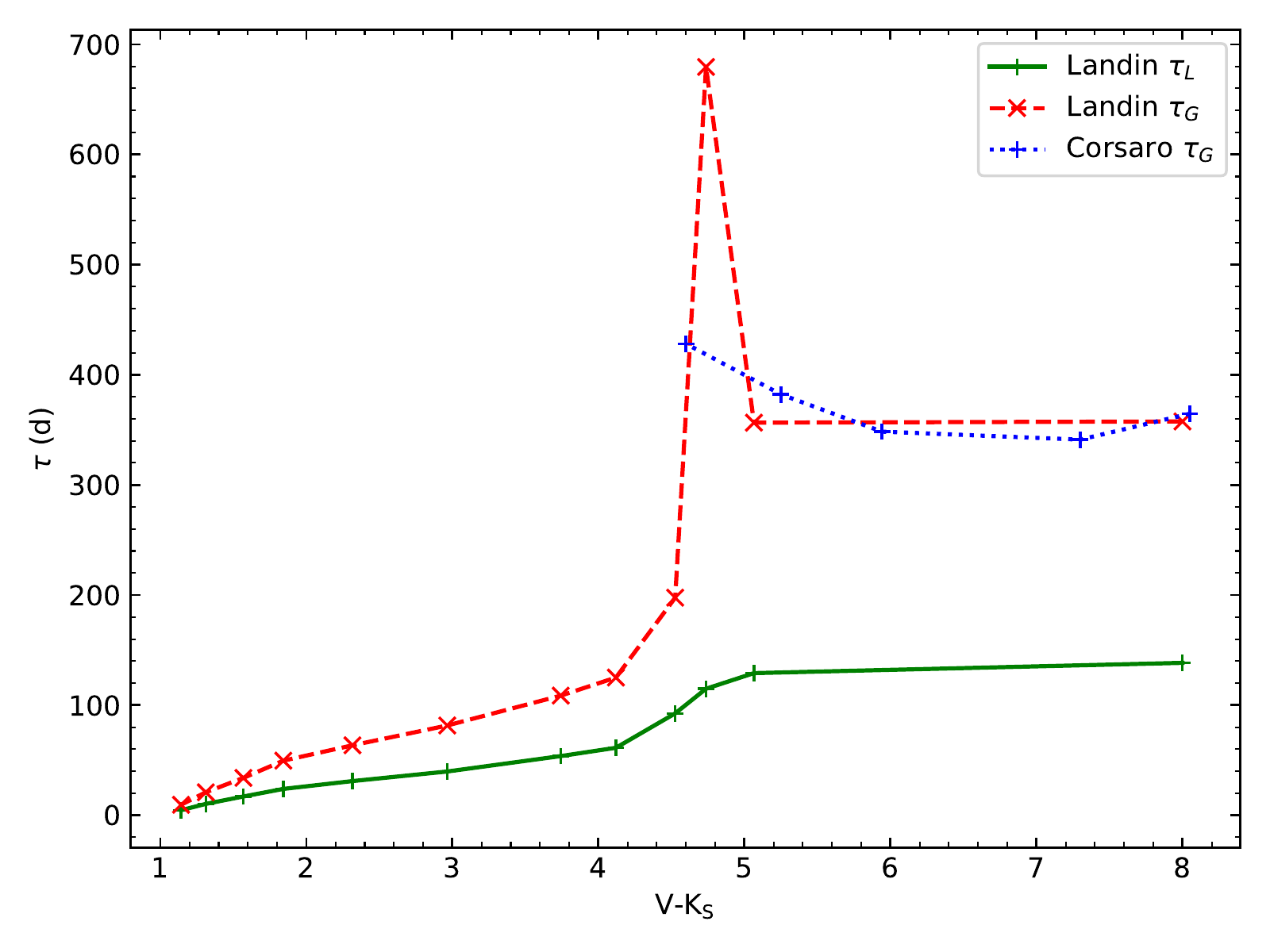}
    \caption{Convective turnover time (in days) vs.~V--K$_{\rm S}$ color index from \protect\cite{Landin2023}. The dashed red line shows the global turnover time ($\tau_G$); the solid green line shows the local turnover time ($\tau_L$); the dotted blue line shows the scaled version of an independent $\tau_G$ calculation for fully convective stars \protect\cite{Corsaro2021} which is in good agreement except in the narrow zone 3250 K $\leq T_{\rm eff} \leq$ 3400 K. We adopt $\tau$(Landin) everywhere except for $\tau_{\rm G}$ for $T_{\rm eff} <$ 3400K (V--K$_{\rm S} \approx$ 4.6), where we transition to the scaled $\tau_{\rm G}$(Corsaro); see text for details. $\tau$(Landin) have been truncated (using linear interpolation) at V--K$_{\rm S} = 8$; see Table \ref{tab: optical cycle periods}.}
    \label{fig: tau vs V-K}
\end{figure}

Unfortunately, theoretically based $\tau$s are also problematic. Due to uncertainties concerning M dwarf internal structure near the fully convective limit \citep[][]{Baraffe2018,Jao2022}, any purely theoretical $\tau$ is at best approximate at low masses \citep[][]{Jao2022}. There are also problems defining a local $\tau$ in the center of fully convective stars. Indeed, with no convection zone bottom, the tachocline concept driving the need for a local $\tau$ requires modification itself. Fortunately, a recent paper has worked to improve $\tau$ models in low mass stars \citep[][]{Landin2023}. We adopt their values here at age $\approx$1Gyr. This is post-ZAMS for all masses here except 0.1 $M_\odot$, but we note from their Figure 1 that $\tau_{\rm G}(0.1 M_\odot)$ evolves negligibly for ages $>$0.3 Gyr. Indeed, over the mass range considered here, $\tau$ varies little along the main sequence except at the highest masses, where differences of $\sim$30\% may accrue at age extremes. We use the V--K$_{\rm S}$ vs.~$T_{\rm eff}$ relation of \cite{Pecaut2013} (and later improvements; Mamajek, E.E. 2021, private communication) to match to our stars. As a cross-check, we compare the calculations at low mass with a different kind of empirical $\tau$ calculation based purely on stellar properties - derived equating the star's output bolometric flux with convective flux \citep[][]{Corsaro2021}. They find $\tau_{\rm G} \propto R [M/(LR)]^{1/3} \propto M^{1/3}T_{\rm eff}^{-4/3}$ for fully convective stars, where $M, L,$ and $R$ are the stellar mass, luminosity and radius, respectively \citep[taken from ][]{Pecaut2013}. We find a scaled version of these physically-based empirical $\tau_{\rm G}$ for the most part compare well with the new $\tau_{\rm G}$ of \citet{Landin2023} (Figure \ref{fig: tau vs V-K}; where we also display their $\tau_{\rm L}$). The exception is in the narrow temperature range 3250K $< T_{\rm eff} < 3400$K, very near the fully convective boundary, where $\tau_{\rm G}$(Landin) shows a spike that $\tau_{\rm G}$(Corsaro) does not. (We also note that $T_{\rm eff}(0.1 M_\odot)$ is cooler in \cite{Landin2023}.) We suggest that $\tau_{\rm G}$ calculations near the stellar center may still have problems \citep[][]{Jao2022}, and in keeping with the idea that $\tau_{\rm L}$ and $\tau_{\rm G}$ should roughly scale with each other \citep[e.g., ][]{Montesinos2001}, we adopt the Corsaro values for $T_{\rm eff} \leq 3400$ K. We use $\tau_{\rm G}$ for fully convective stars, and either $\tau_{\rm G}$ or $\tau_{\rm L}$ otherwise, depending on the circumstances.

With estimated convective turnover times and  Rossby numbers for our 15 M-type stars, we then used results from \cite{Saar1999}, \cite{Lehtinen2016}, and \cite{Olspert2018b} to gather a comparison sample of 40 F, G, and K type stars with robust, well-defined cycles. Of these 40 stars, listed in Table \ref{tab: FGK stars}, 10 have double cycles. Note, however, \cite{Olspert2018b} use multiple different methods to identify stellar cycles; for consistency with our approach, we use results obtained via their harmonic model (see Section \ref{sec: time series analysis techniques} for more details). To estimate the convective turnover times of these stars, we use the method described above.\footnote{Note: while we do not include a plot of $\tau_L$ vs.~B--V, the stars from Table \ref{tab: FGK stars} correspond to roughly the range $1 \leq$~V--K$_{\rm S}$~$\leq 3.5$ in Figure \ref{fig: tau vs V-K}.} We note, however, that we exclude HD 18256 (F6) as it is too hot for our $\tau$ estimations; interestingly, in their study of Ca\,{\sc ii} H and K cycles in 15 stars, \cite{Baliunas2004} found that this star had by the far the largest ``anharmonicity," a measure of deviation from single-period sinusoidal behavior, in their sample.

\begin{table*}
    \centering
    \caption{Activity Cycles For FGK Type Stars Taken From Previous Works.}
    \begin{tabular}{cccccccccccc}
        \hline
        \hline
        Star & Spectral & B-V & $P_{\rm rot}$ & $P_{\rm cyc}$ & Ref. & Star & Spectral & B-V & $P_{\rm rot}$ & $P_{\rm cyc}$ & Ref. \\
         & Type & & (d) & (yr) & & & Type & & (d) & (yr) & \\
        \hline
        HD 1835 & G2.5 & 0.66 & 7.78 & 9.1 & 1 & HD 18256 & F6 & 0.45 & 3 & 6.8 & 1 \\
        HD 20630 & G5 & 0.68 & 9.24 & 5.6 & 1 & HD 76151 & G2 & 0.67 & 15 & 2.52 & 1 \\
        \nodata & \nodata & \nodata & \nodata & \nodata & \nodata & \nodata & \nodata & \nodata & 14.4 & 15.9 & 3 \\
        \nodata & \nodata & \nodata & \nodata & \nodata & \nodata & \nodata & \nodata & \nodata & \nodata & 5.09$^{\rm a}$ & 3 \\
        HD 82443 & G9 & 0.77 & 5.38 & 3.89 & 1 & HD 115404 & K2 & 0.94 & 18.47 & 12.4 & 1 \\
        \nodata & \nodata & \nodata & \nodata & 20 & 2 & \nodata & \nodata & \nodata & \nodata & \nodata & \nodata \\
        HD 149661 & K1 & 0.84 & 21.07 & 16.2 & 1 & HD 165341 & K0 & 0.86 & 19.9 & 15.5 & 1 \\
        \nodata & \nodata & \nodata & \nodata & 4 & \nodata & \nodata & \nodata & \nodata & \nodata & 5.1 & \nodata \\
        HD 100180 & F9.5 & 0.57 & 14 & 3.6 & 1 & HD 190406 & G0 & 0.61 & 13.94 & 2.6 & 1 \\
        \nodata & \nodata & \nodata & \nodata & \nodata & \nodata & \nodata & \nodata & \nodata & \nodata & 16.9 & \nodata \\
        HD 1405 & K2 & 0.95 & 1.756 & 8 & 2 & HD 70573 & G1/2 & 0.62 & 3.314 & 6.9 & 2 \\
        HD 82558 & K1 & 0.93 & 1.604 & 14.5 & 2 & HD 116956 & G9 & 0.80 & 7.86 & 14.7 & 2 \\
        \nodata & \nodata & \nodata & \nodata & 18 & \nodata & \nodata & \nodata & \nodata & \nodata & \nodata & \nodata \\
        HD 135599 & K0 & 0.83 & 5.529 & 14.6 & 2 & HD 171488 & G2 & 0.62 & 1.345 & 9.5 & 2 \\
        Sun & G2 & 0.631 & 26.09 & 10.89 & 3 & HD 103095 & K1 & 0.75 & 34.3 & 7.13 & 3 \\
        HD 10476 & K1 & 0.84 & 35.6 & 10.6 & 3 & HD 10780 & K0 & 0.81 & 22.14 & 7.53 & 3 \\
        HD 114710 & F9.5 & 0.59 & 11.99 & 16.56 & 3 & HD 146233 & G2 & 0.65 & 22.62 & 11.2 & 3 \\
        HD 152391 & G8.5 & 0.76 & 10.62 & 9.03 & 3 & HD 155886 & K2 & 0.85 & 20.58 & 10.44 & 3 \\
        \nodata & \nodata & \nodata & \nodata & 13.73 & \nodata & \nodata & \nodata & \nodata & \nodata & 5.0 & \nodata \\
        HD 156026 & K5 & 1.16 & 16.69 & 17.89 & 3 & HD 160346 & K3 & 0.971 & 32.0 & 7.21 & 3 \\
        HD 16160 & K3 & 0.98 & 48.58 & 12.45 & 3 & HD 165341A & K0 & 0.86 & 19.33 & 5.19 & 3 \\
        HD 166620 & K2 & 0.87 & 42.1 & 16.16 & 3 & HD 185144 & K0 & 0.87 & 27.7 & 6.66 & 3 \\
        HD 201091 & K5 & 1.18 & 35.54 & 7.16 & 3 & HD 201092 & K7 & 1.37 & 34.55 & 11.65 & 3 \\
        \nodata & \nodata & \nodata & \nodata & 21.09 & \nodata & \nodata & \nodata & \nodata & \nodata & \nodata & \nodata \\
        HD 219834B & K2 & 0.91 & 34.78 & 9.32 & 3 & HD 26965 & K0 & 0.82 & 38.65 & 10.66 & 3 \\
        HD 32147 & K3 & 1.06 & 33.7 & 11.13 & 3 & HD 3651 & K0.5 & 0.83 & 37.0 & 16.98 & 3 \\
        HD 37394 & K1 & 0.84 & 11.49 & 5.83 & 3 & HD 4628 & K2.5 & 0.90 & 37.14 & 8.56 & 3 \\
        \nodata & \nodata & \nodata & \nodata & \nodata & \nodata & \nodata & \nodata & \nodata & \nodata & 5.79 & \nodata \\
        HD 78366 & G0 & 0.60 & 9.519 & 14.63 & 3 & HD 81809 & G1.5 & 0.80 & 41.66 & 8.11 & 3 \\
        \hline
    \end{tabular}
    \label{tab: FGK stars}
    \\[0pt]
    \justifying
    \tablecomments{$^{\rm a}$ excluded from plots (see Section \ref{sec: data selection}). References: 1) \protect\cite{Saar1999}, 2) \protect\cite{Lehtinen2016}, 3) \protect\cite{Olspert2018b}.}
\end{table*}

\subsubsection{Data Selection}\label{sec: data selection}

When gathering our sample of FGK stars from other works, we only included cycles in known (or likely) dwarfs. Comparing the results of \cite{Saar1999} with \cite{Olspert2018b}, there are 17 stars in common. Since \cite{Olspert2018b} had longer time series for their analysis, we favor their results in most cases. We note, however, an exception below.

\cite{Saar1999} found a single 2.52-yr cycle in HD 76151, while \cite{Olspert2018b} found 5.0-yr and 15.9-yr cycles using their harmonic model. We interpret this 5.0-yr cycle as a Hale cycle (born of a polarity asymmetry in HD 76151's magnetic field; do Nascimento et al. in preparation). For this reason, we reject this 5.0-yr cycle. We also note that this 5.0-yr cycle resides between the I and A branches when plotted in $P_{\rm cyc}/P_{\rm rot}$--Ro$^{-1}$ space, which we interpret as stronger evidence for this not being a Schwabe cycle (see Section \ref{sec: unified theory}).

\subsection{\texorpdfstring{A\textsubscript{\rm cyc}}{Acyc} versus \texorpdfstring{P\textsubscript{\rm cyc}}{Pcyc}}\label{sec: cycle amplitude vs period}

\begin{figure}
    \centering
    \includegraphics[width=\columnwidth]{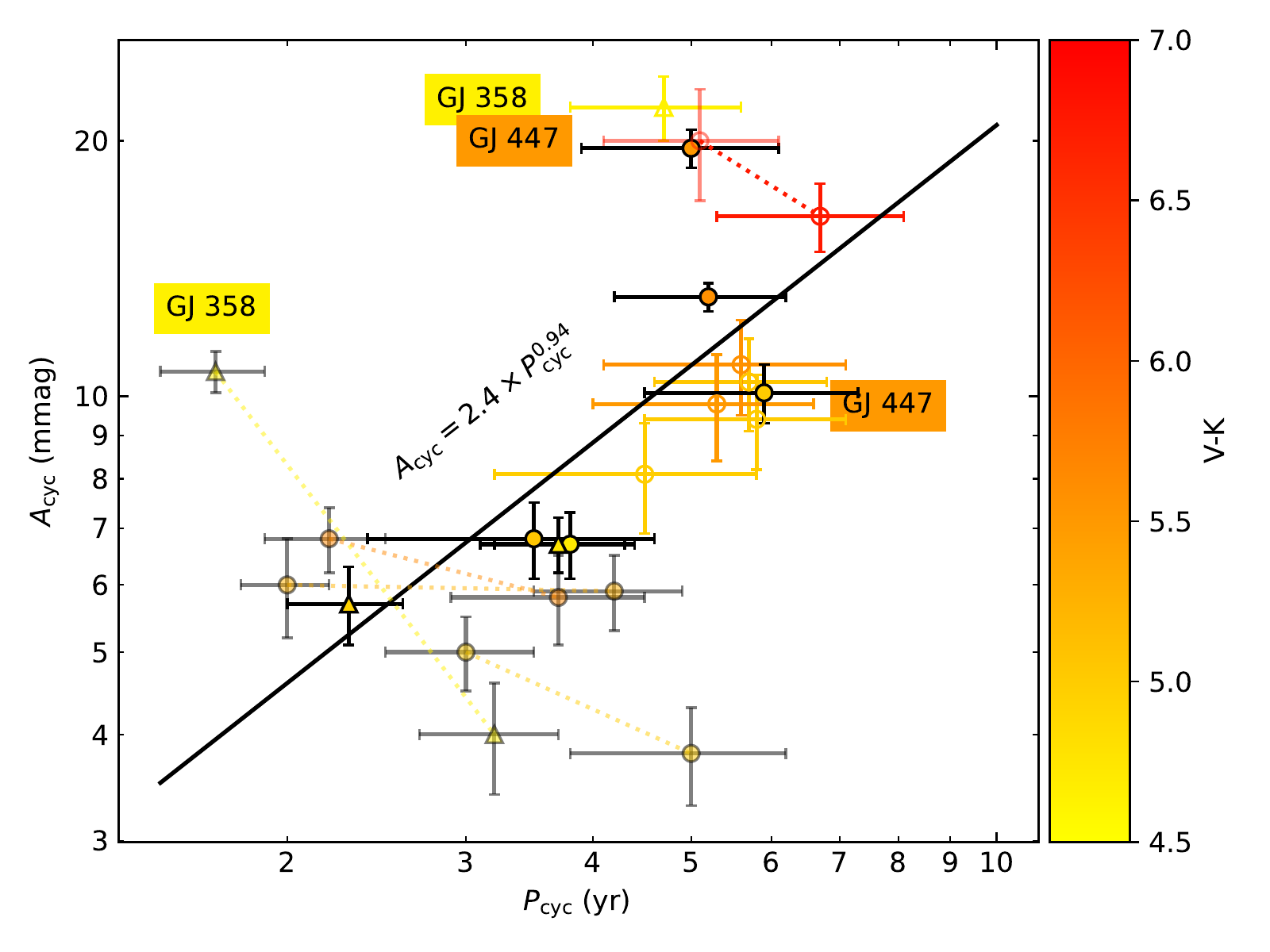}
    \caption{Cycle amplitude (in millimagnitudes) against cycle period (in years) using the well-defined cycles from Table \ref{tab: optical cycle periods}. Color-only markers show cycles inferred from ASAS data; markers with dark outlines are from ASAS-SN data; fully convective stars use circle markers; partially convective stars use triangle markers; faded markers denote that these points were ignored during fitting; concurrent cycles within the same star are connected with dotted lines. The outlier points for GJ 358 and GJ 447 (with colored labels) are discussed in the text.}
    \label{fig: A_cyc vs P_cyc}
\end{figure}

Figure \ref{fig: A_cyc vs P_cyc} shows a plot of cycle amplitude against period for the well-defined cycles in Table \ref{tab: optical cycle periods}, along with a least-squares fit following $A_{\rm cyc} \propto P_{\rm cyc}^{0.94 \pm 0.11}$. When fitting this power law, only stars with a single, well-defined cycle per data set were used (i.e., if a star had a single cycle in both ASAS-3 and ASAS-SN data, then both cycles were used). Proxima's suspect 5.1-yr ASAS-4 cycle was ignored during fitting (see Section \ref{sec: contamination}).

As can be seen in Figure \ref{fig: A_cyc vs P_cyc}, cycles with longer periods generally also have larger amplitudes (ignoring stars with more than one concurrent cycle, whose cycles are connected with dotted lines and shown with faded markers). We speculate that this may be caused by longer periods allowing more build up of magnetic energy, resulting in a more pronounced cycle; this process is interrupted in stars with more than one concurrent cycle, which is why they appear to contradict this trend. This relation may also explain why short period cycles are detected less frequently in ASAS-3 data, which are noisier than their ASAS-SN successor. It may be the case that short period cycles are present in ASAS-3 data, but cannot easily be resolved as their amplitudes are comparable to the ASAS-3 noise level.

Since stars with more than one well-defined cycle per data set were excluded from the power law fit shown in Figure \ref{fig: A_cyc vs P_cyc}, it is interesting to note where these excluded cycles reside in relation to this fit. Indeed, it can be seen that for stars with two concurrent cycles, one is usually closer to this power law than the other; these other cycles may therefore be ``false cycles" (Section \ref{sec: false cycles}) or hints of a multi-branched relationship as seen by some in the $P_{\textrm{cyc}}$--$P_{\textrm{rot}}$ relation \citep[e.g.,][]{Saar1999, B-V2007, Lehtinen2016}. It should also be noted that $A_{\rm cyc}$ can be quite variable (more so in the Sun than $P_{\rm cyc}$), and so some outliers may be expected. That said, the (somewhat tentative) agreement among these excluded cycles is interesting, and we interpret this as further evidence for this relation between cycle amplitude and period.

Multiple branches in the $A_{\rm cyc}$--$P_{\rm cyc}$ relation may explain why, for example, the ASAS-3 cycle of GJ 358 and ASAS-SN cycle of GJ 447 are outliers from the power law shown in Figure \ref{fig: A_cyc vs P_cyc}. As discussed in Section \ref{sec: GJ 358 cycle}, however, GJ 358 already appears to be a highly unusual star. GJ 447, on the other hand, does have a second cycle period in Table \ref{tab: optical cycle periods}, although it is not well constrained. This second cycle has an estimated period of $10.6 \pm 6.9$yr, and estimated amplitude of $21.5 \pm 1.0$mmag, which would put it very close to the power law shown in Figure \ref{fig: A_cyc vs P_cyc}, although with a large period uncertainty.

Using chromospheric {Ca\,{\sc ii}} H and K data on FGK stars, \cite{Saar2002} found stronger evidence for multiple branches in the $A_{\rm cyc}$--$P_{\rm cyc}$ relation owing to their larger sample of stars. However, as the amplitudes from \cite{Saar2002} are from chromospheric indicators, they are not directly comparable to our photometric amplitudes. It would be very interesting to see how the photometric cycle amplitudes of F--M type stars compare, however this is left to future work.

In the case of our Sun, the solar cycle does not have a strict 11-yr periodicity; indeed, the $\sim$4 year range of the solar $P_{\textrm{cyc}}$ \citep[][]{Donahue1992} is fully $\pm$18\% of the average value. Moreover, if a given solar cycle has a shorter than average period, then it will usually have a larger than average amplitude, and vice versa \citep[related to the Waldmeier effect;][]{Waldmeier1935}. For the M-type stars analyzed in this work, the observation intervals cover only a handful of cycles in the best case, so it is difficult to determine if these cycles reflect a natural range of $P_{\rm cyc}$ as seen in the Sun. However, from Figure \ref{fig: A_cyc vs P_cyc}, it is clear that for M-type stars which exhibit two apparent cycles, the shorter cycle usually has a greater amplitude; in cases where the shorter $P_{\rm cyc}$ is approximately half the longer $P_{\rm cyc}$, this is consistent with one polarity being stronger than the other (i.e., a Hale, rather than Schawbe, cycle as do Nascimento et al. in preparation, suggest). Alternatively, if as \cite{Olspert2018b} suggest, these double cycles are the result of a single, quasi-periodic cycle, then perhaps this suggests that M dwarfs behave similarly to the Sun - with shorter than average cycles having greater than average amplitudes, and vice versa.

\subsection{\texorpdfstring{P\textsubscript{\rm cyc}}{Pcyc} versus \texorpdfstring{P\textsubscript{\rm rot}}{Prot} and Ro}\label{sec: P_cyc vs P_rot}

Plotting cycle period against rotation period for the well-defined cycles in Table \ref{tab: optical cycle periods}, along with previously detected cycles in FGK stars presented in Table \ref{tab: FGK stars}, shows high scatter, with no strong correlations. However, convective zones are believed to be a key ingredient of stellar cycles, and these are not accounted for in this plot. We therefore place little emphasis on this plot (and choose to not include it here) as it ignores a major physical parameter of current stellar cycle models: Rossby number.

\begin{figure}
    \centering
    \includegraphics[width=\columnwidth]{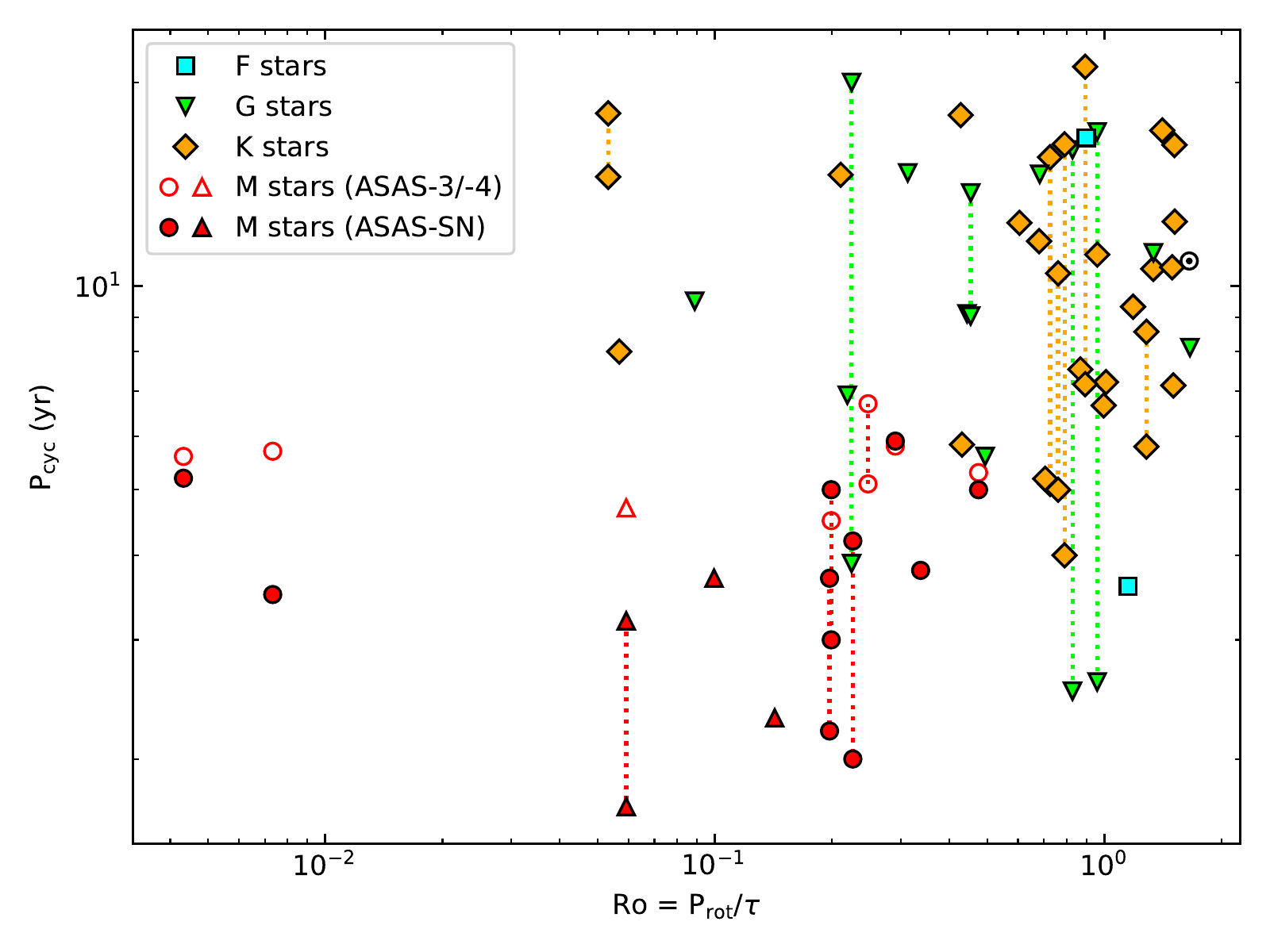}
    \caption{Plot of cycle period (in years) against Rossby number, including FGK type stars from Table \ref{tab: FGK stars}. For the M dwarfs analyzed in this work, filled markers represent cycles inferred from ASAS-SN data; empty markers are from ASAS-3/-4 data; fully convective M stars use circle markers; partially convective M stars use triangle markers. Multiple cycles within the same star (and, for our M dwarfs, found within the same data set) are connected by dotted lines.}
    \label{fig: FGKM P_cyc vs Ro}
\end{figure}

Figure \ref{fig: FGKM P_cyc vs Ro} plots cycle period against Rossby number for the well-defined cycles presented in Table \ref{tab: optical cycle periods}, along with the previously detected cycles in FGK stars presented in Table \ref{tab: FGK stars}. This figure shows that when the convective turnover time is accounted for, in the form of the Rossby number, M-type stars appear to have similar cycles to those of FGK stars at equivalent Ro. However, it is clear that our M dwarfs favor shorter cycle periods, extending the lower cluster of FGK stars to lower $P_{\rm cyc}$. M dwarfs also isolate themselves at low Rossby numbers, though this may be a sampling effect; perhaps M dwarfs extend the upper cluster of FGK stars to lower Ro too.

In Figure \ref{fig: FGKM P_cyc vs Ro}, the longer cycle periods seen in FGK stars likely reflect the fact that these stars have generally been monitored over longer periods than M dwarfs. The evidence for M dwarfs exhibiting shorter $P_{\rm cyc}$ (relative to FGK stars) is strong, but we also found indications of longer, presently unresolvable or poorly constrained cycles in 11 of our M dwarfs. Reproducing these figures with another 10 years of ASAS-SN data may therefore prove interesting.

\subsection{\texorpdfstring{P\textsubscript{cyc}}{Pcyc}/\texorpdfstring{P\textsubscript{\rm rot}}{Prot} versus \texorpdfstring{Ro\textsuperscript{-1}}{Ro-1}}

\begin{figure}
    \centering
    \includegraphics[width=\columnwidth]{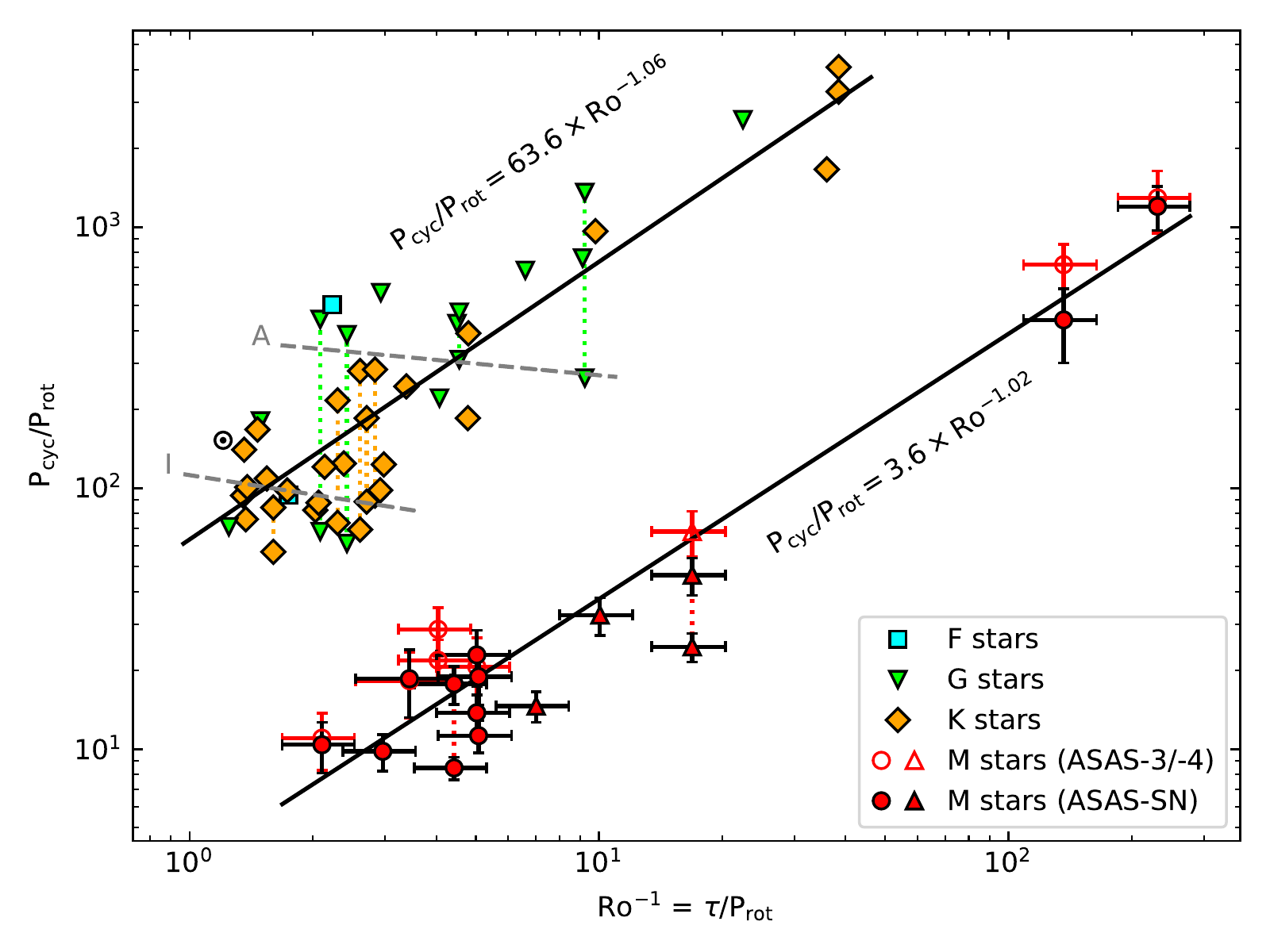}
    \caption{Plot of $P_{\rm cyc}/P_{\rm rot}$ against inverse Rossby number, including FGK type stars from Table \ref{tab: FGK stars}. The dashed grey fits show the A and I branches from \protect\cite{Saar1999} (which have been re-fit using our updated sample), while the black power law fits were found in this work. The global convective turnover time was used to compute Ro for all cycles in this figure. For the FGK stars, since some of our sample did not include uncertainties on their cycle/rotation periods, we weight all FGK cycles equally; for our M dwarfs, we assume a $\pm 20\%$ uncertainty on $\tau$.}
    \label{fig: FGKM P_cyc/P_rot vs Ro}
\end{figure}

Figure \ref{fig: FGKM P_cyc/P_rot vs Ro} shows a  plot of cycle period divided by rotation period against inverse Rossby number for the well-defined cycles presented in Table \ref{tab: optical cycle periods}, along with previously detected cycles in FGK stars presented in Table \ref{tab: FGK stars}.

The power law fit to the M-dwarf data in Figure \ref{fig: FGKM P_cyc/P_rot vs Ro} shows that these data follow the relation:
\begin{equation}
    \frac{P_{\rm cyc}}{P_{\rm rot}} = (3.6 \pm 1.2) \times \textrm{Ro}^{-1.02 \pm 0.06}.
    \label{eq: M-dwarf power law}
\end{equation}
Similarly, Figure \ref{fig: FGKM P_cyc/P_rot vs Ro} shows that FGK stars follow the relation:
\begin{equation}
    \frac{P_{\rm cyc}}{P_{\rm rot}} = (63.6 \pm 1.1) \times \textrm{Ro}^{-1.06 \pm 0.09},
    \label{eq: FGK power law}
\end{equation}
suggesting that $P_{\rm cyc} \propto \tau$ for all stars, in contrast to the results of \cite{Saar1999} - shown by the A and I branches (which we have re-fit using our updated sample).

\cite{Brandenburg1998, Saar1999} interpreted the ratio of $P_{\rm cyc}/P_{\rm rot}$ as a measure of the $\alpha$ effect, based on a simple $\alpha \Omega$ dynamo model proposed by \cite{Robinson1982, Noyes1984}. In this framework, Equations \ref{eq: M-dwarf power law} and \ref{eq: FGK power law} suggest that M dwarfs have weaker $\alpha$ effects than FGK type stars at equivalent Rossby numbers (note the scaling constant in Equation \ref{eq: M-dwarf power law} is 17.7 times smaller than in Equation \ref{eq: FGK power law}). This result could be explained by considering the physical properties of these stars. Since M-type stars have deeper convective zones, and consequently larger turnover times, longer rotation periods are needed to get similar Rossby numbers. Longer rotation periods therefore mean the Coriolis forces acting within these convective zones will be smaller, resulting in less helical turbulent convection. Furthermore, the average convective velocities themselves are lower in these cooler stars; both these properties act to reduce the $\alpha$ effect.

Note, however, that the above explanation would suggest a smooth transition from F to M in Figure \ref{fig: FGKM P_cyc/P_rot vs Ro}. In contrast, this figure shows a gap between FGK and M stars. Perhaps, since M stars are (almost) fully convective, and therefore have (almost) no tachocline, this causes a significant change in the $\alpha$ effect. Alternatively, this gap could simply be the result of our limited sample of stars, which includes two K5s, a K7 but then no stars until M3. Clearly, Figure \ref{fig: FGKM P_cyc/P_rot vs Ro} needs to be extended to cover a wider range of stars, with a wide range of physical properties and rotation rates, to better understand if this gap is real - and if so, and what may be causing it.

\subsection{\texorpdfstring{$\alpha$}{alpha} and \texorpdfstring{$\Omega$}{omega} Effect Scaling}\label{sec: quenching}

The axes of Figure \ref{fig: FGKM P_cyc/P_rot vs Ro} have the benefit of being dimensionless, and Ro in particular is directly related to the mean-field dynamo number, D, as mentioned in Section \ref{sec: turnover time}. Clusters and trends within the data can thus provide insights into the underlying dynamo properties. Indeed, we have already discussed the implications of the scaling constants of Equations \ref{eq: M-dwarf power law} and \ref{eq: FGK power law}; below, we discuss the implications of the power law indices of these equations.

According to most theoretical models \citep[e.g.,][]{Ruediger1993}, the $\alpha$ effect should be quenched with growing magnetic fields, with the reasoning that the stronger a star's magnetic field, the more that cyclonic convection (i.e., the $\alpha$ effect) is suppressed. This paradigm persisted until \cite{Saar1999} found evidence suggesting that the $\alpha$ effect may be \textit{anti-quenched} on their A and I branches. Here, with the benefit of a more focused sample of stars (no binary or evolved stars), we reinvestigate these claims.

Using a 2-D non-linear dynamo model, \cite{Tobias98} found:
\begin{equation}
    P_{\rm cyc} \propto D^\gamma,
    \label{eq: A}
\end{equation}
where D is the dynamo number, and $-0.67 \leq \gamma \leq -0.38$. To explore the role of quenching in a simpler, linear 1-D dynamo model, \cite{Saar1999} defined a quenching index, $q$, such that:
\begin{equation}
    D \propto \alpha \Omega' \propto {\rm Ro}^{-q-2},
    \label{eq: A2}
\end{equation}
where $\Omega'$ is the radial DR and: 
\begin{equation}
    q = q_{\alpha} + q_{\Omega} - 2,
    \label{eq: quenching indices}
\end{equation}
where $q_{\alpha}$ is the quenching index of the $\alpha$ effect (i.e., $\alpha \propto \Omega^{q_\alpha}$, with $\Omega \equiv \frac{2 \pi}{P_{\rm rot}}$), and $q_{\Omega}$ is the quenching index of the $\Omega$ effect (i.e., $\Omega' \propto \Omega^{q_\Omega}$). From solution of the standard dynamo equations (assuming a fixed wavenumber, $k$), \cite{Brandenburg1998} noted that the cycle frequency was given by:
\begin{equation}
    \omega_{\rm cyc} \equiv \frac{2\pi}{P_{\rm cyc}} \propto \Big( \frac{\alpha \Omega' k L}{2} \Big)^{\frac{1}{2}},
    \label{eq: B}
\end{equation}
where $L$ is the dynamo length scale. Combining Equations \ref{eq: A2} and \ref{eq: B}, we have:
\begin{equation}
    P_{\rm cyc} \propto (\alpha \Omega')^{-\frac{1}{2}} \propto {\rm Ro}^{\frac{q}{2}+1},
    \label{eq: C}
\end{equation}
or: 
\begin{equation}
    \frac{P_{\rm cyc}}{P_{\rm rot}} \propto {\rm Ro}^{\frac{q}{2}}.
\end{equation}
Since Figure \ref{fig: FGKM P_cyc/P_rot vs Ro} suggests that the relationship between $P_{\rm cyc}/P_{\rm rot}$-Ro can be described by a power law, where the power law index can take on various values, we can thus write:
\begin{equation}
    {\rm Ro}^{\frac{q}{2}} \propto {\rm Ro}^{\delta},
    \label{eq: delta}
\end{equation}
where $\delta$ is the power law index (i.e., $\delta = - 1.04$ from Equation \ref{eq: M-dwarf power law}). Thus we find (combining Equations \ref{eq: quenching indices} and \ref{eq: delta}):
\begin{equation}
    q_\alpha = 2 \delta - q_\Omega + 2.
\end{equation}
With this equation, we can infer the scaling of the $\alpha$ effect using the power laws plotted in Figure \ref{fig: FGKM P_cyc/P_rot vs Ro}. First, however, we must estimate $q_\Omega$ (the quenching index for radial DR).

DR measurements are difficult, and fraught with misidentifications \citep[][]{Aigrain2015, Basri2018}, but since the good evidence suggests that surface DR scales linearly with rotation period for slower rotators \citep[e.g.,][]{Saar2011}, it is reasonable to assume $q_{\Omega} \approx 1$; for saturated activity stars (discussed in Section \ref{sec: unified theory}), the dependence reverses, and $q_{\Omega} \approx -1.7$ \citep{Saar2011}. We note that we are implicitly assuming that radial DR scales linearly with the measured surface DR.\footnote{Note, also, some observers \citep[e.g.,][]{Barnes2005} argue DR has little/no dependence on rotation period, so our assumed $q_\Omega$ may be incorrect.} With this assumption, and using these $q_\Omega$, we thus have:
\begin{equation}
    q_\alpha = 2 \delta + 1,
\end{equation}
in the unsaturated regime, and:
\begin{equation}
    q_{\alpha} = 2 \delta + 3.7,
\end{equation}
in the saturated regime. Using the power law indices from Equations \ref{eq: M-dwarf power law} and \ref{eq: FGK power law}, we find: $q_{\alpha, M} = -1.03$ and $q_{\alpha, FGK} = -1.13$ in the unsaturated regime, and $q_{\alpha, M} = 1.67$ and $q_{\alpha, FGK} = 1.57$ in the saturated regime. This suggests that the $\alpha$ effect is quenched with faster rotation in the unsaturated regime (but \textit{anti-quenched} in the saturated regime) similarly for FKG and M stars.

As mentioned previously, quenching of the $\alpha$ effect (in the unsaturated regime) is an expected result. In contrast, the $\alpha$ effect \textit{anti-quenching} that we find in the saturated regime is surprising - though anti-quenching is supported by some MHD simulations \citep[e.g.,][]{Chatterjee2011}. We note that our findings in the unsaturated regime disagree with the results of \cite{Saar1999}, whose limited sample of stars (and different assumptions about $\Omega$ effect scaling) led to the opposite conclusion for their A and I branches (note that the fit in Figure \ref{fig: FGKM P_cyc/P_rot vs Ro} is more consistent with their tentative transitional branch for faster rotators; cf. \citealt{Lehtinen2016}). That said, it should be noted that these results are based on a very simple dynamo model \citep{Brandenburg1998}, along with some debated assumptions about DR and $\alpha$ and $\Omega$ effect quenching \citep[e.g.,][]{Barnes2005, Saar2011}.

\subsection{Refining Our Selection of Convective Turnover Times}\label{sec: tau_L I branch}

\begin{figure}
    \centering
    \includegraphics[width=\columnwidth]{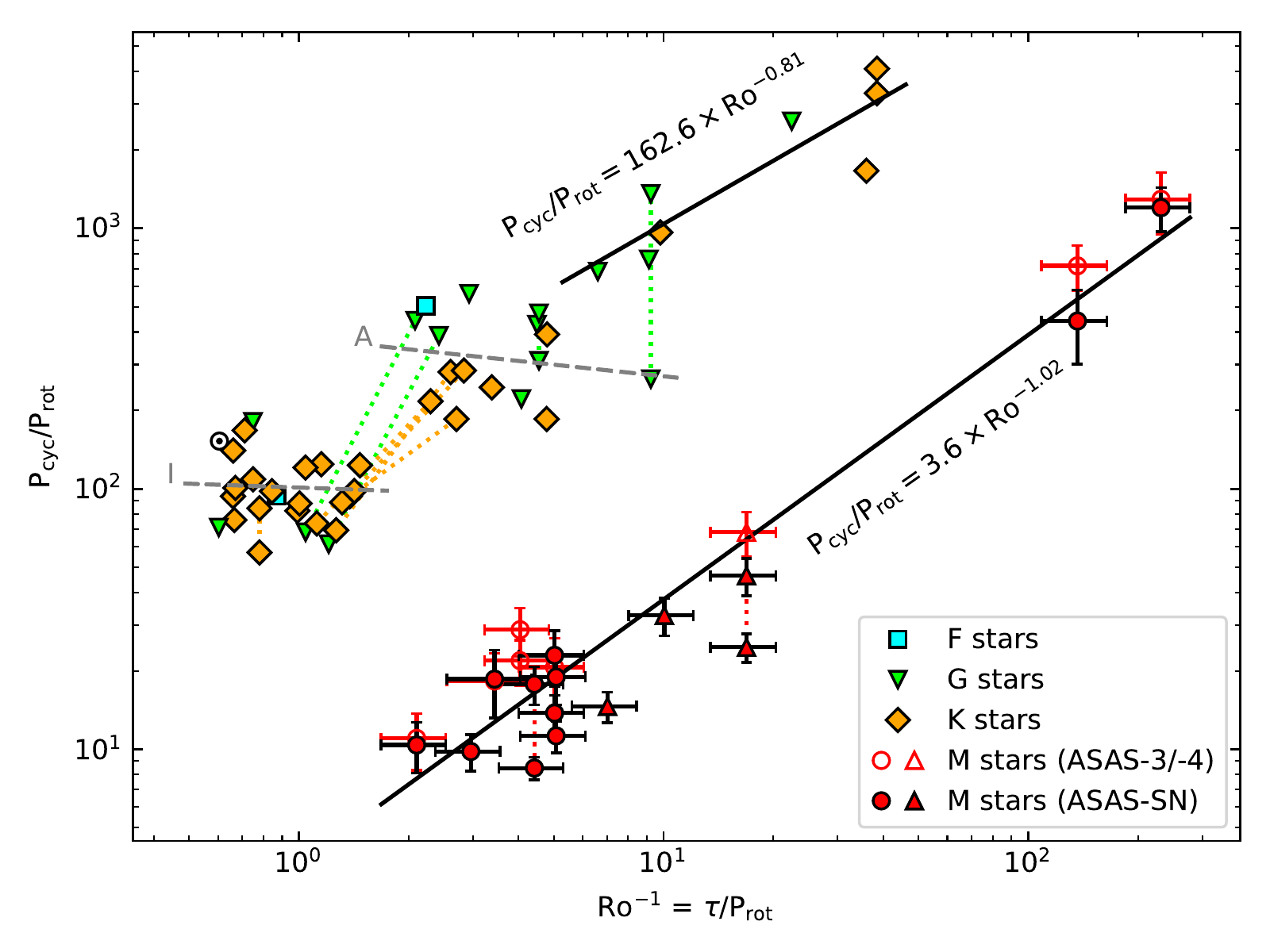}
    \caption{Similar to Figure \ref{fig: FGKM P_cyc/P_rot vs Ro} but with convective turnover times for cycles on the I branch calculated at the base of the convective envelope (i.e., using $\tau_L$ instead of $\tau_G$). The A and I branches were also ignored when fitting a power law to the FGK stars.}
    \label{fig: FGKM P_cyc/P_rot vs Ro tau_L I branch}
\end{figure}

The clustering of stars at the end of the FGK sequence in Figure \ref{fig: FGKM P_cyc/P_rot vs Ro} (particularly around the putative I branch), together with the spread of the sequence in Ro, suggest that perhaps the concept of A and I branches is not completely without merit. If we consider possible A and I branches, though, we must note that we have not been self-consistent in our use of $\tau$. If the I branch is dominated by a tachocline dynamo (as assumed for the Sun), a {\it local} $\tau$, $\tau_L$, specifically computed {\it for that location}, is more suitable. Further, a fit to all the FGK stars together becomes inappropriate; the faster rotators (using $\tau_G$, as before) and the slower, I branch stars (using $\tau_L$) should be treated separately. Using this distinction, we present the results in Figure \ref{fig: FGKM P_cyc/P_rot vs Ro tau_L I branch}.

Due to reasons which will become clear in Section \ref{sec: unified theory}, for stars with cycles on both the A and I branches, we deem the I branch cycle to be the ``primary" and the A branch cycle the ``secondary." In a preliminary version of Figure \ref{fig: FGKM P_cyc/P_rot vs Ro tau_L I branch}, secondary cycles were plotted also using $\tau_L$, but we noted this caused larger scatter in the ``A branch" fit ($\sigma = 1.07$ in log space). If we use $\tau_G$ for these secondary cycles, effectively {\it assuming they arise from a convection zone dynamo still in operation}, the ``A branch" scatter is reduced by $\sim$20\% ($\sigma$= 0.81; Figure \ref{fig: FGKM P_cyc/P_rot vs Ro tau_L I branch}). These new values modify Equation \ref{eq: FGK power law} to be:
\begin{equation}
    \frac{P_{\rm cyc}}{P_{\rm rot}} = (162.6 \pm 0.2) \times \textrm{Ro}^{-0.81 \pm 0.17},
    \label{eq: FGK power law no A or I branch}
\end{equation}
but do little to change the spirit of Section \ref{sec: quenching}'s discussion. A list of different model parameters, their goodness-of-fit metrics, and their implied $q_{\alpha}$ values is provided in Table \ref{tab: P_cyc/Prot-Ro model params}.

\begin{table*}
    \centering
    \caption{Model Parameters for $P_{\rm cyc}/P_{\rm rot} = {\rm a} \times {\rm Ro}^{- \rm b}$ Models.}
    \begin{tabular}{ccccccccccc}
        \hline
        \hline
        \multicolumn{2}{c}{Stars} & $\tau$ & a & $\sigma_{\rm a}$ & b & $\sigma_{\rm b}$ & RMSE & \multicolumn{2}{c}{$q_{\alpha}$} & Used in \\
        Included & Excluded & & & & & & & Unsat. & Sat. & Figures \\
        \hline
        M & \nodata & G & 3.6 & 1.2 & 1.02 & 0.06 & 0.17 & -1.03 & 1.67 & \ref{fig: FGKM P_cyc/P_rot vs Ro},\ref{fig: FGKM P_cyc/P_rot vs Ro tau_L I branch},\ref{fig: dynamo evolution} \\
        FGK & \nodata & G & 63.6 & 1.1 & 1.06 & 0.09 & 0.24 & -1.13 & 1.57 & \ref{fig: FGKM P_cyc/P_rot vs Ro} \\
        I branch & \nodata & G & 112.1 & 1.2 & -0.25 & 0.21 & 0.13 & 1.49 & \nodata & \ref{fig: FGKM P_cyc/P_rot vs Ro} \\
        I branch & \nodata & L & 101.1 & 1.1 & -0.05 & 0.24 & 0.15 & 1.10 & \nodata & \ref{fig: FGKM P_cyc/P_rot vs Ro tau_L I branch},\ref{fig: dynamo evolution} \\
        A branch & \nodata & G & 380.3 & 1.4 & -0.15 & 0.24 & 0.15 & 1.30 & \nodata & \ref{fig: FGKM P_cyc/P_rot vs Ro},\ref{fig: FGKM P_cyc/P_rot vs Ro tau_L I branch},\ref{fig: dynamo evolution} \\
        A branch & \nodata & G,L$^{\rm a}$ & 386.3 & 1.2 & -0.12 & 0.14 & 0.14 & 1.25 & \nodata & \nodata \\
        FKG & I branch & G & 124.0 & 1.2 & 0.83 & 0.11 & 0.20 & -0.67 & 2.03 & \nodata \\
        FKG & A and I branches & G & 162.6 & 1.6 & 0.81 & 0.17 & 0.12 & -0.61 & 2.09 & \ref{fig: FGKM P_cyc/P_rot vs Ro tau_L I branch},\ref{fig: dynamo evolution} \\
        \hline
    \end{tabular}
    \label{tab: P_cyc/Prot-Ro model params}
    \\[0pt]
    \justifying
    \tablecomments{$^{\rm a}$ for stars with two cycles, $\tau_{L}$ was used for both if either was on the I branch. ``$\tau$" column denotes $\tau$ used when estimating convective turnover time (G = global, L = local); ``RMSE" column denotes root mean square error of model (in log space); ``Unsat." column gives the $q_\alpha$ value of the model in the unsaturated regime; ``Sat." column gives the $q_\alpha$ value of the model in the saturated regime.}
\end{table*}

\subsubsection{GJ 358}\label{sec: GJ 358 dual dynamo}

If, as we propose in Section \ref{sec: tau_L I branch}, F--K stars can have convective zone and tachocline dynamos operating simultaneously, then, for consistency, we should also consider this possibility for our partially convective M stars; we note, once again, the exceptional cyclic behavior of GJ 358. GJ 358 is one of (likely) four partially convective M stars in our sample, and is the only such star to show strong evidence for multiple concurrent cycles. Giving this star the same treatment as the F--K stars from the previous section, we might conclude that these cycles are from different dynamos. However, from a sample size of one, the division between the ``A" and ``I branches" for M stars is somewhat unclear.

Assuming GJ 358's cyclic behavior is the result of separate tachocline and convective zone dynamos, then its changing (notably, decreasing) cycle period makes some sense. Since these two dynamos are (in our simple scenario) independent, then the cyclic behavior that we see is actually a superposition of two independent cycles; if these cycles have different (quasi)periods,\footnote{Note: we know of no theoretical or observational motivation for these cycles to have the same period.} then observing a ``cycle with a changing period" is not totally surprising.

Using the I branch and M dwarf fits from Figure \ref{fig: FGKM P_cyc/P_rot vs Ro tau_L I branch}, we can infer theoretical cycle periods for GJ 358's tachocline and convective zone dynamos. Starting with the tachocline dynamo (i.e., the I branch), we find $P_{\rm cyc} = P_{\rm rot} \times 101.1 \times {\rm Ro}^{0.05}$, which, when plugging in the values (using $\tau_L$ to compute Ro), gives $P_{\rm cyc} = 6.5{\rm yr}$. This is $2 \sigma$ above the $4.7 \pm 0.9$yr period we infer from GJ 358's ASAS-3 data, which is the longest period we were able to constrain for this star, and shifts it left in Figure \ref{fig: FGKM P_cyc/P_rot vs Ro tau_L I branch} to Ro$^{-1} \approx 4$ (extending the I branch to larger Ro$^{-1}$). Now considering the convective zone dynamo, using the fit to all M stars, we find $P_{\rm cyc} = P_{\rm rot} \times 3.6 \times {\rm Ro}^{-1.02}$, which gives $P_{\rm cyc} = 4.4 {\rm yr}$. This is consistent with the aforementioned ASAS-3 cycle period, and with the same Ro as shown in Figure \ref{fig: FGKM P_cyc/P_rot vs Ro tau_L I branch}. Perhaps these two dynamos are not totally independent, and interactions between the two cause more significant deviations from true periodic behavior than is typically observed in other stars, resulting in the decreasing cycle period we observe in Figure \ref{fig: GJ 358 GP fit}.

\subsection{A New Interpretation of Dynamo Evolution}\label{sec: unified theory}

Indeed, our results here and those of \cite{Saar1999} may not necessarily be at odds; perhaps we are seeing a switch in dominance between two different types of dynamos with different properties. We largely set aside discussion of the A branch here, since more recent work casts doubt on its robustness/reality \citep[e.g.,][]{Saikai2018}. We also skip discussion of the ``superactive'' (S) branch, which is occupied almost solely by binaries where tidal rotational locking may alter the dynamo fundamentally and place it beyond the scope of the presumed single star dynamos considered here.

We propose the following scenario. The most rapid rotators seem to have nearly solid-body rotation, for example GJ 1243 (M4, $P_{\rm rot}=0.59$d, $k/k_\odot = (\Delta \Omega/\Omega)/(\Delta \Omega_\odot/\Omega_\odot) \approx 0.01$; \citealt{Davenport2020}), LQ Hya (K2, $P_{\rm rot}=1.597$d, $k/k_\odot = 0.028$; \citealt{Kovari2004}), or V889 Her (G2, $P_{\rm rot}=1.337$d, $k/k_\odot = 0.045$; \citealt{Kovari2011}). With so little DR, their dynamos are likely a turbulent $\alpha^2$ type or at best $\alpha^2\Omega$, with the dynamo extant throughout the convection zone. This concept is consistent with models \citep[e.g.,][]{Guerrero2019}, and some detailed MHD models get roughly the same power law index for $P_{\rm cyc}/P_{\rm rot} \propto $ Ro$^{\delta}$: for example, \cite{Warnecke2018} obtain $-0.89 \pm 0.04$, and if one does a least-squares fit of the five long cycle models in \cite{Brun2022} (which are clearly not on the I branch) using their ``stellar" Ro, the slope is $-0.90 \pm 0.47$. These both compare well with our values of $-1.06 \pm 0.09$ and $-0.81 \pm 0.17$.

\begin{figure}
    \centering
    \includegraphics[width=\columnwidth]{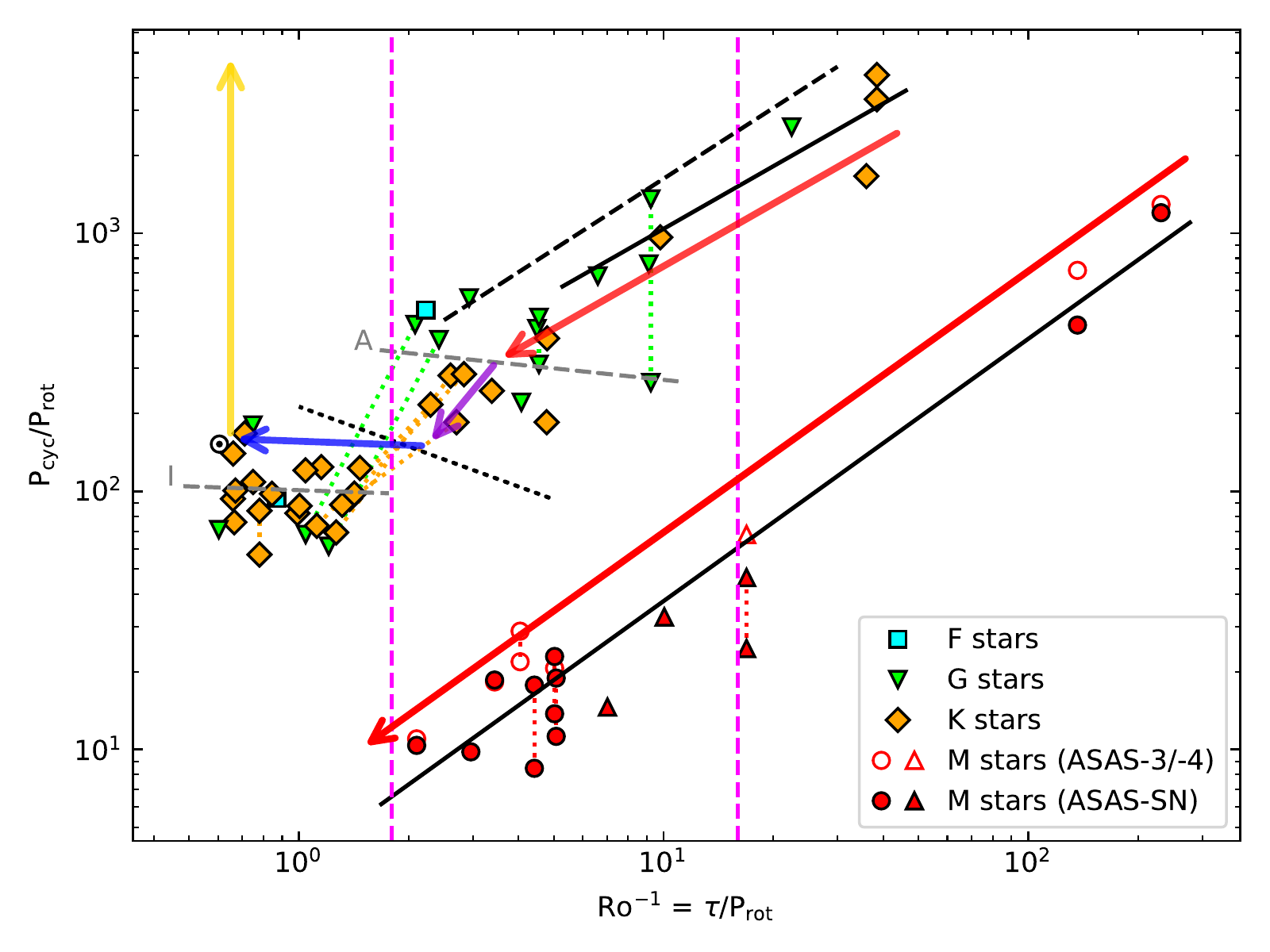}
    \caption{Symbols and gray lines as in Figure \ref{fig: FGKM P_cyc/P_rot vs Ro tau_L I branch}, with sketch of proposed dynamo evolution and MHD model trends overlayed. Evolution begins on the ZAMS at high Ro$^{-1}$ in the upper right. Both FGK and M stars begin with $\alpha$ anti-quenching $\alpha^2$ or $\alpha^2\Omega$ dynamos and initially low but increasing DR (red arrows). Core/envelope decoupling in FGK stars (possibly at at a peak in DR and the end of activity saturation; right magenta dashed line) starts a phase of $\alpha$ quenching, and the beginning of the formation of a tachocline, and likely an $\alpha$ anti-quenched, tachocline based $\alpha\Omega$ dynamo as well. This dynamo does not dominate, however, until the star reaches the I branch (blue), though there may be a regime of mixed dynamo dominance first (purple), around the time that large-scale poloidal fields become dominant (left magenta dashed line). After further evolution on the I branch, eventually the star slows to the point that the dynamo sputters, Maunder-like magnetic minima appear, $P_{\rm cyc}$ lengthens, and in the end the cycle fails entirely, rotational evolution ceases and the star evolves vertically off the diagram (gold). In M stars, once activity saturation ends (right magenta dashed line), an $\alpha$-quenched $\alpha^2$ or $\alpha^2\Omega$ dynamo takes over. 3D MHD models generally agree with the trends found here for the Ms and fast rotating FGKs \citep[e.g.,][dashed black]{Brun2022} and the I branch \citep[][dotted black]{Warnecke2018}. See text for more details.}
    \label{fig: dynamo evolution}
\end{figure}

We therefore identify the fits found here as representing the initial, more turbulent, $\alpha$-anti-quenched, full convection zone dynamos. Stars on the ZAMS start with rapid rotation, low DR and saturated magnetic activity, likely with predominantly poloidal large scale field structure \citep[][]{Donati2009}. As stars lose angular momentum, they evolve from upper right to lower left in Figure \ref{fig: FGKM P_cyc/P_rot vs Ro}, with $P_{\rm cyc}/P_{\rm rot}$ decreasing, and (if \citealt{Saar2011} is correct) increasing DR (sketched in Figure \ref{fig: dynamo evolution}; red arrows). DR increases to a maximum point, at which point saturated magnetic activity stops, the $\alpha$ effect starts becoming quenched, and further spindown leads to activity reduction along the well-known rotation-activity relations.  

(As a brief aside, we note that an interesting alternative to the full-convection zone dynamo, particularly in M stars, is the idea of dynamo cycles driven by thermal fluctuations \citep[e.g.,][]{Nigro2022}. This could potentially explain the lack of rotational dependence in M stars, as the distribution of thermal fluctuations would primarily depend on mass/$T_{\rm eff}$, echoed in the trend seen here of $P_{\rm cyc} \propto \tau$.  
The similar trend in faster rotating FGK stars could also be influenced, at least in part, 
by such a thermally driven dynamo. More work on these models would be useful.)

At some point in partially convective stars, the radiative core and convective envelope decouple \citep[e.g.,][]{MacGregor1991}, allowing a region of strong shear to form at the interface (the tachocline). The Sun and slower FGK rotators are thought to have $\alpha\Omega$ dynamos with the primary generation region in this strong shear layer. M stars, lacking significant (or any) tachoclines, cannot make this transition. 

However, once the tachocline dynamo dominates in older FGK stars, which happens at least by the time the stars reach the I branch around Ro$^{-1}$ = 1 to 3, further cycle evolution proceeds along the I branch, with $P_{\rm cyc}/P_{\rm rot}$ slightly increasing, or maybe remaining roughly constant (shown by a blue arrow in Figure \ref{fig: dynamo evolution}). Perhaps there is a transitional phase of mixed dynamo dominance before this, explaining the diffuse A ``branch" and abundance of stars with multiple cycles in this region. The recently noted ``bend" in the chromospheric rotation-activity relation \citep{Lehtinen2021} may be another reflection of this change. ZDI shows that large scale toroidal fields already dominate well before the I branch is reached, at around Ro$^{-1} \approx 15$ \citep[on the $\tau_C$ scale used here, converted from][]{Donati2009}, about where DR peaks and activity saturation ends (shown by the right magenta dashed line in Figure \ref{fig: dynamo evolution}). Perhaps this is where core/envelope decoupling also takes place. Large scale toroidal field dominance only ends at Ro$^{-1} \approx 2$Ro$^{-1}_\odot \approx 1$ \citep[][]{Donati2009} - i.e., at about the time the star first reaches the I branch. Note that no further stars are seen along the $-1.06$ slope in Figure \ref{fig: FGKM P_cyc/P_rot vs Ro} among the FGK stars once the I branch is reached. We view this as supporting the final dominance of the tachocline dynamo.

Following the arguments using the same simple dynamo model, \cite{Saar1999} suggest the change in slope at the I branch is due to $\alpha$ anti-quenching, which implies (surprisingly) that here $\alpha$ \textit{increases} with the magnetic field ($q_\alpha$ = 1.10 on the I branch in Figure \ref{fig: FGKM P_cyc/P_rot vs Ro tau_L I branch}; Table \ref{tab: P_cyc/Prot-Ro model params}). This might be due to $\alpha$ in this regime being driven by magnetic buoyancy and pumping effects \citep[][]{Chatterjee2011}. Here again, some detailed MHD models are in reasonably good agreement: a power law fit to the results of models M1--4 of \cite{Warnecke2018} (their slow-moderate rotators, ignoring the very slowest) yields $P_{\rm cyc}/P_{\rm rot} = 213 $Ro$^{0.514}$, which lies between the A and I branches and is of similar slope (shown by a dotted black line in Figure \ref{fig: dynamo evolution}). As noted above, we further suggest that some secondary cycles are remnant convection zone dynamos still functioning as the tachocline dynamo ramps up to dominance. This dual $P_{\rm cyc}$ - dual dynamo transitional regime corresponds to the purple transitional epoch arrow in the dynamo evolution in Figure \ref{fig: dynamo evolution}.

In the end, though, after further evolution along the I branch, below a critical Ro$^{-1}$, the dynamo may begin to sputter \citep[][]{vanSaders2016}, the star begins to experience magnetic grand minima \citep[][]{Metcalfe2016}, differential rotation weakens \citep[][]{Metcalfe2022} or may even reverse sign \citep[][]{Gastine2014,Karak2020}, and the dynamo dies, with $P_{\rm
cyc}$ going to infinity \citep[see, e.g., the lengthening $P_{\rm cyc}$ in the evolutionary sequence of G2 stars 18 Sco, the Sun, and $\alpha$ Cen A][]{Judge2017}. With the cycling dynamo shut down, large-scale fields driving spin-down vanish, stopping further rotational evolution, and the star evolves vertically off the diagram at constant Ro \citep[e.g.,][shown by a vertical gold arrow in Figure \ref{fig: dynamo evolution}]{Metcalfe2017}.

\section{Conclusions}\label{sec: conclusions}

We have found evidence for magnetic activity cycles in at least 12 M-type stars, and traces of cyclic behavior in three more, using photometric time series. Of these 15 M dwarfs, approximately 12 are type M3.5 or later and likely fully convective. For these cycles, we investigated the correlation between cycle amplitude and period, and found that this is reasonably well described by a power law with an exponent of $0.94 \pm 0.11$. We also found indications of multiple branches in this relation, potentially in agreement with the findings of \cite{Saar2002}. However, the unclear nature of these additional ``cycles" (see Section \ref{sec: false cycles}) makes the latter less certain.

We further investigated correlations between cycle and rotation period, cycle period and Rossby number, and the ratio of cycle to rotation period and inverse Rossby number. Our results suggest that the dynamo processes acting within M dwarfs may not be as different from those acting in FGK type stars (at least the more rapidly rotating ones with Ro$^{-1} < 3$) as some previous models \citep[e.g.,][]{Kitchatinov1999} suggest. Some more recent models \citep[e.g.,][shown in Figure \ref{fig: dynamo evolution}]{Warnecke2018,Brun2022} appear to capture the overall trend for faster rotating FGK stars in Figure \ref{fig: FGKM P_cyc/P_rot vs Ro tau_L I branch}; if similar models in fully convective M stars show similar trends, theory and observations may be in reasonable agreement.

Using Rossby number to parameterize magnetic activity, and a simple dynamo model \citep{Saar1999}, we find, at equivalent Rossby numbers, the $\alpha$ effect is similarly quenched with rotation ($q_\alpha \approx -1$ in the unsaturated regime), but is of reduced amplitude in M-dwarfs compared to that of FGK stars. These findings are in agreement with some models \citep[e.g.,][]{Ruediger1993}, however are in apparent disagreement with the results of \cite{Saar1999}, whose limited sample suggested a contrary relation (at least for their A and I branches). However, we propose that what we are actually seeing is the result of one type of dynamo (an $\alpha$-quenched $\alpha^2$ or $\alpha^2\Omega$ full convection zone dynamo), present in all fast rotators (a thermally driven dynamo is an alternative possibility). In FGK stars, once they have spun down sufficiently, a tachocline-based $\alpha$ anti-quenched $\alpha\Omega$ dynamo takes over \citep[forming the I branch of][]{Brandenburg1998}. Fully convective M stars, lacking tachoclines, spin down retaining their full convective zone dynamos. We also find that the $\alpha$ effect is again anti-quenched in the saturated regime, which some theoretical models support under certain conditions \citep[e.g.,][]{Chatterjee2011}.

Clearly, stellar cycles and the underlying physical mechanisms are not yet fully understood. To follow up this work, we suggest these stars be monitored using ZDI, which can be used to determine large-scale magnetic field topology and polarity, and in the UV or X-rays, which are much more sensitive and directly related to magnetic activity. These follow-up observations would reveal whether the cycles presented in Table \ref{tab: optical cycle periods} are a result of global magnetic activity, or some other phenomena (i.e. Rossby waves, etc.). It will be particularly useful to search for more cycles in K7-M2 stars, which are lacking in our sample. These cycles may reveal whether the $\alpha$ effect transition from FGK to fully convective M stars (Figure \ref{fig: FGKM P_cyc/P_rot vs Ro}) is sharp or gradual. It is also intriguing that $\delta$ (i.e, the power law slopes) appears to remain constant through the joint $\alpha$ and DR transition when activity saturation ends (Section \ref{sec: unified theory}); we propose that the implications of this for the underlying dynamo should be explored.

\section*{Acknowledgments}

This paper includes data collected by the TESS mission, which are publicly available from the Mikulski Archive for Space Telescopes (MAST). Funding for the TESS mission is provided by NASA’s Science Mission directorate. We gratefully acknowledge use of data provided by the ASAS and ASAS-SN collaborations.

Z.A.I thanks the University of Southampton's Center for Astrophysics $|$ Harvard \& Smithsonian exchange program, with special thanks to program coordinators Diego Altamirano and Jeremy Drake. S.H.S. gratefully acknowledges support from NASA Heliophysics LWS grant NNX16AB79G, NASA XRP grant 80NSSC21K0607, and NASA EPRV grant 80NSSC21K1037. We thank Axel Brandenburg for helpful discussions. This work was also supported by NASA's Swift Guest Investigator program under Grant 80NSSC20K1111, and B.J.W was supported by NASA contract NAS8-03060 to the Chandra X-Ray Center. We thank the anonymous referee for many helpful suggestions which have improved the paper.

This research also made use of Lightkurve, a Python package for Kepler and TESS data analysis \citep{Lightkurve2018}; NumPy \citep{Harris2020}; Astropy,\footnote{\url{http://www.astropy.org}} a community-developed core Python package for Astronomy \citep{astropy2013, astropy2018}; SciPy \citep{SciPy2020}; Matplotlib, a 2D graphics environment \citep{Hunter2007}.

\section*{Data Availability}

Data analyzed herein are publicly available via the ASAS All-Star Catalog,\footnote{\url{http://www.astrouw.edu.pl/asas}} the ASAS-SN light curve server,\footnote{\url{https://asas-sn.osu.edu}} and the Mikulski Space Archive for Space Telescopes (MAST): \dataset[10.17909/xdwe-q663]{http://dx.doi.org/10.17909/xdwe-q663}. ASAS-4 data on Proxima were provided by G.~Pojma{\'n}ski.


\bibliography{bibliography}
\bibliographystyle{aasjournal}



\end{document}